\documentclass[a4paper,11pt]{article}
\pdfoutput=1 

\usepackage{jheppub} 
\usepackage{tikz}
\usepackage{dsfont}
\usepackage{extarrows}
\usepackage{hyperref}
\usepackage{cleveref}
\usepackage{supertabular}
\usepackage{todonotes}
\usepackage{mathrsfs}
\usepackage{array}
\usepackage{lipsum}
\usepackage{breqn}
\usepackage{physics}
\usepackage{tikz-cd}
\usepackage{slashed}
\usetikzlibrary{decorations.markings}
\usepackage{bbm}
\newcommand{\ii}{\mathbbm{i}}

\tikzset{
  line/.style={line width=0.25mm},
  curve/.style={line,smooth,tension=1},
  ->-/.style={
    decoration={
      markings,
      mark=at position #1 with {\arrow[>=stealth]{>}}
    },
    postaction={decorate}
  },
  -<-/.style={
    decoration={
      markings,
      mark=at position #1 with {\arrow[>=stealth]{<}}
    },
    postaction={decorate}
  },
  -tri-/.style={
    decoration={
      markings,
      mark=at position #1 with {
        \fill[blue] (0:4.5pt) -- (140:4.5pt) -- (220:4.5pt) -- cycle;
      }
    },
    postaction={decorate}
  }
}

\usepackage{helvet} 
\usepackage{amsmath}
\usepackage{amssymb} 

\newcommand{\newreptheorem}[2]{%
\newenvironment{rep#1}[1]{%
 \def\rep@title{#2 \ref{##1}}%
 \begin{rep@theorem}}%
 {\end{rep@theorem}}}
\makeatother

\newreptheorem{lemma}{Lemma}

\newreptheorem{conj}{Conjecture}

\usepackage{lscape} 
\usepackage{braket}

\usepackage[T1]{fontenc} 

\newcommand{\be}{\begin{equation}}
\newcommand{\ee}{\end{equation}}
\newcommand{\ba}{\begin{aligned}}
\newcommand{\ea}{\end{aligned}}
\newcommand{\bea}{\begin{eqnarray}}
\newcommand{\eea}{\end{eqnarray}}

\def\bp{\begin{pmatrix}}
\def\ep{\end{pmatrix}}

\preprint{USTC-ICTS/PCFT-26-57}

\title{\boldmath On general background of quantum non-invertible symmetry in 2D}

\author{Jin Chen$^{a,b}$}
\author{and Qiang Jia$^{c}$}
\affiliation{$^{a}$Department of Physics, Xiamen University, Xiamen, 361005, China}
\affiliation{$^{b}$Peng Huanwu Center for Fundamental Theory, Hefei, Anhui 230026, China}
\affiliation{$^{c}$Department of Physics, Korea Advanced Institute of Science \& Technology,
Daejeon 34141, Korea}

\abstract{We study the dual quantum symmetry $\mathrm{Rep}(G)$ in the two-dimensional theory $\widetilde{\mathfrak{T}}_{\mathrm{Rep}(G)}$ obtained by gauging a non-abelian symmetry $G$ of $\mathfrak{T}_G$, where for Lie groups $G$ the gauging is understood as flat gauging. We develop a general framework for computing partition functions of $\widetilde{\mathfrak{T}}_{\mathrm{Rep}(G)}$ in arbitrary non-invertible symmetry backgrounds, represented by topological defect networks of $\mathrm{Rep}(G)$, in terms of the partition functions of the original theory $\mathfrak{T}_G$. We also derive the inverse transformation, expressing partition functions of $\mathfrak{T}_G$ in $G$ backgrounds in terms of those of $\widetilde{\mathfrak{T}}_{\mathrm{Rep}(G)}$. We test the framework in several examples, including finite groups with multiplicity-free and higher-multiplicity fusion rules, and discuss a formal extension to compact
Lie groups, focusing on $\mathrm{Rep}(SU(2))$.}

\begin{document}
\maketitle
\flushbottom

\section{Introduction}

Symmetry is a fundamental organizing principle in quantum field theory.  It
constrains the spectrum and correlation functions of a theory, and provides
information about renormalization group flows and possible valuable infrared phases even when the dynamics are strongly coupled.  The modern notion of
generalized symmetry extends this principle from ordinary group actions to
topological operators of various codimensions \cite{Gaiotto:2014kfa}, for comprehensive reviews on generalized and non-invertible symmetries and
their applications, see e.g.~\cite{Chang:2018iay,Thorngren:2019iar,Komargodski:2020mxz,Thorngren:2021yso,Cordova:2022ieu,Apte:2022xtu, Chang:2022hud, Lin:2023uvm,Kaidi:2023maf,Zhang:2023wlu,Damia:2023ses,Cordova:2023bja,Choi:2023pdp,Antinucci:2023ezl,Bhardwaj:2024kvy,Bhardwaj:2024qiv,Cordova:2024ypu,Copetti:2024rqj,DelZotto:2024arv,Nakayama:2024msv,Cordova:2024goh,Cordova:2024nux,Chen:2023qnv,Bhardwaj:2025piv,Seiberg:2025bqy,Antinucci:2025fjp,Chen:2025qub, Zhang:2026gqp}. Coupling a symmetry to background fields then turns these structural constraints into physical observables.  For an ordinary group symmetry, this is achieved by turning on a non-dynamical background gauge field.  Such a coupling has two closely related uses.  First, when the coupling is gauge invariant, summing over the background gauge fields defines a new theory in which the symmetry has been gauged.  Second, keeping the background fixed probes the response of the theory: a failure of background gauge invariance signals a 't Hooft anomaly, which obstructs gauging and must be matched along renormalization group flows.

For a finite symmetry group, a background has no local connection data and is
specified by a flat $G$-bundle, so gauging reduces to a sum over its holonomy
sectors. The resulting finite gauge sector is topological and has no
propagating local degrees of freedom.  For a continuous Lie group, ordinary
dynamical gauging generally also includes connections with nonzero curvature.
One may instead restrict the integral to flat connections. Throughout this
paper we refer to this operation as \emph{flat gauging}.  Continuous orbifolds in two-dimensional CFT can be viewed as realizations of this restriction.  In particular, the twisted sector of the $SO(3)$ continuous orbifold of the self-dual free boson was shown to agree with the non-rational $c=1$ limit theory constructed by Runkel and Watts
\cite{Runkel:2001ng,Gaberdiel:2011aa,Restuccia:2013tba}.  Flat backgrounds in
two dimensions admit a natural description in terms of topological defect
lines.  Ordinary zero-form symmetries correspond to invertible lines, while
allowing non-group-like fusion leads to fusion-category symmetries
\cite{Bhardwaj:2017xup,Chang:2018iay,Thorngren:2021yso}.  Their backgrounds
are networks of topological defect lines (TDLs) equipped with junction morphisms, and a generalized orbifold is obtained by choosing a suitable Frobenius algebra object and condensing its defect network over spacetime.  Different algebra objects, or different algebra structures on the same underlying object, can lead to different orbifold phases and topological interfaces; successive gaugings are naturally organized into an orbifold groupoid \cite{Gaiotto:2020iye,Diatlyk:2023fwf,Chen:2024ulc}.  This language also makes contact with symmetry-protected topological phases in condensed matter systems.  Stacking with an invertible SPT phase changes the topological weights assigned to flat symmetry backgrounds and can therefore modify a subsequent gauging without changing the local degrees of freedom
\cite{Dijkgraaf:1989pz,Chen:2012ctz,Else:2014vma}.  In a defect-network
description these choices enter through junction phases, while an obstruction
to gauging is encoded in the associators and $F$-symbol data
\cite{Kawagoe:2021gqi}.

A canonical example of generalized orbifolding begins with a
two-dimensional theory $\mathfrak{T}_G$ carrying an anomaly-free finite
symmetry group $G$.  Gauging $G$ by the finite-group prescription above
gives the ordinary orbifold
$\widetilde{\mathfrak{T}}_{\text{Rep}(G)}=\mathfrak{T}_G/G$, whose Wilson
lines furnish the dual quantum symmetry $\text{Rep}(G)$
\cite{Dijkgraaf:1989hb,Bhardwaj:2017xup}.  When $G$ is abelian, all
irreducible representations are one-dimensional and this reduces to the
familiar group-like quantum symmetry of an abelian orbifold.  For non-abelian
$G$, $\text{Rep}(G)$ contains lines of quantum dimension larger than one
and is genuinely non-invertible.  Conversely, gauging the regular algebra of
$\text{Rep}(G)$ returns the original theory.  The pair
$G\leftrightarrow\text{Rep}(G)$ therefore provides a controlled setting in
which ordinary flat backgrounds and fusion-category defect networks can be
related explicitly.

The categorical definition of gauging is conceptually complete, but it does
not by itself provide a convenient formula for a partition function in a
specified non-invertible background.  Even on a torus, a resolved
$\text{Rep}(G)$ background is labeled not only by the lines wrapping the two
cycles but also by their fusion channel and, when fusion multiplicities are
present, by a basis at each junction.  Conversely, recovering a particular
flat $G$-background requires more information than the fully gauged sum.  A
useful transform must therefore remember representation indices, junction
gauges, multiplicity labels, and the simultaneous-conjugacy relation on a
commuting pair of holonomies.

Recently there have been many progress on these questions made in the study of generalized
orbifold partition functions and their modular properties
\cite{Perez-Lona:2023djo,Perez-Lona:2024sds}.  In later work,
an explicit procedure was developed for the quantum
$\text{Rep}(G)$ symmetry of non-abelian orbifolds
\cite{Banerjee:2026mnx}.  Their multiplicity-free examples begin with an
ansatz for a partial trace containing a single quantum-symmetry line.  Modular
$S$ and $T$ transformations generate further partial traces, and modular
consistency, together with the requirement that gauging the quantum symmetry
returns the original theory, constrains the remaining coefficients.  This is
a powerful bootstrap because it extracts a large set of amplitudes from a
small initial input.  At the same time, it leaves open a complementary
question: can the coefficients be evaluated directly from the Wilson-line
network in each flat $G$-background, without using modular covariance to
propagate an ansatz from one partial trace to another?

In this paper we answer this question by constructing the transform directly
from the representation theory of $G$.  We denote the theory obtained after
gauging by
$\widetilde{\mathfrak{T}}_{\text{Rep}(G)}=\mathfrak{T}_G/G$, and consider a
resolved trivalent $\text{Rep}(G)$ network on the torus.  If $\rho$ and
$\sigma$ label the lines along the two cycles, $\lambda$ labels the
intermediate channel, and $i,j$ label the two junctions, its partition
function takes the form
\begin{equation}
    \widetilde Z_{\rho,\sigma}^{\lambda;i,j}
    =\frac{1}{|G|}
      \sum_{\substack{g,h\in G\\ gh=hg}}
      W_{\rho,\sigma}^{\lambda;i,j}(g,h)\,Z_{g,h}\,.
\end{equation}
Here $Z_{g,h}$ is the partition function of $\mathfrak{T}_G$ in the flat
background with commuting holonomies $(g,h)$.  The kernel $W$ is obtained
by transporting the Wilson lines through this background and contracting the
resulting representation matrices with the chosen intertwiner and
cointertwiner.  In this way the holonomy dependence, the fusion channel, and
all multiplicity indices enter the same local calculation.  No modular
transformation is needed to determine an additional network amplitude once
its junction data have been specified. We also construct the inverse transform from the regular Frobenius algebra
\begin{equation}
    \mathcal{A}=\text{Fun}(G)
    \simeq \bigoplus_{\rho\in\widehat{G}}d_\rho\,\rho
\end{equation}
in $\text{Rep}(G)$.  The ordinary condensation network gives the trivial
$G$-background.  Dressing its multiplication and comultiplication by
commuting right-regular actions isolates instead the simultaneous-conjugacy
orbit of a general pair $(g,h)$.  Expanding these dressed maps in the same
junction bases produces coefficients
$C_{\rho,\sigma}^{\lambda;i,j}(g,h)$ such that
\begin{equation}
    Z_{g,h}
    =\frac{1}{|G|}
      \sum_{\rho,\sigma,\lambda;i,j}
      C_{\rho,\sigma}^{\lambda;i,j}(g,h)\,
      \widetilde Z_{\rho,\sigma}^{\lambda;i,j}\,.
\end{equation}
The forward and inverse constructions therefore have a common microscopic
origin.  The first evaluates a $\text{Rep}(G)$ Wilson network in a flat
$G$-background, whereas the second decorates the regular-algebra network so
as to resolve that background again. Moreover, we show that in a suitable gauge, the inverse kernel can be written as
    \begin{equation}
        C^{\lambda;i,j}_{\rho,\sigma}(x,y) = \frac{d_{\rho}d_{\sigma}}{d_{\lambda}} W^{\lambda;j,i}_{\sigma,\rho}(y,x)\,.
    \end{equation}

For finite groups we work out both transforms in examples of increasing
complexity.  The categories $\text{Rep}(S_3)$, $\text{Rep}(D_4)$, and
$\text{Rep}(Q_8)$ provide multiplicity-free tests and permit a direct
comparison with the modular-bootstrap results of
\cite{Banerjee:2026mnx}.  The example $\text{Rep}(A_4)$, for which
$Z\otimes Z$ contains two copies of $Z$, shows how the construction keeps
the full multiplicity space rather than compressing it into an unresolved
partial trace.  These examples verify that regauging the regular
$\text{Rep}(G)$ algebra recovers the original torus backgrounds, with the
understanding that the resolved defect traces can form a redundant set and
the inverse acts on their physical image.  We finally explore the formal
extension to compact Lie groups through $\text{Rep}(SU(2))$, where the
finite sums are replaced by Peter-Weyl expansions and integrals over flat
connections. 

The paper is organized as follows.  In section~2 we review
$\text{Rep}(G)$ as a fusion category, fix the intertwiner conventions, and
describe generalized gauging by a Frobenius algebra object.  In section~3 we
derive the fixed-cut torus Fourier kernel from Wilson-line transport and then
construct its inverse by dressing the regular Frobenius algebra.  Section~4
applies the construction to the multiplicity-free categories
$\text{Rep}(S_3)$, $\text{Rep}(D_4)$, and $\text{Rep}(Q_8)$.
Section~5 treats $\text{Rep}(A_4)$ and makes the multiplicity indices
explicit.  In section~6 we discuss the formal $\text{Rep}(SU(2))$
extension.

\section{Review of $\text{Rep}(G)$ symmetry and non-invertible gauging}

In this section, we briefly review $\text{Rep}(G)$ for a finite group $G$ as a fusion category, together with the categorical data that will be needed later. We also review the gauging of a non-invertible symmetry described by a fusion category $\mathscr{C}$, formulated as the condensation of a Frobenius algebra object $\mathcal{A}$ in $\mathscr{C}$. We then specialize to the case $\mathscr{C}=\text{Rep}(G)$ and work out the ingredients required for our following analysis. 

\subsection{$\text{Rep}(G)$ as a fusion category}
Throughout this section, we take $G$ to be a finite group, and all representations in $\text{Rep}(G)$ to be finite-dimensional unitary complex representations. We denote by $\widehat{G}\equiv \text{Irr}(G)\subset \text{Rep}(G)$ the set of irreducible representations of $G$. For each $\rho\in\widehat{G}$, let $E_{\rho}$ be the corresponding representation space, with complex dimension $d_{\rho}=\dim E_{\rho}$, and choose an orthonormal basis
\begin{equation}
    e^{\rho}_{\alpha}=\left\{e^{\rho}_{1},e^{\rho}_{2},\cdots,e^{\rho}_{d_{\rho}}\right\}\,,\qquad \alpha=1,\cdots,d_{\rho}\,.
\end{equation}
For any $g\in G$, the representation matrix $\rho(g):E_{\rho}\to E_{\rho}$ is taken to be unitary, so that $\rho(g)^{\dagger}=\rho(g^{-1})$, and its action on the basis vectors is defined to be
\begin{equation}
    \rho(g)\circ e^{\rho}_{\alpha}=\sum_{\beta=1}^{d_{\rho}} e^{\rho}_{\beta}\,\rho(g)_{\beta\alpha}\,.
\end{equation}

The representation category $\text{Rep}(G)$ is a fusion category, whose fusion product is given by the tensor product of representations. In particular, for $\rho,\sigma\in\widehat{G}$ we write
\begin{equation}\label{eq:tensor-product-irr}
    \rho\otimes\sigma=\sum_{\lambda\in\widehat{G}} N_{\rho\sigma}^{\lambda}\,\lambda\,,
\end{equation}
where
\begin{equation}
    N_{\rho\sigma}^{\lambda}=\dim \text{Hom}_{\text{Rep}(G)}(E_{\rho}\otimes E_{\sigma},E_{\lambda})\in\mathbb{Z}_{\geq 0}
\end{equation}
is the fusion coefficient specifying the multiplicity of the irreducible representation $\lambda$ in the decomposition of $\rho\otimes\sigma$. When $N_{\rho\sigma}^{\lambda}=0,1$ for all $\rho,\sigma,\lambda\in\widehat{G}$, each irreducible representation appears at most once in the tensor product decomposition, and we refer to this as the \emph{multiplicity-free} case. In this section, however, we do not restrict to this case, and allow $N_{\rho\sigma}^{\lambda}$ to be an arbitrary non-negative integer. In categorical language, the irreducible representations are the simple objects of $\text{Rep}(G)$. In the corresponding two-dimensional theory $\mathfrak{T}_{\text{Rep}(G)}$, they may also be regarded as simple topological defect lines (TDLs), with $\text{Rep}(G)$ interpreted as the global symmetry category of $\mathfrak{T}_{\text{Rep}(G)}$. When $G$ is non-abelian, the symmetry category $\text{Rep}(G)$ is non-invertible.

The fusion rule \eqref{eq:tensor-product-irr} alone is not sufficient to specify a fusion category, since distinct fusion categories may share the same fusion rules. In the case of $\text{Rep}(G)$, the additional structure is encoded in the intertwiners $I$ and co-intertwiners $I^{\vee}$ associated with tensor products. The co-intertwiner
\begin{equation}
    I_{\lambda;i}^{\vee \rho,\sigma}:E_{\lambda}\to E_{\rho}\otimes E_{\sigma}\,,\qquad i=1,\cdots,N_{\rho\sigma}^{\lambda}\,,
\end{equation}
embeds the irreducible representation $\lambda$ into the $i$-th copy of $\lambda$ appearing in the decomposition of $\rho\otimes\sigma$. It satisfies the intertwining relation
\begin{equation}\label{eq:defining_property_cointertwiner}
    I_{\lambda;i}^{\vee \rho,\sigma}\circ \lambda(g)=\bigl(\rho(g)\otimes \sigma(g)\bigr)\circ I_{\lambda;i}^{\vee \rho,\sigma}\,,\qquad \forall g\in G\,.
\end{equation}
In the chosen bases, this map can be expanded as
\begin{equation}
    I_{\lambda;i}^{\vee \rho,\sigma}\circ e^{\lambda}_{\gamma}
    =\sum_{\alpha,\beta} e^{\rho}_{\alpha}\otimes e^{\sigma}_{\beta}\,
    \mathcal{I}^{\vee(\rho,\alpha),(\sigma,\beta)}_{(\lambda,\gamma);i}\,,
\end{equation}
where the coefficients $\mathcal{I}^{\vee(\rho,\alpha),(\sigma,\beta)}_{(\lambda,\gamma);i}$ form a $(d_{\rho}d_{\sigma})\times d_{\lambda}$ matrix encoding the embedding of the basis vectors of $E_{\lambda}$ into the tensor-product basis of $E_{\rho}\otimes E_{\sigma}$.

On the other hand, the intertwiner $I$ is defined by
\begin{equation}
    I^{\lambda;i}_{\rho,\sigma}:E_{\rho}\otimes E_{\sigma}\to E_{\lambda}\,,\qquad i=1,\cdots,N_{\rho\sigma}^{\lambda}\,,
\end{equation}
and projects the tensor product $\rho\otimes\sigma$ onto the $i$-th copy of the irreducible representation $\lambda$ appearing in its decomposition. It satisfies the intertwining relation
\begin{equation}\label{eq:defining_property_intertwiner}
    I^{\lambda;i}_{\rho,\sigma}\circ \bigl(\rho(g)\otimes \sigma(g)\bigr)=\lambda(g)\circ I^{\lambda;i}_{\rho,\sigma}\,,\qquad \forall g\in G\,.
\end{equation}
Similarly, in the chosen bases we may expand
\begin{equation}
    I^{\lambda;i}_{\rho,\sigma}\circ \bigl(e^{\rho}_{\alpha}\otimes e^{\sigma}_{\beta}\bigr)
    =\sum_{\gamma} e^{\lambda}_{\gamma}\,
    \mathcal{I}^{(\lambda,\gamma);i}_{(\rho,\alpha),(\sigma,\beta)}\,,
\end{equation}
where the coefficients $\mathcal{I}^{(\lambda,\gamma);i}_{(\rho,\alpha),(\sigma,\beta)}$ form the dual $d_{\lambda}\times(d_{\rho}d_{\sigma})$ matrix, encoding the projection of the tensor-product basis vectors of $E_{\rho}\otimes E_{\sigma}$ onto the basis vectors of $E_{\lambda}$.

These maps are represented diagrammatically as
\begin{equation}
    \begin{gathered}
        \begin{tikzpicture}
            \draw[line,thick,->-=0.75] (0,0)--(0,2);
            \draw[line,thick,->-=0.3] (-1.414,-1.414)--(0,0);
            \draw[line,thick,->-=0.3] (1.414,-1.414)--(0,0);
            \filldraw[black] (0,0) circle (2pt);
            \node at (0,-0.4) {$i$};
            \node at (0.25,2) {$\lambda$};
            \node at (-1.1,-1.414) {$\rho$};
            \node at (1.1,-1.414) {$\sigma$};
        \end{tikzpicture}
    \end{gathered}
    \qquad\qquad
    \begin{gathered}
        \begin{tikzpicture}
            \draw[line,thick,-<-=0.75] (0,0)--(0,-2);
            \draw[line,thick,-<-=0.3] (-1.414,1.414)--(0,0);
            \draw[line,thick,-<-=0.3] (1.414,1.414)--(0,0);
            \filldraw[black] (0,0) circle (2pt);
            \node at (0,0.5) {$i^{\vee}$};
            \node at (0.25,-2) {$\lambda$};
            \node at (-1.1,1.414) {$\rho$};
            \node at (1.1,1.414) {$\sigma$};
        \end{tikzpicture}
    \end{gathered}
\end{equation}
where each line is labeled by an element of $\widehat{G}$ and represents the corresponding complex vector space $E$. The junctions are labeled by $i$ and $i^{\vee}$, which distinguish the different channels of the intertwiner $I^{\lambda;i}_{\rho,\sigma}$ and the co-intertwiner $I^{\vee \rho,\sigma}_{\lambda;i}$, respectively, with $i,i^{\vee}=1,\cdots,N_{\rho\sigma}^{\lambda}$.

The fusion category $\text{Rep}(G)$ is associative only up to an isomorphism. For $\rho,\sigma,\lambda\in\widehat{G}$, one has
\begin{equation}
    \left(E_{\rho}\otimes E_{\sigma}\right)\otimes E_{\lambda}
    =\alpha_{\rho,\sigma,\lambda}\circ \Bigl(E_{\rho}\otimes \left(E_{\sigma}\otimes E_{\lambda}\right)\Bigr)\,,
\end{equation}
where
\begin{equation}
    \alpha_{\rho,\sigma,\lambda}:E_{\rho}\otimes \left(E_{\sigma}\otimes E_{\lambda}\right)\to \left(E_{\rho}\otimes E_{\sigma}\right)\otimes E_{\lambda}
\end{equation}
is the associator. Concretely, let $\delta$ be an irreducible representation appearing in the decomposition of $\rho\otimes\sigma\otimes\lambda$. A basis vector $e^{\delta}_{\chi}$ can be expanded as a linear combination of tensor-product basis vectors $e^{\rho}_{\alpha}\otimes e^{\sigma}_{\beta}\otimes e^{\lambda}_{\gamma}$. This expansion depends on how the tensor product is parenthesized, namely on whether one first fuses $\rho$ with $\sigma$ or $\sigma$ with $\lambda$, and on the choice of intermediate channel. The associator $\alpha_{\rho,\sigma,\lambda}$ relates these two decompositions. Diagrammatically, this relation is expressed as
\begin{equation}
    \begin{gathered}
        \begin{tikzpicture}
            \draw[line,thick,-<-=0.55] (0,1)--(0,0);
            \draw[line,thick,-<-=0.55] (0,0)--(-0.707,-0.707);
            \draw[line,thick,-<-=0.55] (-0.707,-0.707)--(-1.414,-1.414);
            \draw[line,thick,-<-=0.55] (0,0)--(1.414,-1.414);
            \draw[line,thick,-<-=0.55] (-0.707,-0.707)--(0,-1.414);
            \filldraw[black] (0,0) circle (2pt);
            \filldraw[black] (-0.707,-0.707) circle (2pt);
            \node at (-0.3,0.2) {$j$};
            \node at (-1.007,-0.507) {$i$};
            \node at (-0.15,-0.55) {$\mu$};
            \node at (-1.414,-1.714) {$\rho$};
            \node at (0,-1.714) {$\sigma$};
            \node at (1.414,-1.714) {$\lambda$};
            \node at (0,1.3) {$\delta$};
        \end{tikzpicture}
    \end{gathered}
    \quad = \quad
    \sum_{\nu,k,l} \left[F^{\delta;\mu\nu}_{\rho\sigma\lambda}\right]_{kl}^{ij}
    \begin{gathered}
        \begin{tikzpicture}
            \draw[line,thick,-<-=0.55] (0,1)--(0,0);
            \draw[line,thick,-<-=0.55] (0,0)--(-1.414,-1.414);
            \draw[line,thick,-<-=0.55] (0,0)--(0.707,-0.707);
            \draw[line,thick,-<-=0.55] (0.707,-0.707)--(1.414,-1.414);
            \draw[line,thick,-<-=0.55] (0.707,-0.707)--(0,-1.414);
            \filldraw[black] (0,0) circle (2pt);
            \filldraw[black] (0.707,-0.707) circle (2pt);
            \node at (0.15,-0.55) {$\nu$};
            \node at (0.3,0.2) {$l$};
            \node at (1.007,-0.507) {$k$};
            \node at (-1.414,-1.714) {$\rho$};
            \node at (0,-1.714) {$\sigma$};
            \node at (1.414,-1.714) {$\lambda$};
            \node at (0,1.3) {$\delta$};
        \end{tikzpicture}
    \end{gathered}
\end{equation}
where the $F$-symbol is the component of the associator in the outgoing channel $\delta$. In terms of the chosen intertwiners, this relation may be written as
\begin{equation}
    I^{\delta;j}_{\mu,\lambda}\circ \bigl(I^{\mu;i}_{\rho,\sigma}\otimes \text{id}_{E_{\lambda}}\bigr)
    =\sum_{\nu,k,l} \left[F^{\delta;\mu\nu}_{\rho\sigma\lambda}\right]_{kl}^{ij}
    \, I^{\delta;l}_{\rho,\nu}\circ \bigl(\text{id}_{E_{\rho}}\otimes I^{\nu;k}_{\sigma,\lambda}\bigr)\,.
\end{equation}

Similarly, one may introduce the co-associator and the corresponding $F^{\vee}$-symbol, defined diagrammatically by
\begin{equation}
    \begin{gathered}
        \begin{tikzpicture}
            \draw[line,thick,->-=0.55] (0,-1)--(0,0);
            \draw[line,thick,->-=0.55] (0,0)--(-0.707,0.707);
            \draw[line,thick,->-=0.55] (-0.707,0.707)--(-1.414,1.414);
            \draw[line,thick,->-=0.55] (0,0)--(1.414,1.414);
            \draw[line,thick,->-=0.55] (-0.707,0.707)--(0,1.414);
            \filldraw[black] (0,0) circle (2pt);
            \filldraw[black] (-0.707,0.707) circle (2pt);
            \node at (-0.15,0.55) {$\mu$};
            \node at (-0.3,-0.2) {$j^{\vee}$};
            \node at (-1.007,0.507) {$i^{\vee}$};
            \node at (-1.414,1.714) {$\rho$};
            \node at (0,1.714) {$\sigma$};
            \node at (1.414,1.714) {$\lambda$};
            \node at (0,-1.3) {$\delta$};
        \end{tikzpicture}
    \end{gathered}
    \quad = \quad
    \sum_{\nu,k,l} \left[F^{\vee\rho\sigma\lambda}_{\delta;\mu\nu}\right]_{kl}^{ij}
    \begin{gathered}
        \begin{tikzpicture}
            \draw[line,thick,->-=0.55] (0,-1)--(0,0);
            \draw[line,thick,->-=0.55] (0,0)--(-1.414,1.414);
            \draw[line,thick,->-=0.55] (0,0)--(0.707,0.707);
            \draw[line,thick,->-=0.55] (0.707,0.707)--(1.414,1.414);
            \draw[line,thick,->-=0.55] (0.707,0.707)--(0,1.414);
            \filldraw[black] (0,0) circle (2pt);
            \filldraw[black] (0.707,0.707) circle (2pt);
            \node at (0.15,0.55) {$\nu$};
            \node at (0.3,-0.2) {$l^{\vee}$};
            \node at (1.007,0.507) {$k^{\vee}$};
            \node at (-1.414,1.714) {$\rho$};
            \node at (0,1.714) {$\sigma$};
            \node at (1.414,1.714) {$\lambda$};
            \node at (0,-1.3) {$\delta$};
        \end{tikzpicture}
    \end{gathered}
\end{equation}
and satisfying
\begin{equation}
    \bigl(I^{\vee\rho,\sigma}_{\mu;i}\otimes \text{id}_{E_{\lambda}}\bigr)\circ I^{\vee\mu,\lambda}_{\delta;j}
    =\sum_{\nu,k,l} \left[F^{\vee\rho\sigma\lambda}_{\delta;\mu\nu}\right]_{kl}^{ij}
    \,\bigl(\text{id}_{E_{\rho}}\otimes I^{\vee\sigma,\lambda}_{\nu;k}\bigr)\circ I^{\vee\rho,\nu}_{\delta;l}\,.
\end{equation}
Both $F$ and $F^{\vee}$ depend on the choice of intertwiners and co-intertwiners, and hence are not uniquely defined. In this paper, however, we will not need their explicit expressions.

\subsection{Gauging non-invertible symmetry and Frobenius algebra}

We begin with a brief review of the categorical description of gauging in two dimensions. For an ordinary finite group symmetry $G$ of a 2D theory $\mathfrak{T}_G$, gauging may be described as a sum over flat gauge configurations labeled by holonomies along nontrivial 1-cycles. For example, on a torus a flat gauge configuration is specified by a commuting pair $(g,h)$ with $g,h\in G$ and $gh=hg$, where $g$ and $h$ denote the $G$-valued holonomies along the spatial and temporal cycles, respectively. The gauged partition function is therefore
\begin{equation}
    Z/G=\frac{1}{|G|}\sum_{\substack{g,h\in G\\ gh=hg}} Z[g,h]\,,
\end{equation}
where $Z[g,h]$ denotes the partition function of $\mathfrak{T}_G$ in the background with spatial holonomy $g$ and temporal holonomy $h$ turned on.

In the language of topological defect lines (TDLs), turning on a holonomy along a 1-cycle may be understood as inserting a TDL along the dual 1-cycle. For example, on the torus, inserting a symmetry operator $g$ along the spatial cycle produces a temporal holonomy, acting on the Hilbert space as $\mathcal{H}\to g\circ\mathcal{H}$. Conversely, inserting a symmetry defect $h$ along the temporal cycle produces a spatial holonomy, which twists the spatial boundary condition and defines the twisted Hilbert space $\mathcal{H}_h$. In this paper, we will refer to both viewpoints simply as topological defects. Accordingly, we write the partition function as $Z_{g,h}$, where $g$ and $h$ denote defects along the spatial and temporal directions, respectively, and adopt the convention
\begin{equation}
    Z[g,h]=Z_{h,g^{-1}}\,.
\end{equation}

The defect-network description of symmetry backgrounds is particularly useful for non-invertible symmetries, for which it is generally difficult to formulate an ordinary gauge-field description. To gauge a general finite non-invertible symmetry $\mathscr{C}$ at the level of partition function, one must first construct the partition functions associated with arbitrary symmetry backgrounds, represented by defect networks, and then sum over all such networks with appropriate weights. In categorical terms, this gauging procedure is implemented by condensing a separable Frobenius algebra object $\mathcal{A}$ in $\mathscr{C}$ over spacetime. We will not give a full discussion of Frobenius algebras here, but only summarize the data that will be needed later.

In the symmetry category $\mathscr{C}$, an algebra object may be written as a direct sum of simple TDLs
\begin{equation}
    \mathcal{A}=\bigoplus_{a\in\mathscr{C}} n_a\, a\,,\qquad n_a\in\mathbb{Z}_{\geq 0}\,.
\end{equation}
In general, $\mathcal{A}$ is not simple. There are therefore topological junctions describing the embedding of a simple TDL $a$ into the algebra object $\mathcal{A}$ and the corresponding projection from $\mathcal{A}$ onto $a$. We denote these junctions by blue triangles
\begin{equation}
\begin{gathered}
    \begin{tikzpicture}
        \draw[thick,line,-tri-=.55,->-=0.3] (0,0)--(0,2);
        \node at (-0.25,0.25) {$a$};
        \node at (-0.25,1.75) {$\mathcal{A}$};
        \node at (1.6,1.1) {$J_{a,r}\in \text{Hom}(a,\mathcal{A})$};
    \end{tikzpicture}
\end{gathered}
\quad
\begin{gathered}
    \begin{tikzpicture}
        \draw[thick,line,-tri-=.55,->-=0.85] (0,0)--(0,2);
        \node at (-0.25,0.25) {$\mathcal{A}$};
        \node at (-0.25,1.75) {$a$};
        \node at (1.6,1.1) {$J^{\vee}_{a,r}\in \text{Hom}(\mathcal{A},a)$};
    \end{tikzpicture}
\end{gathered}
\end{equation}
Here $J_{a,r}$ encodes the embedding of the simple line $a$ into the algebra object $\mathcal{A}$, while $J^{\vee}_{a,r}$ encodes the projection from the algebra object $\mathcal{A}$ onto $a$, with $r=1,\cdots,n_a$.

The algebra object is furthermore equipped with a multiplication map $m:\mathcal{A}\otimes\mathcal{A}\to\mathcal{A}$. In terms of simple lines, this map may be expanded in components $M_{(a,r),(b,s)}^{(c,t),i}$ as
\begin{equation}\label{eq:multiplication_Frobenius}
    \begin{gathered}
        \begin{tikzpicture}
            \draw[thick,line,-tri-=.45,->-=0.75] (0,0)--(0,2);
            \draw[thick,line,-tri-=.55,->-=0.3] (-1.414,-1.414)--(0,0);
            \draw[thick,line,-tri-=.55,->-=0.3] (1.414,-1.414)--(0,0);
            \filldraw[black] (0,0) circle (2pt);
            \node at (0,-0.4) {$m$};
            \node at (0.25,0.4) {$\mathcal{A}$};
            \node at (0.5,-0.15) {$\mathcal{A}$};
            \node at (-0.5,-0.15) {$\mathcal{A}$};
            \node at (-0.45,-1) {$J_{a,r}$};
            \node at (0.5,-1) {$J_{b,s}$};
            \node at (0.4,0.9) {$J^{\vee}_{c,t}$};
            \node at (0.25,2) {$c$};
            \node at (-1.1,-1.414) {$a$};
            \node at (1.1,-1.414) {$b$};
        \end{tikzpicture}
    \end{gathered}
    \quad = \quad
    \sum_{i=1}^{N_{ab}^{c}} M_{(a,r),(b,s)}^{(c,t),i}
    \begin{gathered}
        \begin{tikzpicture}
            \draw[thick,line,->-=0.75] (0,0)--(0,2);
            \draw[thick,line,->-=0.3] (-1.414,-1.414)--(0,0);
            \draw[thick,line,->-=0.3] (1.414,-1.414)--(0,0);
            \filldraw[black] (0,0) circle (2pt);
            \node at (0,-0.4) {$i$};
            \node at (0.25,2) {$c$};
            \node at (-1.1,-1.414) {$a$};
            \node at (1.1,-1.414) {$b$};
        \end{tikzpicture}
    \end{gathered}
\end{equation}
where $i$ labels a basis of morphisms from $E_a\otimes E_b$ to $E_c$, with $i=1,\cdots,N_{ab}^{c}$. In the special case $\mathscr{C}=\text{Rep}(G)$, this label $i$ corresponds to the intertwiner introduced above. Similarly, one defines the comultiplication $m^{\vee}:\mathcal{A}\to\mathcal{A}\otimes\mathcal{A}$ by
\begin{equation}\label{eq:comultiplication_Frobenius}
    \begin{gathered}
        \begin{tikzpicture}
            \draw[thick,line,-tri-=.45,->-=0.25] (0,-2)--(0,0);
            \draw[thick,line,-tri-=.55,->-=0.85] (0,0)--(-1.414,1.414);
            \draw[thick,line,-tri-=.55,->-=0.85] (0,0)--(1.414,1.414);
            \filldraw[black] (0,0) circle (2pt);
            \node at (0,0.4) {$m^{\vee}$};
            \node at (0.25,-0.4) {$\mathcal{A}$};
            \node at (0.5,0.15) {$\mathcal{A}$};
            \node at (-0.5,0.15) {$\mathcal{A}$};
            \node at (-0.45,1.1) {$J^{\vee}_{a,r}$};
            \node at (0.5,1.1) {$J^{\vee}_{b,s}$};
            \node at (0.4,-1.1) {$J_{c,t}$};
            \node at (0.25,-2) {$c$};
            \node at (-1.1,1.414) {$a$};
            \node at (1.1,1.414) {$b$};
        \end{tikzpicture}
    \end{gathered}
    \quad = \quad
    \sum_{i=1}^{N_{ab}^{c}} M^{\vee (a,r),(b,s)}_{(c,t),i}
    \begin{gathered}
        \begin{tikzpicture}
            \draw[thick,line,-<-=0.55] (0,0)--(0,-2);
            \draw[thick,line,-<-=0.55] (-1.414,1.414)--(0,0);
            \draw[thick,line,-<-=0.55] (1.414,1.414)--(0,0);
            \filldraw[black] (0,0) circle (2pt);
            \node at (0,0.4) {$i^{\vee}$};
            \node at (0.25,-2) {$c$};
            \node at (-1.1,1.414) {$a$};
            \node at (1.1,1.414) {$b$};
        \end{tikzpicture}
    \end{gathered}
\end{equation}
There also exist a unit morphism $\eta:1\to\mathcal{A}$, which embeds the tensor unit into $\mathcal{A}$, and a counit morphism $\eta^{\vee}:\mathcal{A}\to 1$, although we will not use them in this paper. We will denote the algebra object by $(\mathcal{A},m,\eta,m^{\vee},\eta^{\vee})$, or simply by $(\mathcal{A},m,m^{\vee})$ when no confusion can arise. The multiplication and comultiplication data are essential: two algebra objects may have the same underlying decomposition into simple components but different multiplication or comultiplication maps, and should therefore be regarded as distinct.

The algebra object of interest to us is a separable Frobenius algebra. Its multiplication is associative and its comultiplication is coassociative
    \begin{equation}
        \begin{gathered}
            \begin{tikzpicture}
                \draw[line,thick,-<-=0.55] (0,1)--(0,0);
                \draw[line,thick,-<-=0.55] (0,0)--(-0.707,-0.707);
                \draw[line,thick,-<-=0.55] (-0.707,-0.707)--(-1.414,-1.414);
                \draw[line,thick,-<-=0.55] (0,0)--(1.414,-1.414);
                \draw[line,thick,-<-=0.55] (-0.707,-0.707)--(0,-1.414);
                \filldraw[black] (0,0) circle (2pt);
                \filldraw[black] (-0.707,-0.707) circle (2pt);
                \node at (-0.5,0) {$m$};
                \node at (-0.707-0.5,-0.707) {$m$};
                \node at (-0.15,-0.55) {$\mathcal{A}$};
                \node at (-1.414,-1.414-0.3) {$\mathcal{A}$};
                \node at (0,-1.414-0.3) {$\mathcal{A}$};
                \node at (1.414,-1.414-0.3) {$\mathcal{A}$};
                \node at (0,1.3) {$\mathcal{A}$};
            \end{tikzpicture}
        \end{gathered}\quad = \quad
        \begin{gathered}
            \begin{tikzpicture}
                \draw[line,thick,-<-=0.55] (0,1)--(0,0);
                \draw[line,thick,-<-=0.55] (0,0)--(-1.414,-1.414);
                \draw[line,thick,-<-=0.55] (0,0)--(0.707,-0.707);
                \draw[line,thick,-<-=0.55] (0.707,-0.707)--(1.414,-1.414);
                \draw[line,thick,-<-=0.55] (0.707,-0.707)--(0,-1.414);
                \filldraw[black] (0,0) circle (2pt);
                \filldraw[black] (0.707,-0.707) circle (2pt);
                \node at (0.5,0) {$m$};
                \node at (0.707+0.5,-0.707) {$m$};
                \node at (0.15,-0.55) {$\mathcal{A}$};
                \node at (-1.414,-1.414-0.3) {$\mathcal{A}$};
                \node at (0,-1.414-0.3) {$\mathcal{A}$};
                \node at (1.414,-1.414-0.3) {$\mathcal{A}$};
                \node at (0,1.3) {$\mathcal{A}$};
            \end{tikzpicture}
        \end{gathered}
    \end{equation}
and
    \begin{equation}
        \begin{gathered}
            \begin{tikzpicture}
                \draw[line,thick,->-=0.55] (0,-1)--(0,0);
                \draw[line,thick,->-=0.55] (0,0)--(-0.707,0.707);
                \draw[line,thick,->-=0.55] (-0.707,0.707)--(-1.414,1.414);
                \draw[line,thick,->-=0.55] (0,0)--(1.414,1.414);
                \draw[line,thick,->-=0.55] (-0.707,0.707)--(0,1.414);
                \filldraw[black] (0,0) circle (2pt);
                \filldraw[black] (-0.707,0.707) circle (2pt);
                \node at (-0.5,0) {$m^{\vee}$};
                \node at (-0.707-0.5,0.707) {$m^{\vee}$};
                \node at (-0.15,0.55) {$\mathcal{A}$};
                \node at (-1.414,1.414+0.3) {$\mathcal{A}$};
                \node at (0,1.414+0.3) {$\mathcal{A}$};
                \node at (1.414,1.414+0.3) {$\mathcal{A}$};
                \node at (0,-1.3) {$\mathcal{A}$};
            \end{tikzpicture}
        \end{gathered}\quad = \quad
        \begin{gathered}
            \begin{tikzpicture}
                \draw[line,thick,->-=0.55] (0,-1)--(0,0);
                \draw[line,thick,->-=0.55] (-0,0)--(-1.414,1.414);
                \draw[line,thick,->-=0.55] (0,0)--(0.707,0.707);
                \draw[line,thick,->-=0.55] (0.707,0.707)--(1.414,1.414);
                \draw[line,thick,->-=0.55] (0.707,0.707)--(0,1.414);
                \filldraw[black] (0,0) circle (2pt);
                \filldraw[black] (0.707,0.707) circle (2pt);
                \node at (0.5,0) {$m^{\vee}$};
                \node at (0.707+0.5,0.707) {$m^{\vee}$};
                \node at (0.15,0.55) {$\mathcal{A}$};
                \node at (-1.414,1.414+0.3) {$\mathcal{A}$};
                \node at (0,1.414+0.3) {$\mathcal{A}$};
                \node at (1.414,1.414+0.3) {$\mathcal{A}$};
                \node at (0,-1.3) {$\mathcal{A}$};
            \end{tikzpicture}
        \end{gathered}
    \end{equation}
The separablity condition is
    \begin{equation}
        \begin{gathered}
            \begin{tikzpicture}
                \draw[line,thick,->-=0.55] (0,-1)--(0,0);
            \draw[line,thick,->-=0.55] (0,0) arc (-90:-270:0.5);
            \draw[line,thick,->-=0.55] (0,0) arc (-90:90:0.5);
                \draw[line,thick,->-=0.55] (0,1)--(0,2);
                \node at (-0.5,-0.25) {$m^{\vee}$};
                \node at (-0.5,1.25) {$m$};
                \filldraw[black] (0,0) circle (2pt);
                \filldraw[black] (0,1) circle (2pt);
                \node at (-0.8,0.5) {$\mathcal{A}$};
                \node at (0.8,0.5) {$\mathcal{A}$};
                \node at (0,-1.3) {$\mathcal{A}$};
                \node at (0,2.3) {$\mathcal{A}$};
            \end{tikzpicture}
        \end{gathered}\quad = \quad 
        \begin{gathered}
        \begin{tikzpicture}
            \draw[line,thick,->-=0.55] (0,-1)--(0,2);
            \node at (0,-1.3) {$\mathcal{A}$};
        \end{tikzpicture}
        \end{gathered}
    \end{equation}
so that we can shrink such bubbles freely. These two properties allow the $\mathcal{A}$-network to be deformed topologically. As a result, condensing $\mathcal{A}$ on a two-dimensional spacetime, represented by a sufficiently fine mesh of $\mathcal{A}$-lines, defines a well-posed operation. For example, on the torus, if we represent the condensation of $\mathcal{A}$ by a sufficiently fine $\mathcal{A}$-network, the network can always be simplified to the form
\begin{equation}\label{eq:frobenius-algebra-torus}
    \begin{gathered}
        \begin{tikzpicture}[scale=2]
            \draw[very thick] (0,0) rectangle (1.2,1.2);
            \draw[line,thick,->-=.55] (.6,0) arc (0:45:.6);
            \draw[line,thick,->-=.55] (0,0.6) arc (90:45:.6);
            \draw[line,thick,-<-=.55] (.6,1.2) arc (-180:-135:.6);
            \draw[line,thick,-<-=.55] (1.2,0.6) arc (-90:-135:.6);
            \draw[line,thick,->-=.55] (0.6/1.414,0.6/1.414)--(1.2-0.6/1.414,1.2-0.6/1.414);
            \node at (1.4,0.6) {$\mathcal{A}$};
            \node at (0.6,1.4) {$\mathcal{A}$};
            \node at (-0.2,0.6) {$\mathcal{A}$};
            \node at (0.6,-0.2) {$\mathcal{A}$};
            \node at (0.7,0.5) {$\mathcal{A}$};
            \filldraw[black] (0.6/1.414,0.6/1.414) circle (1pt);
            \filldraw[black] (1.2-0.6/1.414,1.2-0.6/1.414) circle (1pt);
            \node at (0.3,0.3) {$m$};
            \node at (0.95,0.95) {$m^{\vee}$};
        \end{tikzpicture}
    \end{gathered}
\end{equation}
where the square represents the torus, with opposite edges identified. Gauging a non-invertible symmetry $\mathscr{C}$ is then defined by condensing the associated Frobenius algebra object $\mathcal{A}$ on spacetime.

In the remainder of this section, we specialize to the case $\mathscr{C}=\text{Rep}(G)$. The Frobenius algebra object associated with gauging $\text{Rep}(G)$ is given by the regular representation, decomposed as
\begin{equation}
    \mathcal{A}=\bigoplus_{\sigma\in\widehat{G}} d_{\sigma}\,\sigma\,,
\end{equation}
whose total dimension is
\begin{equation}
    \sum_{\sigma\in\widehat{G}} d_{\sigma}^{2}=|G|\,.
\end{equation}
We denote a basis of the regular representation by $v_g\in\mathcal{A}$, with $g\in G$. Under the left and right actions of $G$, these basis vectors transform as
\begin{equation}
    L_g\circ v_h=v_{gh}\,,\qquad R_g\circ v_h=v_{hg}\,.
\end{equation}
The intertwiners and co-intertwiners $J_{\rho,r}$ and $J^{\vee}_{\rho,r}$ are defined with respect to the left action by
\begin{equation}\label{eq:J_as_intertwiner}
    J_{\rho,r}\circ \rho(g)=L_g\circ J_{\rho,r}\,,\qquad
    J^{\vee}_{\rho,r}\circ L_g=\rho(g)\circ J^{\vee}_{\rho,r}\,.
\end{equation}
In terms of the basis vectors $e^{\rho}_{\alpha}$ of the irreducible representation $\rho$ and $v_g$ of the regular representation, we may expand
\begin{equation}
    J_{\rho,r}\circ e^{\rho}_{\alpha}
    =\sum_{g\in G} v_g\,\mathcal{J}^{g}_{(\rho,\alpha),r}\,,\qquad
    J^{\vee}_{\rho,r}\circ v_g
    =\sum_{\alpha} e^{\rho}_{\alpha}\,\mathcal{J}^{\vee g}_{(\rho,\alpha),r}\,.
\end{equation}

The Frobenius structure is given by
    \begin{equation}
        m(v_g \otimes v_h) = \delta_{g,h} v_g\,,\quad \eta(1) = \sum_{g\in G} v_g\,,
    \end{equation}
and 
    \begin{equation}
        m^{\vee}(v_g) = v_g \otimes v_g\,,\quad \eta^{\vee}(v_g)=1\,,
    \end{equation}
where $1\in \widehat{G}$ is the trivial representation. In particular, $m\circ m^{\vee} = \text{id}_{\mathcal{A}}$.

In terms of the decomposition $\mathcal{A}= \bigoplus_{\sigma \in \widehat{G}} d_{\sigma} \sigma$, we can also decompose the multiplication $m$ into morphism $E_{\rho}\otimes E_{\sigma} \rightarrow E_{\lambda}$ as shown on the RHS of \eqref{eq:multiplication_Frobenius}
    \begin{equation}
        m_{(\rho,r),(\sigma,s)}^{(\lambda,t)} = J^{\vee}_{\lambda,t} \circ m\circ  (J_{\rho,r} \otimes J_{\sigma,s})= \sum_{i=1}^{N_{\rho \sigma}^{\lambda}} M_{(\rho,r),(\sigma,s)}^{(\lambda,t),i} I_{\rho,\sigma}^{\lambda;i}\,,
    \end{equation}
where we expand $m_{(\rho,r),(\sigma,s)}^{(\lambda,t)}$ in terms of the intertwiner basis $I_{\rho,\sigma}^{\lambda;i}$, with the coefficient $M_{(\rho,r),(\sigma,s)}^{(\lambda,t),i}$. Similarly, we can decompose the comultiplication $m^{\vee}$ into morphism $E_{\lambda} \rightarrow E_{\rho} \otimes E_{\sigma}$ as shown on the RHS of \eqref{eq:comultiplication_Frobenius}
    \begin{equation}
        m^{\vee (\rho,r),(\sigma,s)}_{(\lambda,t)} = (J^{\vee}_{\rho,r}\otimes J^{\vee}_{\sigma,s}) \circ m^{\vee} \circ J_{\lambda,t} = \sum_{i=1}^{N_{\rho \sigma}^{\lambda}} M^{\vee (\rho,r),(\sigma,s)}_{(\lambda,t),i} I^{\vee \rho,\sigma}_{\lambda;i}\,,
    \end{equation}
where we expand $m^{\vee (\rho,r),(\sigma,s)}_{(\lambda,t)}$ in terms of the cointertwiner basis $I^{\vee \rho,\sigma}_{\lambda;i}$, with the coefficient $M^{\vee (\rho,r),(\sigma,s)}_{(\lambda,t),i}$. 

Finally, we may decompose the $\mathcal{A}$-network on the torus in \eqref{eq:frobenius-algebra-torus} into channels of the form $\rho\otimes \sigma \to \lambda$
\begin{equation}
        \begin{gathered}
\begin{tikzpicture}[scale=2]
\draw[very thick] (0,0) rectangle (1.2,1.2);
\draw[line,thick,->-=.55]  (.6,0) arc (0:45:.6); 
\draw[line,thick,->-=.55]  (0,0.6) arc (90:45:.6); 
\draw[line,thick,-<-=.55] (.6,1.2) arc (-180:-135:.6);
\draw[line,thick,-<-=.55] (1.2,0.6) arc (-90:-135:.6);
\draw[line,thick,->-=.55] (0.6/1.414,0.6/1.414)--(1.2-0.6/1.414,1.2-0.6/1.414);
\node at (1.2+0.2,0.6) {$\mathcal{A}$};
\node at (0.6,1.2+0.2) {$\mathcal{A}$};
\node at (-0.2,0.6) {$\mathcal{A}$};
\node at (0.6,-0.2) {$\mathcal{A}$};
\node at (0.6+0.1,0.6-0.1) {$\mathcal{A}$};
\filldraw[black] (0.6/1.414,0.6/1.414) circle (1pt);
\filldraw[black] (1.2-0.6/1.414,1.2-0.6/1.414) circle (1pt);
 \node at (0.3,0.3) {$m$};
 \node at (0.95,0.95) {$m^{\vee}$};
\end{tikzpicture}
\end{gathered}\quad = \quad \sum_{\rho,\sigma,\lambda,i,j}\sum_{r,s,t} M^{\vee (\sigma,s),(\rho,r)}_{(\lambda,t),j} M_{(\rho,r),(\sigma,s)}^{(\lambda,t),i}
        \begin{gathered}
\begin{tikzpicture}[scale=2]
\draw[very thick] (0,0) rectangle (1.2,1.2);
\draw[line,thick,->-=.55]  (.6,0) arc (0:45:.6); 
\draw[line,thick,->-=.55]  (0,0.6) arc (90:45:.6); 
\draw[line,thick,-<-=.55] (.6,1.2) arc (-180:-135:.6);
\draw[line,thick,-<-=.55] (1.2,0.6) arc (-90:-135:.6);
\draw[line,thick,->-=.55] (0.6/1.414,0.6/1.414)--(1.2-0.6/1.414,1.2-0.6/1.414);
\node at (1.2+0.2,0.6) {$\rho$};
\node at (0.6,1.2+0.2) {$\sigma$};
\node at (-0.2,0.6) {$\rho$};
\node at (0.6,-0.2) {$\sigma$};
\node at (0.6+0.1,0.6-0.1) {$\lambda$};
\filldraw[black] (0.6/1.414,0.6/1.414) circle (1pt);
\filldraw[black] (1.2-0.6/1.414,1.2-0.6/1.414) circle (1pt);
 \node at (0.25,0.25) {$I_{\rho,\sigma}^{\lambda\text{;}i}$};
 \node at (0.95,0.95) {$I^{\vee \sigma,\rho}_{\lambda\text{;}j}$};
\end{tikzpicture}
\end{gathered}
\nonumber
    \end{equation}
Let us denote the torus partition function $Z$ with a general background of non-invertible symmetry as\footnote{Those backgrounds are not independent in general.}
    \begin{equation}
        Z_{\rho,\sigma}^{\lambda;i,j} \quad \equiv \quad Z\ 
        \begin{gathered}
\begin{tikzpicture}[scale=0.9]
\node at (0, 1.45) {};
\draw[line] (-.15-.4,-.2-.4)--(-.25-.4, -.2-.4) -- (-.25-.4, 1.4)-- (-.15-.4,1.4);
\draw[very thick] (0,0) rectangle (1.2,1.2);
\draw[line,thick]  (.6,0) arc (0:90:.6); 
\draw[line,thick] (.6,1.2) arc (-180:-90:.6);
\draw[line,thick] (0.6/2^0.5,0.6/2^0.5)--(1.2-0.6/2^0.5,1.2-0.6/2^0.5);
\draw[line] (1.35, -.2-.4)--(1.45, -.2-.4) -- (1.45, 1.4)--(1.35, 1.4);
\node at (-.15-.2,-.2+0.8) {$\rho$};
\node at (.6,-.35) {$\sigma$};
\node at (0.6+0.15,0.6-0.15) {$\lambda$};
\node at (0.25,0.25) {$i$};
\node at (1.2-0.25,1.2-0.25) {$j$};
\end{tikzpicture}
\end{gathered}\quad \,,
\end{equation}
where $i,j$ are the labels of intertwiner and cointertwiner inserted at the two junctions, and we omit the arrow on the diagram. Then the gauged partition by condensing the Frobenius algebra $\mathcal{A}$ is 
    \begin{equation}\label{eq:gauge_Rep_tivial}
        Z/\text{Rep}(G) = \frac{1}{\dim \mathcal{A}}\sum_{\rho,\sigma,\lambda;i,j} C_{\rho,\sigma}^{\lambda;i,j} Z_{\rho,\sigma}^{\lambda;i,j}\,,
    \end{equation}
where $\dim \mathcal{{A}}=|G|$ is the total dimension of the Frobenius algebra, and we sum over all non-invertible symmetry backgrounds with the Fourier coefficient $C_{\rho,\sigma}^{\lambda;i,j}$ given by the sum
    \begin{equation}\label{eq:gauge_Rep_tivial_coefficient}
        C_{\rho,\sigma}^{\lambda;i,j} = \dim \mathcal{A}\sum_{r,s,t} M^{\vee (\sigma,s),(\rho,r)}_{(\lambda,t),j} M_{(\rho,r),(\sigma,s)}^{(\lambda,t),i}\,.
    \end{equation}

\section{Fourier kernel for a general $\text{Rep}(G)$ defect network}

Consider a two-dimensional theory $\mathfrak{T}_G$ with an anomaly-free finite $G$-symmetry. Gauging the $G$-symmetry produces a dual theory
\begin{equation}
    \widetilde{\mathfrak{T}}_{\text{Rep}(G)}=\mathfrak{T}_G/G\,,
\end{equation}
equipped with the dual symmetry $\text{Rep}(G)$. When $G$ is non-abelian, $\text{Rep}(G)$ is non-invertible, and we will restrict to this case in what follows. Conversely, gauging the $\text{Rep}(G)$ symmetry by condensing the Frobenius algebra $\mathcal{A}$ associated with $\text{Rep}(G)$ returns the original theory
\begin{equation}
    \widetilde{\mathfrak{T}}_{\text{Rep}(G)}/\text{Rep}(G)=\mathfrak{T}_G\,.
\end{equation}

In this section, we study the coupling of the theory $\widetilde{\mathfrak{T}}_{\text{Rep}(G)}$ to a general non-invertible symmetry background, represented by a topological defect network. At the level of partition functions on a closed manifold, the partition function $\widetilde{Z}$ in a $\text{Rep}(G)$ background is related to the partition function $Z$ in a $G$ background by a generalized Fourier transformation. We will develop a general method for computing the Fourier kernel for an arbitrary $\text{Rep}(G)$ background in terms of $G$ backgrounds, and also the inverse kernel expressing a $G$ background in terms of $\text{Rep}(G)$ backgrounds. For illustration, we will focus primarily on defect networks on the torus of the form \eqref{eq:frobenius-algebra-torus}, although the construction extends straightforwardly to higher-genus Riemann surfaces and to defect networks of arbitrary shape.

\subsection{Fourier kernel : From $G$ to $\text{Rep}(G)$}

Consider a $\text{Rep}(G)$ defect network on torus represented as
\begin{equation}
            \begin{gathered}
\begin{tikzpicture}[scale=2]
\draw[very thick] (0,0) rectangle (1.2,1.2);
\draw[line,thick,->-=.55]  (.6,0) arc (0:45:.6); 
\draw[line,thick,->-=.55]  (0,0.6) arc (90:45:.6); 
\draw[line,thick,-<-=.55] (.6,1.2) arc (-180:-135:.6);
\draw[line,thick,-<-=.55] (1.2,0.6) arc (-90:-135:.6);
\draw[line,thick,->-=.55] (0.6/1.414,0.6/1.414)--(1.2-0.6/1.414,1.2-0.6/1.414);
\node at (1.2+0.2,0.6) {$\rho$};
\node at (0.6,1.2+0.2) {$\sigma$};
\node at (-0.2,0.6) {$\rho$};
\node at (0.6,-0.2) {$\sigma$};
\node at (0.6+0.1,0.6-0.1) {$\lambda$};
\filldraw[black] (0.6/1.414,0.6/1.414) circle (1pt);
\filldraw[black] (1.2-0.6/1.414,1.2-0.6/1.414) circle (1pt);
\node at (0.25,0.25) {$I_{\rho,\sigma}^{\lambda\text{;}i}$};
\node at (0.95,0.95) {$I^{\vee \sigma,\rho}_{\lambda\text{;}j}$};
\end{tikzpicture}
\end{gathered}\qquad \Longleftrightarrow \qquad
    \begin{gathered}
        \begin{tikzpicture}
                \draw[line] (0,0.7)--(0,1.5);
                \draw[line] (3.6,0.7+1.2)--(3.6,1.5+1.2);
                \draw[line] (0.9,0)--(1.7,0);
                \draw[line] (0.9,0-0.2)--(1.7,0-0.2);
                \draw[line] (0.9+1.2,3.6)--(1.7+1.2,3.6);
                \draw[line] (0.9+1.2,0+0.2+3.6)--(1.7+1.2,0+0.2+3.6);
                \draw[line,thick,->-=0.55] (0,1)--(1.2+0.2,1);
                \draw[line,thick,->-=0.55] (1.2+0.2,0)--(1.2+0.2,1);
                \draw[line,thick,->-=0.55] (1.2+0.2,1)--(1.2+0.2+1.2,1+1.2);
                \draw[line,thick,->-=0.55] (1.2+0.2+1.2,1+1.2)--(1.2+0.2+1.2+1,1+1.2);
                \draw[line,thick,->-=0.55] (1.2+0.2+1.2,1+1.2)--(1.2+0.2+1.2,1+1.2+1.4);
                \filldraw[black] (1.2+0.2+1.2,1+1.2) circle (1.5pt);
                \filldraw[black] (1.2+0.2,1) circle (1.5pt);
                \node at (0.4,0.8) {$\rho$};
                \node at (1.6,0.4) {$\sigma$};
                \node at (1.8+0.4,1.8-0.3) {$\lambda$};
                \node at (2.8,3.3) {$\sigma$};
                \node at (3.3,2.0) {$\rho$};
                \node at (2.7,1.9) {$j$};
                \node at (1.6,0.9) {$i$};
            \end{tikzpicture}
        \end{gathered}
\end{equation}
For our purposes it is more convenient to use the second diagram, in which the endpoints of the horizontal and vertical lines are identified separately. Since the TDLs in $\text{Rep}(G)$ are realized as Wilson lines in the $G$ gauge theory, the Fourier kernel may be obtained by turning on a $G$-background and evaluating the Wilson line network in that background. More precisely, we include the $G$-background as
    
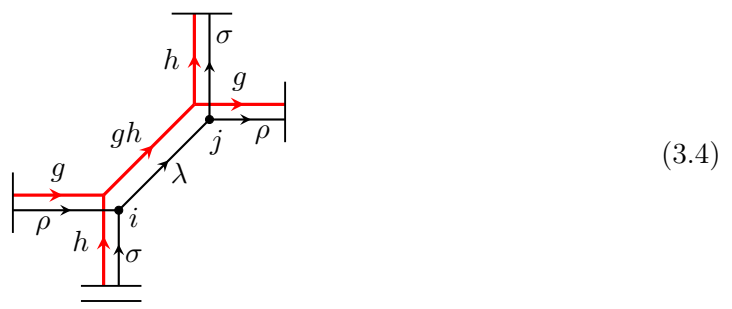
\begin{equation}\label{eq:Rep_G_and_G_background}
    \begin{gathered}
        \begin{tikzpicture}
                \draw[line,very thick,red,->-=0.55] (0,1.2)--(1.2,1.2);
                \draw[line,very thick,red,->-=0.55] (1.2,0)--(1.2,1.2);
                \draw[line,very thick,red,->-=0.55] (1.2,1.2)--(2.4,2.4);
                \draw[line,very thick,red,->-=0.55] (2.4,2.4)--(3.6,2.4);
                \draw[line,very thick,red,->-=0.55] (2.4,2.4)--(2.4,3.6);
                \node at (0.6,1.5) {$g$};
                \node at (0.9,0.6) {$h$};
                \node at (1.8-0.3,1.8+0.2) {$gh$};
                \node at (2.1,3) {$h$};
                \node at (3,2.7) {$g$};
                \draw[line] (0,0.7)--(0,1.5);
                \draw[line] (3.6,0.7+1.2)--(3.6,1.5+1.2);
                \draw[line] (0.9,0)--(1.7,0);
                \draw[line] (0.9,0-0.2)--(1.7,0-0.2);
                \draw[line] (0.9+1.2,3.6)--(1.7+1.2,3.6);
                \draw[line] (0.9+1.2,0+0.2+3.6)--(1.7+1.2,0+0.2+3.6);
                \draw[line,thick,->-=0.55] (0,1)--(1.2+0.2,1);
                \draw[line,thick,->-=0.55] (1.2+0.2,0)--(1.2+0.2,1);
                \draw[line,thick,->-=0.55] (1.2+0.2,1)--(1.2+0.2+1.2,1+1.2);
                \draw[line,thick,->-=0.55] (1.2+0.2+1.2,1+1.2)--(1.2+0.2+1.2+1,1+1.2);
                \draw[line,thick,->-=0.55] (1.2+0.2+1.2,1+1.2)--(1.2+0.2+1.2,1+1.2+1.4);
                \filldraw[black] (1.2+0.2+1.2,1+1.2) circle (1.5pt);
                \filldraw[black] (1.2+0.2,1) circle (1.5pt);
                \node at (0.4,0.8) {$\rho$};
                \node at (1.6,0.4) {$\sigma$};
                \node at (1.8+0.4,1.8-0.3) {$\lambda$};
                \node at (2.8,3.3) {$\sigma$};
                \node at (3.3,2.0) {$\rho$};
                \node at (2.7,1.9) {$j$};
                \node at (1.6,0.9) {$i$};
            \end{tikzpicture}
        \end{gathered}
    \end{equation}
where the red lines represent the $G$-defect network. The $\text{Rep}(G)$ defects are drawn slightly displaced from the $G$-defects, as shown in the figure. As we will explain below, the resulting Fourier kernel is in fact independent of the positions of the junctions in the fixed $G$-background.

The rule is as follows. We start with the vector space $E_{\rho}\otimes E_{\sigma}$ at the lower-left corner of \eqref{eq:Rep_G_and_G_background}, and move along the arrows on the $\text{Rep}(G)$ lines. Whenever a $\text{Rep}(G)$ line labeled by $\rho$ crosses a $G$-defect labeled by $h$, we act on the corresponding vector space by the representation matrix $\rho(h)$ or $\rho(h^{-1})$, depending on the relative orientation of the two lines. When two $\text{Rep}(G)$ lines labeled by $\rho$ and $\sigma$ meet at a junction and fuse into $\lambda$, we act with the intertwiner $I^{\lambda;i}_{\rho,\sigma}$, which projects onto the vector space $E_{\lambda}$. Conversely, when a $\text{Rep}(G)$ line labeled by $\lambda$ splits into lines labeled by $\sigma$ and $\rho$, we act with the co-intertwiner $I^{\vee \lambda;j}_{\sigma,\rho}$, which embeds $E_{\lambda}$ into $E_{\sigma}\otimes E_{\rho}$. We then apply the swap map $S_{\sigma,\rho}$ to restore the ordering of the tensor factors from $E_{\sigma}\otimes E_{\rho}$ to $E_{\rho}\otimes E_{\sigma}$. Finally, after returning to the starting point, we take the trace. The entire procedure is illustrated in the following diagram.
    
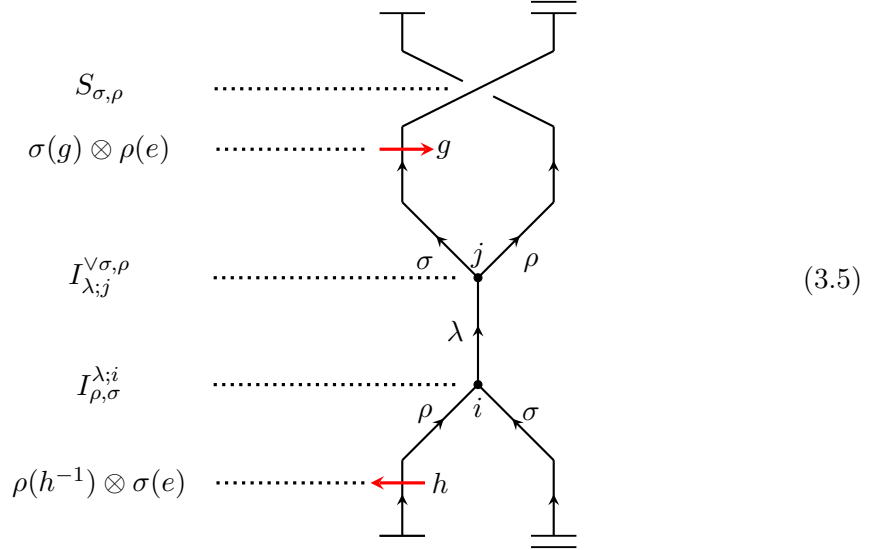
\begin{equation}
        \begin{gathered}
        \begin{tikzpicture}
            \draw[line,thick] (-1,2)--(1,3);
            \draw[line,thick] (1,2)--(0.2,2+0.4);
            \draw[line,thick] (-0.2,2+0.6)--(-1,3);
            \draw[line,thick] (-1,3)--(-1,3.5);
            \draw[line,thick] (1,3)--(1,3.5);
            \draw[line,thick,->-=0.55] (-1,1)--(-1,2);
            \draw[line,thick,->-=0.55] (1,1)--(1,2);
            \draw[line,thick,-<-=0.55] (-1,1)--(0,0);
            \draw[line,thick,-<-=0.55] (1,1)--(0,0);
            \draw[line,thick,-<-=0.55] (0,0)--(0,-1.414);
            \draw[line,thick,->-=0.55] (-1,-1.414-1)--(0,-1.414);
            \draw[line,thick,->-=0.55] (1,-1.414-1)--(0,-1.414);
            \draw[line,thick,->-=0.55] (-1,-1.414-1-1)--(-1,-1.414-1);
            \draw[line,thick,->-=0.55] (1,-1.414-1-1)--(1,-1.414-1);   
            \filldraw[black] (0,0) circle (1.5pt);
            \filldraw[black] (0,-1.414) circle (1.5pt);
            \node at (0,0.3) {$j$};
            \node at (0,-1.414-0.3) {$i$};
            \node at (-0.3,-0.707) {$\lambda$};
            \node at (-0.7,-1.8) {$\rho$};
            \node at (0.7,-1.8) {$\sigma$};
            \node at (-0.7,0.2) {$\sigma$};
            \node at (0.7,0.2) {$\rho$};
            \node at (-0.5,-1.414-1-0.3) {$h$};
            \node at (-0.45,1.7) {$g$};
            \draw[line,very thick,red,-<-=0.1] (-1.3,-1.414-1-0.3)--(-0.7,-1.414-1-0.3);
            \draw[line,very thick,red,-<-=0.1] (-0.7,1.7)--(-1.3,1.7);
            \draw[thick,line] (0.7-2,-1.414-1-1)--(1.3-2,-1.414-1-1);
            \draw[thick,line] (0.7,-1.414-1-1)--(1.3,-1.414-1-1);
            \draw[thick,line] (0.7,-1.414-1-1-0.15)--(1.3,-1.414-1-1-0.15);
            \draw[thick,line] (0.7-2,-1.414-1-1)--(1.3-2,-1.414-1-1);
            \draw[thick,line] (0.7,2+1.5)--(1.3,2+1.5);
            \draw[thick,line] (0.7,2+0.15+1.5)--(1.3,2+0.15+1.5);
            \draw[thick,line] (0.7-2,2+1.5)--(1.3-2,2+1.5);
            \draw[dotted,very thick] (-0.4,2.5)--(-3.5,2.5);
            \draw[dotted,very thick] (-1.5,-1.414-1-0.3)--(-3.5,-1.414-1-0.3);
            \draw[dotted,very thick] (-1.5,1.7)--(-3.5,1.7);
            \draw[dotted,very thick] (-0.3,0)--(-3.5,0);
            \draw[dotted,very thick] (-0.3,-1.414)--(-3.5,-1.414);
            \node at (-5,2.5) {$S_{\sigma,\rho}$};
            \node at (-5,-1.414-1-0.3) {$\rho(h^{-1})\otimes \sigma(e)$};
            \node at (-5,-1.414) {$I_{\rho,\sigma}^{\lambda\text{;}i}$};
            \node at (-5,0) {$I^{\vee \sigma,\rho}_{\lambda\text{;}j}$};
            \node at (-5,1.7) {$ \sigma(g)\otimes \rho(e)$};
        \end{tikzpicture}
        \end{gathered}
    \end{equation}

Therefore the VEV of the $\text{Rep}(G)$ defect network in a $G$-background is given by
\begin{equation}\label{eq:Rep(G)_G_kernel_raw}
    W_{\rho,\sigma}^{\lambda;i,j}(g,h)
    =
    \textrm{Tr}_{E_{\rho}\otimes E_{\sigma}}
    \left[
        S_{\sigma,\rho}
        \circ
        \left(\sigma(g)\otimes \rho(e)\right)
        \circ
        I^{\vee \sigma,\rho}_{\lambda;j}
        I_{\rho,\sigma}^{\lambda;i}
        \circ
        \left(\rho(h^{-1})\otimes \sigma(e)\right)
    \right] .
\end{equation}
Using
\begin{equation}
    S_{\sigma,\rho}\circ \left(\sigma(g)\otimes \rho(e)\right)
    =
    \left(\rho(e)\otimes \sigma(g)\right)\circ S_{\sigma,\rho}\,,
\end{equation}
we may rewrite this as
\begin{equation}\label{eq:Rep(G)_G_kernel}
    W_{\rho,\sigma}^{\lambda;i,j}(g,h)
    =
    \textrm{Tr}_{E_{\rho}\otimes E_{\sigma}}
    \left[
        S_{\sigma,\rho}
        \circ
        I^{\vee \sigma,\rho}_{\lambda;j}
        I_{\rho,\sigma}^{\lambda;i}
        \circ
        \left(\rho(h^{-1})\otimes \sigma(g)\right)
    \right] .
\end{equation}
Accordingly, the partition function $\widetilde{Z}_{\rho,\sigma}^{\lambda;i,j}$ in a general $\text{Rep}(G)$ background is given by
\begin{equation}
    \widetilde{Z}_{\rho,\sigma}^{\lambda;i,j}
    =
    \frac{1}{|G|}
    \sum_{\substack{g,h\in G\\ gh=hg}}
    W_{\rho,\sigma}^{\lambda;i,j}(g,h)\, Z_{g,h}\,.
\end{equation}
One may further show that $W_{\rho,\sigma}^{\lambda;i,j}(g,h)$ depends only on the conjugacy class of the commuting pair $(g,h)$
\begin{equation}
    W_{\rho,\sigma}^{\lambda;i,j}(kgk^{-1},khk^{-1})
    =
    W_{\rho,\sigma}^{\lambda;i,j}(g,h)\,,
\end{equation}
due to the intertwining properties \eqref{eq:defining_property_intertwiner} and \eqref{eq:defining_property_cointertwiner}.

Before proceeding, we comment that the coefficient \eqref{eq:Rep(G)_G_kernel} is in fact invariant under deformations of the defect network, owing to the intertwining property \eqref{eq:defining_property_intertwiner} and \eqref{eq:defining_property_cointertwiner} of the junction operators. As an illustration, suppose we move the junction $i$ to the left so that it crosses the $h$-line in the diagram \eqref{eq:Rep_G_and_G_background}, while keeping the junction $j$ and the endpoints fixed. Topologically, the resulting diagram takes the form
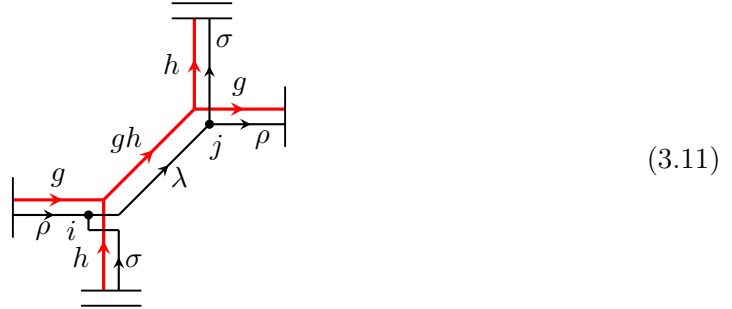
\begin{equation}
    \begin{gathered}
        \begin{tikzpicture}
                \draw[line,very thick,red,->-=0.55] (0,1.2)--(1.2,1.2);
                \draw[line,very thick,red,->-=0.55] (1.2,0)--(1.2,1.2);
                \draw[line,very thick,red,->-=0.55] (1.2,1.2)--(2.4,2.4);
                \draw[line,very thick,red,->-=0.55] (2.4,2.4)--(3.6,2.4);
                \draw[line,very thick,red,->-=0.55] (2.4,2.4)--(2.4,3.6);
                \node at (0.6,1.5) {$g$};
                \node at (0.9,0.45) {$h$};
                \node at (1.8-0.3,1.8+0.2) {$gh$};
                \node at (2.1,3) {$h$};
                \node at (3,2.7) {$g$};
                \draw[line] (0,0.7)--(0,1.5);
                \draw[line] (3.6,0.7+1.2)--(3.6,1.5+1.2);
                \draw[line] (0.9,0)--(1.7,0);
                \draw[line] (0.9,0-0.2)--(1.7,0-0.2);
                \draw[line] (0.9+1.2,3.6)--(1.7+1.2,3.6);
                \draw[line] (0.9+1.2,0+0.2+3.6)--(1.7+1.2,0+0.2+3.6);
                \draw[line,thick,->-=0.55] (0,1)--(1.2-0.2,1);
                \draw[line,thick,->-=0.55] (1.2+0.2,0)--(1.2+0.2,0.8);
                \draw[line,thick] (1.2+0.2,0.8)--(1.2-0.2,0.8);
                \draw[line,thick] (1.2-0.2,0.8)--(1.2-0.2,1);
                \draw[line,thick] (1.2-0.2,1)--(1.2+0.2,1);
                \draw[line,thick,->-=0.55] (1.2+0.2,1)--(1.2+0.2+1.2,1+1.2);
                \draw[line,thick,->-=0.55] (1.2+0.2+1.2,1+1.2)--(1.2+0.2+1.2+1,1+1.2);
                \draw[line,thick,->-=0.55] (1.2+0.2+1.2,1+1.2)--(1.2+0.2+1.2,1+1.2+1.4);
                \filldraw[black] (1.2+0.2+1.2,1+1.2) circle (1.5pt);
                \filldraw[black] (1.2-0.2,1) circle (1.5pt);
                \node at (0.4,0.8) {$\rho$};
                \node at (1.6,0.4) {$\sigma$};
                \node at (1.8+0.4,1.8-0.3) {$\lambda$};
                \node at (2.8,3.3) {$\sigma$};
                \node at (3.3,2.0) {$\rho$};
                \node at (2.7,1.9) {$j$};
                \node at (1-0.225,0.8) {$i$};
            \end{tikzpicture}
        \end{gathered}
    \end{equation}
then the VEV of the $\text{Rep}(G)$-defect network becomes
    \begin{equation}
    \begin{split}
        &\widehat{W}_{\rho,\sigma}^{\lambda;i,j} (g,h)\\
        =& \textrm{Tr}_{E_{\rho}\otimes E_{\sigma}} \left[ S_{\sigma,\rho}\circ \left(\sigma(g)\otimes \rho(e)\right)\circ I^{\vee \sigma,\rho}_{\lambda\text{;}j} \circ \lambda(h^{-1}) \circ I_{\rho,\sigma}^{\lambda\text{;}i}\circ (\rho(e)\otimes \sigma(h)) \right]\,.
    \end{split}
    \end{equation}
This expression is in fact equivalent to \eqref{eq:Rep(G)_G_kernel_raw} and \eqref{eq:Rep(G)_G_kernel}, since
    \begin{equation}
        \lambda(h^{-1}) \circ I_{\rho,\sigma}^{\lambda\text{;}i} = I_{\rho,\sigma}^{\lambda\text{;}i}\circ (\rho(h^{-1})\otimes \sigma(h^{-1}))\,,
    \end{equation}
and 
\begin{equation}
    \bigl(\rho(h^{-1})\otimes \sigma(h^{-1})\bigr)\circ \bigl(\rho(e)\otimes \sigma(h)\bigr)
    =
    \rho(h^{-1})\otimes \sigma(e)\,.
\end{equation}
Therefore, we have
    \begin{equation}
        \widehat{W}_{\rho,\sigma}^{\lambda;i,j} (g,h)=W_{\rho,\sigma}^{\lambda;i,j} (g,h)\,,
    \end{equation}
showing that the coefficient is invariant under the deformation described above. By similar arguments, one may check that $W_{\rho,\sigma}^{\lambda;i,j}(g,h)$ is also invariant under other deformations of the defect network, and is therefore well defined in the $G$-gauge background.

We would like to emphasis that although we work with torus, the method can be generalized to general closed Riemann surfaces. After resolving the defect network into trivalent junctions and choosing an orientation for each edge, we assign the corresponding representation space to every edge and an intertwiner or cointertwiner to every junction. For a flat $G$-background, we move along the network, multiply the representation matrices along the edges and contract all indices at the junctions, and then take the trace.

\subsection{Projector for multiplicity free case}

Before turning to the inverse Fourier kernel, let us comment further on the Fourier kernel in the multiplicity-free case. Introduce the projector
\begin{equation}
    P_{\rho,\sigma}^{\lambda}
    =
    S_{\sigma,\rho}\circ I^{\vee \sigma,\rho}_{\lambda} I_{\rho,\sigma}^{\lambda}\,,
\end{equation}
so that the Fourier kernel may be written as
\begin{equation}
    W_{\rho,\sigma}^{\lambda}(g,h)
    =
    \textrm{Tr}_{E_{\rho}\otimes E_{\sigma}}
    \left[
        P_{\rho,\sigma}^{\lambda}\circ
        \left(\rho(h^{-1})\otimes \sigma(g)\right)
    \right]\,.
\end{equation}
There is a canonical basis of projectors given by the character projectors
\begin{equation}\label{eq:character_projector}
    \Pi_{\rho,\sigma}^{\lambda}
    =
    \frac{d_{\lambda}}{|G|}
    \sum_{k\in G}
    \chi_{\lambda}(k^{-1})\,\bigl(\rho(k)\otimes \sigma(k)\bigr)\,,
\end{equation}
and hence the projectors $P_{\rho,\sigma}^{\lambda}$ are proportional to the character projectors
\begin{equation}
    P_{\rho,\sigma}^{\lambda}
    =
    p_{\rho,\sigma}^{\lambda}\,\Pi_{\rho,\sigma}^{\lambda}\,.
\end{equation}
It follows that the Fourier kernel is
\begin{equation}\label{eq:kernel_coefficient_character_projector}
\begin{split}
    W_{\rho,\sigma}^{\lambda}(g,h)
    &=
    p_{\rho,\sigma}^{\lambda}\,
    \textrm{Tr}_{E_{\rho}\otimes E_{\sigma}}
    \left[
        \Pi_{\rho,\sigma}^{\lambda}\circ
        \left(\rho(h^{-1})\otimes \sigma(g)\right)
    \right] \\
    &=
    \frac{p_{\rho,\sigma}^{\lambda}d_{\lambda}}{|G|}
    \sum_{k\in G}
    \chi_{\lambda}(k^{-1})\,
    \chi_{\rho}(kh^{-1})\,
    \chi_{\sigma}(kg)\,.
\end{split}
\end{equation}
Accordingly, the dual partition function takes the general form
\begin{equation}\label{eq:dual_partition_function_multiplicity_free}
    \widetilde{Z}_{\rho,\sigma}^{\lambda}
    =
    \frac{p_{\rho,\sigma}^{\lambda}d_{\lambda}}{|G|^2}
    \sum_{\substack{g,h\in G\\ gh=hg}}
    \sum_{k\in G}
    \chi_{\lambda}(k^{-1})\,
    \chi_{\rho}(kh^{-1})\,
    \chi_{\sigma}(kg)\,
    Z_{g,h}\,.
\end{equation}

We will determine the coefficients $p_{\rho,\sigma}^{\lambda}$ explicitly in several examples below. In the higher-multiplicity case, however, the projector $\Pi_{\rho,\sigma}^{\lambda}$ is not sufficient to resolve the degeneracy of $\lambda$ in $\rho\otimes \sigma$.

\subsection{Inverse Fourier kernel : From $\text{Rep}(G)$ to $G$}
We now consider the inverse Fourier transformation, expressing the partition function $Z_{g,h}$ of $\mathfrak{T}_G$ in terms of the partition functions $\widetilde{Z}_{\rho,\sigma}^{\lambda;i,j}$ of the dual theory $\widetilde{\mathfrak{T}}_{\text{Rep}(G)}$. As a first step, the trivial-sector partition function $Z_{e,e}$ is obtained by condensing the Frobenius algebra
\begin{equation}
    \mathcal{A}=\bigoplus_{\sigma\in\widehat{G}} d_{\sigma}\,\sigma\,,
\end{equation}
which was discussed in the previous section. From \eqref{eq:gauge_Rep_tivial} and \eqref{eq:gauge_Rep_tivial_coefficient}, we obtain
\begin{equation}\label{eq:RepG_to_G_trivial_sector}
    Z_{e,e}
    =
    \frac{1}{\dim\mathcal{A}}
    \sum_{\rho,\sigma,\lambda;i,j}
    C_{\rho,\sigma}^{\lambda;i,j}\,
    \widetilde{Z}_{\rho,\sigma}^{\lambda;i,j}\,,
\end{equation}
where the coefficient $C_{\rho,\sigma}^{\lambda;i,j}$ is given by the contraction of the multiplication and comultiplication coefficients
\begin{equation}
    C_{\rho,\sigma}^{\lambda;i,j}
    =
    \dim\mathcal{A}
    \sum_{r,s,t}
    M^{\vee(\sigma,s),(\rho,r)}_{(\lambda,t),j}
    M_{(\rho,r),(\sigma,s)}^{(\lambda,t),i}\,.
\end{equation}
Our goal in this section is to determine the more general coefficient $C_{\rho,\sigma}^{\lambda;i,j}(g,h)$ such that
\begin{equation}
    Z_{g,h}
    =
    \frac{1}{\dim\mathcal{A}}
    \sum_{\rho,\sigma,\lambda;i,j}
    C_{\rho,\sigma}^{\lambda;i,j}(g,h)\,
    \widetilde{Z}_{\rho,\sigma}^{\lambda;i,j}\,.
\end{equation}

To begin, we rewrite the Frobenius defect network \eqref{eq:frobenius-algebra-torus} in the form
\begin{equation}
\begin{gathered}
\begin{tikzpicture}[scale=2]
\draw[very thick] (0,0) rectangle (1.2,1.2);
\draw[line,thick,->-=.55]  (.6,0) arc (0:45:.6); 
\draw[line,thick,->-=.55]  (0,0.6) arc (90:45:.6); 
\draw[line,thick,-<-=.55] (.6,1.2) arc (-180:-135:.6);
\draw[line,thick,-<-=.55] (1.2,0.6) arc (-90:-135:.6);
\draw[line,thick,->-=.55] (0.6/1.414,0.6/1.414)--(1.2-0.6/1.414,1.2-0.6/1.414);
\node at (1.2+0.2,0.6) {$\mathcal{A}$};
\node at (0.6,1.2+0.2) {$\mathcal{A}$};
\node at (-0.2,0.6) {$\mathcal{A}$};
\node at (0.6,-0.2) {$\mathcal{A}$};
\node at (0.6+0.1,0.6-0.1) {$\mathcal{A}$};
\filldraw[black] (0.6/1.414,0.6/1.414) circle (1pt);
\filldraw[black] (1.2-0.6/1.414,1.2-0.6/1.414) circle (1pt);
 \node at (0.3,0.3) {$m$};
 \node at (0.95,0.95) {$m^{\vee}$};
\end{tikzpicture}
\end{gathered} \qquad \Longleftrightarrow \qquad
    \begin{gathered}
        \begin{tikzpicture}
                \draw[line] (0,0.7)--(0,1.5);
                \draw[line] (3.6,0.7+1.2)--(3.6,1.5+1.2);
                \draw[line] (0.9,0)--(1.7,0);
                \draw[line] (0.9,0-0.2)--(1.7,0-0.2);
                \draw[line] (0.9+1.2,3.6)--(1.7+1.2,3.6);
                \draw[line] (0.9+1.2,0+0.2+3.6)--(1.7+1.2,0+0.2+3.6);
                \draw[line,thick,->-=0.55] (0,1)--(1.2+0.2,1);
                \draw[line,thick,->-=0.55] (1.2+0.2,0)--(1.2+0.2,1);
                \draw[line,thick,->-=0.55] (1.2+0.2,1)--(1.2+0.2+1.2,1+1.2);
                \draw[line,thick,->-=0.55] (1.2+0.2+1.2,1+1.2)--(1.2+0.2+1.2+1,1+1.2);
                \draw[line,thick,->-=0.55] (1.2+0.2+1.2,1+1.2)--(1.2+0.2+1.2,1+1.2+1.4);
                \filldraw[black] (1.2+0.2+1.2,1+1.2) circle (1.5pt);
                \filldraw[black] (1.2+0.2,1) circle (1.5pt);
                \node at (0.4,0.8) {$\mathcal{A}$};
                \node at (1.6,0.4) {$\mathcal{A}$};
                \node at (1.8+0.4,1.8-0.3) {$\mathcal{A}$};
                \node at (2.8,3.3) {$\mathcal{A}$};
                \node at (3.3,2.0) {$\mathcal{A}$};
                \node at (2.7,1.9) {$m^{\vee}$};
                \node at (1.7,0.9) {$m$};
            \end{tikzpicture}
        \end{gathered}
\end{equation}
where the endpoints of the horizontal and vertical lines are identified separately. Including the $G$-gauge background, the defect network on the torus takes the form
    \begin{equation}
    \begin{gathered}
        \begin{tikzpicture}
                \draw[line,very thick,red,->-=0.55] (0,1.2)--(1.2,1.2);
                \draw[line,very thick,red,->-=0.55] (1.2,0)--(1.2,1.2);
                \draw[line,very thick,red,->-=0.55] (1.2,1.2)--(2.4,2.4);
                \draw[line,very thick,red,->-=0.55] (2.4,2.4)--(3.6,2.4);
                \draw[line,very thick,red,->-=0.55] (2.4,2.4)--(2.4,3.6);
                \node at (0.6,1.5) {$g$};
                \node at (0.9,0.6) {$h$};
                \node at (1.8-0.3,1.8+0.2) {$gh$};
                \node at (2.1,3) {$h$};
                \node at (3,2.7) {$g$};
                \draw[line] (0,0.7)--(0,1.5);
                \draw[line] (3.6,0.7+1.2)--(3.6,1.5+1.2);
                \draw[line] (0.9,0)--(1.7,0);
                \draw[line] (0.9,0-0.2)--(1.7,0-0.2);
                \draw[line] (0.9+1.2,3.6)--(1.7+1.2,3.6);
                \draw[line] (0.9+1.2,0+0.2+3.6)--(1.7+1.2,0+0.2+3.6);
                \draw[line,thick,->-=0.55] (0,1)--(1.2+0.2,1);
                \draw[line,thick,->-=0.55] (1.2+0.2,0)--(1.2+0.2,1);
                \draw[line,thick,->-=0.55] (1.2+0.2,1)--(1.2+0.2+1.2,1+1.2);
                \draw[line,thick,->-=0.55] (1.2+0.2+1.2,1+1.2)--(1.2+0.2+1.2+1,1+1.2);
                \draw[line,thick,->-=0.55] (1.2+0.2+1.2,1+1.2)--(1.2+0.2+1.2,1+1.2+1.4);
                \filldraw[black] (1.2+0.2+1.2,1+1.2) circle (1.5pt);
                \filldraw[black] (1.2+0.2,1) circle (1.5pt);
                \node at (0.4,0.8) {$\mathcal{A}$};
                \node at (1.6,0.4) {$\mathcal{A}$};
                \node at (1.8+0.4,1.8-0.3) {$\mathcal{A}$};
                \node at (2.8,3.3) {$\mathcal{A}$};
                \node at (3.3,2.0) {$\mathcal{A}$};
                \node at (2.7,1.9) {$m^{\vee}$};
                \node at (1.7,0.9) {$m$};
            \end{tikzpicture}
        \end{gathered}
    \end{equation}
Equation \eqref{eq:RepG_to_G_trivial_sector} then states that this Frobenius defect network projects onto the $(e,e)$-sector, namely the partition function $Z_{e,e}$ with trivial $G$-background.

To see how this works, we read off the action of the $G$-defect from the following diagram
    \begin{equation}
        \begin{gathered}
        \begin{tikzpicture}
            \draw[line,thick] (-1,2)--(1,3);
            \draw[line,thick] (1,2)--(0.2,2+0.4);
            \draw[line,thick] (-0.2,2+0.6)--(-1,3);
            \draw[line,thick] (-1,3)--(-1,3.5);
            \draw[line,thick] (1,3)--(1,3.5);
            \draw[line,thick,->-=0.55] (-1,1)--(-1,2);
            \draw[line,thick,->-=0.55] (1,1)--(1,2);
            \draw[line,thick,-<-=0.55] (-1,1)--(0,0);
            \draw[line,thick,-<-=0.55] (1,1)--(0,0);
            \draw[line,thick,-<-=0.55] (0,0)--(0,-1.414);
            \draw[line,thick,->-=0.55] (-1,-1.414-1)--(0,-1.414);
            \draw[line,thick,->-=0.55] (1,-1.414-1)--(0,-1.414);
            \draw[line,thick,->-=0.55] (-1,-1.414-1-1)--(-1,-1.414-1);
            \draw[line,thick,->-=0.55] (1,-1.414-1-1)--(1,-1.414-1);   
            \filldraw[black] (0,0) circle (1.5pt);
            \filldraw[black] (0,-1.414) circle (1.5pt);
            \node at (0,0.45) {$m^{\vee}$};
            \node at (0,-1.414-0.3) {$m$};
            \node at (-0.3,-0.707) {$\mathcal{A}$};
            \node at (-0.8,-1.8) {$\mathcal{A}$};
            \node at (0.8,-1.8) {$\mathcal{A}$};
            \node at (-0.8,0.3) {$\mathcal{A}$};
            \node at (0.8,0.3) {$\mathcal{A}$};
            \node at (-0.5,-1.414-1-0.3) {$h$};
            \node at (-0.45,1.7) {$g$};
            \draw[line,very thick,red,-<-=0.1] (-1.3,-1.414-1-0.3)--(-0.7,-1.414-1-0.3);
            \draw[line,very thick,red,-<-=0.1] (-0.7,1.7)--(-1.3,1.7);
            \draw[thick,line] (0.7-2,-1.414-1-1)--(1.3-2,-1.414-1-1);
            \draw[thick,line] (0.7,-1.414-1-1)--(1.3,-1.414-1-1);
            \draw[thick,line] (0.7,-1.414-1-1-0.15)--(1.3,-1.414-1-1-0.15);
            \draw[thick,line] (0.7-2,-1.414-1-1)--(1.3-2,-1.414-1-1);
            \draw[thick,line] (0.7,2+1.5)--(1.3,2+1.5);
            \draw[thick,line] (0.7,2+0.15+1.5)--(1.3,2+0.15+1.5);
            \draw[thick,line] (0.7-2,2+1.5)--(1.3-2,2+1.5);
            \draw[dotted,very thick] (-0.4,2.5)--(-3.5,2.5);
            \draw[dotted,very thick] (-1.5,-1.414-1-0.3)--(-3.5,-1.414-1-0.3);
            \draw[dotted,very thick] (-1.5,1.7)--(-3.5,1.7);
            \draw[dotted,very thick] (-0.3,0)--(-3.5,0);
            \draw[dotted,very thick] (-0.3,-1.414)--(-3.5,-1.414);
            \node at (-5,2.5) {$S_{\mathcal{A,\mathcal{A}}}$};
            \node at (-5,-1.414-1-0.3) {$L_{h^{-1}}\otimes L_e$};
            \node at (-5,-1.414) {$m$};
            \node at (-5,0) {$m^{\vee}$};
            \node at (-5,1.7) {$ L_{g}\otimes L_e$};
        \end{tikzpicture}
        \end{gathered}
    \end{equation}
thus it maps a basis vector $v_k\otimes v_\ell$ according to
\begin{equation}
\begin{gathered}
    v_k\otimes v_\ell
    \xrightarrow{L_{h^{-1}}\otimes L_e}
    v_{h^{-1}k}\otimes v_\ell
    \xrightarrow{m}
    \delta_{\ell,h^{-1}k}\, v_\ell
    \xrightarrow{m^{\vee}}
    \delta_{\ell,h^{-1}k}\, v_\ell\otimes v_\ell \\
    \xrightarrow{L_g\otimes L_e}
    \delta_{\ell,h^{-1}k}\, v_{g\ell}\otimes v_\ell
    \xrightarrow{S_{\mathcal{A},\mathcal{A}}}
    \delta_{\ell,h^{-1}k}\, v_\ell\otimes v_{g\ell}\,.
\end{gathered}
\end{equation}
For this to contribute to the trace, we must have $k=\ell$, and in addition $g$ and $h$ must satisfy
\begin{equation}
    k=h^{-1}k\,,\qquad k=gk\,.
\end{equation}
This implies $h=g=e$.

We now explain how the multiplication and comultiplication may be modified so as to extract a nontrivial commuting pair $(g,h)$. Before doing so, it is useful to consider a simpler example in which there is only a single $\mathcal{A}$-loop along the temporal direction, together with a $G$-gauge defect inserted along the spatial direction
\begin{equation}
    \begin{gathered}
        \begin{tikzpicture}[scale=2]
            \draw[very thick] (0,0) rectangle (1.2,1.2);
            \draw[line,thick,->-=0.85] (0.6,0)--(0.6,1.2);
            \draw[line,very thick,red,->-=0.85] (0,0.6)--(1.2,0.6);
            \node at (0.45,0.3) {$\mathcal{A}$};
            \node at (0.3,0.75) {$g$};
        \end{tikzpicture}
    \end{gathered}
\end{equation}
This configuration is represented by
\begin{equation}
    \begin{gathered}
        \begin{tikzpicture}
            \draw[line,thick,->-=0.55] (0,-1.5)--(0,1.5);
            \draw[line,thick] (-0.3,-1.5)--(0.3,-1.5);
            \draw[line,thick] (-0.3,1.5)--(0.3,1.5);
            \draw[line,very thick,red,-<-=0.15] (0.3,-0.75)--(-0.3,-0.75);
            \node at (0.5,-0.75) {$g$};
            \draw[dotted,very thick] (-0.5,-0.75)--(-2.5,-0.75);
            \node at (-3,-0.75) {$L_g$};
        \end{tikzpicture}
    \end{gathered}
\end{equation}
and simply yields
\begin{equation}
    \text{Tr}_{\mathcal{A}} L_g = \dim\mathcal{A}\,\delta_{g,e}\,,
\end{equation}
thereby projecting onto the trivial sector $g=e$. In order to project onto the non-trivial sector labeled by a general element $g$, we can insert a morphism operator $\mathcal{P}_{x}$ along the $\mathcal{A}$ line and consider 
    \begin{equation}
        \begin{gathered}
            \begin{tikzpicture}
                \draw[line,thick,->-=0.55] (0,-1.5)--(0,1.5);
                \draw[line,thick] (-0.3,-1.5)--(0.3,-1.5);
                \draw[line,thick] (-0.3,1.5)--(0.3,1.5);
                \draw[line,very thick,red,-<-=0.15] (0.3,-0.75)--(-0.3,-0.75);
                \filldraw[black] (0,0.75) circle (2pt);
                \node at (0.5,-0.75) {$g$};
                \draw[dotted,very thick] (-0.5,-0.75)--(-2.5,-0.75);
                \node at (-3,-0.75) {$L_g$};
                \draw[dotted,very thick] (-0.3,0.75)--(-2.5,0.75);
                \node at (-3,0.75) {$\mathcal{P}_x$}; 
            \end{tikzpicture}
        \end{gathered}
    \end{equation}
The simplest choices are $\mathcal{P}_x=L_{x^{-1}}$ and $\mathcal{P}_x=R_{x^{-1}}$. The two choices give, respectively
    \begin{equation}
        \text{Tr}_{\mathcal{A}}\left( L_{x^{-1}}\circ L_g\right) = |G|\delta_{x,g}\,,\quad \text{Tr}_{\mathcal{A}}\left(R_{x^{-1}}\circ L_g \right) = |C_G(g)|\delta_{x,[g]}\,,
    \end{equation}
where the second delta function is defined by $\delta_{x,[g]}=\sum_{k\in [g]}\delta_{x,k}$, and is non-zero when $x$ lies in the conjugacy class $[g]$ of $g$. To see this, note that for any basis vector $v_h$, the operator $R_{x^{-1}}\circ L_g$ acts as
    \begin{equation}
        v_h \xrightarrow{L_g} v_{gh} \xrightarrow{R_{x^{-1}}} v_{ghx^{-1}}\,,
    \end{equation}
which contributes to the trace only when $h=ghx^{-1}$, or equivalently when $x=h^{-1}gh$. Thus $x$ must belong to the conjugacy class $[g]$, and the number of such $h$ is $|C_G(g)|$, where $C_G(g)$ denotes the centralizer group of $g$. In summary, the choice $\mathcal{P}_x=L_{x^{-1}}$ projects onto the sector $x=g$, whereas $\mathcal{P}_x=R_{x^{-1}}$ projects onto the entire conjugacy class $x\in [g]$.

The discussion of a single $\mathcal{A}$-loop motivates the following modification of the multiplication
\begin{equation}
    m_{(\rho,r),(\sigma,s)}^{(\lambda,t)}(y)
    =
    J^{\vee}_{\lambda,t}
    \circ
    m
    \circ
    (\mathcal{P}_{y^{-1}}\otimes \text{id}_{\mathcal{A}})
    \circ
    (J_{\rho,r}\otimes J_{\sigma,s})\,,
\end{equation}
and comultiplication
\begin{equation}
    m^{\vee(\rho,r),(\sigma,s)}_{(\lambda,t)}(x)
    =
    (J^{\vee}_{\rho,r}\otimes J^{\vee}_{\sigma,s})
    \circ
    (\mathcal{P}_{x}\otimes \text{id}_{\mathcal{A}})
    \circ
    m^{\vee}
    \circ
    J_{\lambda,t}\,.
\end{equation}
In this way, the operator $\mathcal{P}$ is inserted along the $\mathcal{A}$-line, and the corresponding diagram takes the form
    \begin{equation}\label{eq:diagram_for_general_coefficient}
        \begin{gathered}
        \begin{tikzpicture}
            \draw[line,thick] (-1,2)--(1,3);
            \draw[line,thick] (1,2)--(0.2,2+0.4);
            \draw[line,thick] (-0.2,2+0.6)--(-1,3);
            \draw[line,thick] (-1,3)--(-1,3.5);
            \draw[line,thick] (1,3)--(1,3.5);
            \draw[line,thick,->-=0.55] (-1,1)--(-1,2);
            \draw[line,thick,->-=0.55] (1,1)--(1,2);
            \draw[line,thick,-<-=0.55] (-1,1)--(0,0);
            \draw[line,thick,-<-=0.55] (1,1)--(0,0);
            \draw[line,thick,-<-=0.55] (0,0)--(0,-1.414);
            \draw[line,thick,->-=0.55] (-1,-1.414-1)--(0,-1.414);
            \draw[line,thick,->-=0.55] (1,-1.414-1)--(0,-1.414);
            \draw[line,thick,->-=0.55] (-1,-1.414-1-1)--(-1,-1.414-1);
            \draw[line,thick,->-=0.55] (1,-1.414-1-1)--(1,-1.414-1);
            \filldraw[black] (0,0) circle (1.5pt);
            \filldraw[black] (0,-1.414) circle (1.5pt);
            \filldraw[black] (-1,1) circle (2pt);
            \filldraw[black] (-0.8,-1.414-0.8) circle (2pt);
            \node at (0,0.45) {$m^{\vee}$};
            \node at (0,-1.414-0.3) {$m$};
            \node at (-0.3,-0.707) {$\mathcal{A}$};
            \node at (-0.8,-1.8) {$\mathcal{A}$};
            \node at (0.8,-1.8) {$\mathcal{A}$};
            \node at (-0.8,0.3) {$\mathcal{A}$};
            \node at (0.8,0.3) {$\mathcal{A}$};
            \node at (-0.5,-1.414-1-0.3) {$h$};
            \node at (-0.45,1.7) {$g$};
            \draw[line,very thick,red,-<-=0.1] (-1.3,-1.414-1-0.3)--(-0.7,-1.414-1-0.3);
            \draw[line,very thick,red,-<-=0.1] (-0.7,1.7)--(-1.3,1.7);
            \draw[thick,line] (0.7-2,-1.414-1-1)--(1.3-2,-1.414-1-1);
            \draw[thick,line] (0.7,-1.414-1-1)--(1.3,-1.414-1-1);
            \draw[thick,line] (0.7,-1.414-1-1-0.15)--(1.3,-1.414-1-1-0.15);
            \draw[thick,line] (0.7-2,-1.414-1-1)--(1.3-2,-1.414-1-1);
            \draw[thick,line] (0.7,2+1.5)--(1.3,2+1.5);
            \draw[thick,line] (0.7,2+0.15+1.5)--(1.3,2+0.15+1.5);
            \draw[thick,line] (0.7-2,2+1.5)--(1.3-2,2+1.5);
            \draw[dotted,very thick] (-0.4,2.5)--(-3.5,2.5);
            \draw[dotted,very thick] (-1.5,-1.414-1-0.3)--(-3.5,-1.414-1-0.3);
            \draw[dotted,very thick] (-1.5,1.7)--(-3.5,1.7);
            \draw[dotted,very thick] (-0.3,0)--(-3.5,0);
            \draw[dotted,very thick] (-0.3,-1.414)--(-3.5,-1.414);
            \draw[dotted,very thick] (-1.3,1)--(-3.5,1);
            \draw[dotted,very thick] (-0.8-0.3,-1.414-0.8)--(-3.5,-1.414-0.8);
            \node at (-5,1) {$\mathcal{P}_{x}$};
            \node at (-5,-1.414-0.8) {$\mathcal{P}_{y^{-1}}$};
            \node at (-5,2.5) {$S_{\mathcal{A,\mathcal{A}}}$};
            \node at (-5,-1.414-1-0.3) {$L_{h^{-1}}\otimes L_e$};
            \node at (-5,-1.414) {$m$};
            \node at (-5,0) {$m^{\vee}$};
            \node at (-5,1.7) {$ L_{g}\otimes L_e$};
        \end{tikzpicture}
        \end{gathered}
    \end{equation}
At first sight, the left action $\mathcal{P}_x=L_{x^{-1}}$ appears to be the most natural choice, since it projects directly onto the sector $x=g$. However, the modified maps $m(y)$ and $m^{\vee}(x)$ defined using the left action fail to be an intertwiner and a co-intertwiner, respectively. As a result, they cannot be expanded in the intertwiner basis $I$ and co-intertwiner basis $I^{\vee}$ so as to isolate the contribution from each $\text{Rep}(G)$ defect channel. The reason is simply that $J$ and $J^{\vee}$ are themselves intertwiners defined with respect to the left action, as in \eqref{eq:J_as_intertwiner}, and left actions do not commute in general.

Therefore, in order to expand $m(y)$ and $m^{\vee}(x)$ in terms of the same intertwiner and co-intertwiner bases $I$ and $I^{\vee}$, we must choose $\mathcal{P}_{y^{-1}}=R_y$ on the multiplication side and $\mathcal{P}_x=R_{x^{-1}}$ on the comultiplication side, since the right action automatically commutes with the left action. Indeed, one may check that the diagram \eqref{eq:diagram_for_general_coefficient} acts on a basis vector $v_k\otimes v_\ell$ as
\begin{equation}
\begin{gathered}
    v_k\otimes v_\ell
    \xrightarrow{L_{h^{-1}}\otimes L_e}
    v_{h^{-1}k}\otimes v_\ell
    \xrightarrow{\mathcal{P}_{y^{-1}}}
    v_{h^{-1}ky}\otimes v_\ell
    \xrightarrow{m}
    \delta_{\ell,h^{-1}ky}\,v_\ell
    \xrightarrow{m^{\vee}}
    \delta_{\ell,h^{-1}ky}\,v_\ell\otimes v_\ell \\
    \xrightarrow{\mathcal{P}_x}
    \delta_{\ell,h^{-1}ky}\,v_{\ell x^{-1}}\otimes v_\ell
    \xrightarrow{L_g\otimes L_e}
    \delta_{\ell,h^{-1}ky}\,v_{g\ell x^{-1}}\otimes v_\ell
    \xrightarrow{S_{\mathcal{A},\mathcal{A}}}
    \delta_{\ell,h^{-1}ky}\,v_\ell\otimes v_{g\ell x^{-1}}\,.
\end{gathered}
\end{equation}
For this to contribute to the trace, we must have $k=\ell$, and moreover $g$ and $h$ must satisfy
\begin{equation}
    k=h^{-1}ky\,,\qquad k=gkx^{-1}\,,
\end{equation}
or equivalently,
\begin{equation}
    x=k^{-1}gk\,,\qquad y=k^{-1}hk\,.
\end{equation}
Thus the insertion of $\mathcal{P}$ picks out all commuting pairs conjugate to $(g,h)$ under
\begin{equation}
    (g,h)\sim (k^{-1}gk,k^{-1}hk)\,.
\end{equation}
Since the partition function is also invariant under conjugation,
\begin{equation}
    Z_{g,h}=Z_{k^{-1}gk,k^{-1}hk}\,,
\end{equation}
this indeed yields the desired inverse transformation.

We may then expand the modified multiplication as
\begin{equation}\label{eq:M-coefficient}
\begin{split}
    m_{(\rho,r),(\sigma,s)}^{(\lambda,t)} (y) =&  J^{\vee}_{\lambda,t} \circ m\circ (R_{y}\otimes \text{id}_{\mathcal{A}})\circ   (J_{\rho,r} \otimes J_{\sigma,s})\\
    =&\sum_{i=1}^{N_{\rho \sigma}^{\lambda}} M_{(\rho,r),(\sigma,s)}^{(\lambda,t),i} (y) I_{\rho,\sigma}^{\lambda;i}\,,
\end{split}
\end{equation}
and the modified comultiplication as
\begin{equation}\label{eq:Mvee-coefficient}
\begin{split}
m^{\vee (\rho,r),(\sigma,s)}_{(\lambda,t)}(x) =& (J^{\vee}_{\rho,r}\otimes J^{\vee}_{\sigma,s})\circ \left( R_{x^{-1}}\otimes \text{id}_{\mathcal{A}} \right) \circ m^{\vee} \circ J_{\lambda,t}\\
=&\sum_{i=1}^{N_{\rho \sigma}^{\lambda}} M^{\vee (\rho,r),(\sigma,s)}_{(\lambda,t),i}(x) I^{\vee \rho,\sigma}_{\lambda;i}\,.
\end{split}
\end{equation}
We then introduce the coefficient
\begin{equation}\label{eq:gauge_Rep_general_coefficient}
        C_{\rho,\sigma}^{\lambda;i,j}(x,y) = \dim \mathcal{A}\sum_{r,s,t} M^{\vee (\sigma,s),(\rho,r)}_{(\lambda,t),j}(x) M_{(\rho,r),(\sigma,s)}^{(\lambda,t),i}(y)\,,
    \end{equation}
with $\dim \mathcal{A} = |G|$. With this notation, the inverse Fourier transformation is given by
    \begin{equation}\label{eq:gauge_Rep_general}
        Z_{g,h} = \frac{1}{|G|} \sum_{\rho,\sigma,\lambda;i,j} C_{\rho,\sigma}^{\lambda;i,j}(g,h)  \widetilde{Z}_{\rho,\sigma}^{\lambda;i,j}\,.
    \end{equation}

\subsection{A nice choice of gauge}\label{sec:nice_gauge}
Up to this point, we have not chosen any specific intertwiners $I,J$ and cointertwiners $I^{\vee},J^{\vee}$. In this subsection, we will pick a convenient choice of gauge such that the inverse kernel will simplify significantly. We choose $I$ and $I^{\vee}$ to satisfy the orthogonality and completeness
    \begin{equation}
        I^{\mu;i}_{\rho,\sigma} \circ I^{\vee \rho,\sigma}_{\lambda;j} = \delta_{\mu,\lambda} \delta_{i,j} \mathbf{1}_{V_{\lambda}}\,,\quad \sum_{\lambda,i}I^{\vee \rho,\sigma}_{\lambda;i} \circ I^{\lambda;i}_{\rho,\sigma} = \mathbf{1}_{\rho}\otimes \mathbf{1}_{\sigma}\,.
    \end{equation} 
For $J$ and $J^{\vee}$, notice that the Schur orthogonality states
\begin{equation}
        \frac{1}{|G|}\sum_{g}\overline{\rho(g)_{\alpha r}} \sigma(g)_{\beta s} = \frac{\delta_{\rho,\sigma} \delta_{\alpha,\beta}\delta_{r,s}}{d_{\rho}}\,,
    \end{equation}
and we can expand any function on $G$ using the matrix elements of the irreducible representation. We will choose the $J$ and $J^{\vee}$ according to
    \begin{equation}
        J_{\rho,r} \circ e^{\rho}_{\alpha} = \sqrt{\frac{d_{\rho}}{|G|}}\sum_{g\in G} \overline{\rho(g)}_{\alpha r} v_g\,,\quad J^{\vee}_{\rho,r}\circ v_g = \sqrt{\frac{d_{\rho}}{|G|}} \sum_{\alpha} \rho(g)_{\alpha r} e_{\alpha}^{\rho}\,,
    \end{equation}
which satisfies
    \begin{equation}
        J^{\vee}_{\sigma,s}\circ J_{\rho,r} = \delta_{\rho,\sigma} \delta_{r,s} \mathbf{1}_{E_{\rho}}\,,\quad \sum_{\rho,r} J_{\rho,r}\circ J^{\vee}_{\rho,r} = \mathbf{1}_{\mathcal{A}}\,.
    \end{equation}

We first consider the multiplication
\begin{equation}
    m_{(\rho,r),(\sigma,s)}^{(\lambda,t)}(y)
    =
    J^{\vee}_{\lambda,t}
    \circ
    m
    \circ
    (R_{y}\otimes \text{id}_{\mathcal{A}})
    \circ
    (J_{\rho,r}\otimes J_{\sigma,s})\,,
\end{equation}
acting on the basis $e^{\rho}_{\alpha}\otimes e^{\sigma}_{\beta}$ gives
\begin{equation}
\begin{split}
    &J^{\vee}_{\lambda,t}
    \circ
    m
    \circ
    (R_{y}\otimes \text{id}_{\mathcal{A}})
    \circ
    (J_{\rho,r}\otimes J_{\sigma,s})\circ e^{\rho}_{\alpha}\otimes e^{\sigma}_{\beta} \\
    =&J^{\vee}_{\lambda,t}
    \circ
    m
    \circ
    (R_{y}\otimes \text{id}_{\mathcal{A}})
    \circ \left(\frac{\sqrt{d_{\rho} d_{\sigma}}}{|G|} \sum_{g,h} \overline{\rho(g)}_{\alpha r} \overline{\sigma(h)}_{\beta s} v_g\otimes v_h \right)\\
    =&J^{\vee}_{\lambda,t}
    \circ
    m
    \circ \left(\frac{\sqrt{d_{\rho} d_{\sigma}}}{|G|} \sum_{g,h} \overline{\rho(gy^{-1})}_{\alpha r} \overline{\sigma(h)}_{\beta s} v_g\otimes v_h \right)\\
    =&J^{\vee}_{\lambda,t}
    \circ \left(\frac{\sqrt{d_{\rho} d_{\sigma}}}{|G|} \sum_{g} \overline{\rho(gy^{-1})}_{\alpha r} \overline{\sigma(g)}_{\beta s} v_g \right)\\
    =& \frac{\sqrt{d_{\rho} d_{\sigma} d_{\lambda}}}{|G|^{3/2}} \sum_g \sum_{\gamma} \lambda(g)_{\gamma t}\overline{\rho(gy^{-1})}_{\alpha r} \overline{\sigma(g)}_{\beta s} e^{\lambda}_{\gamma}\\
    =& \frac{\sqrt{d_{\rho} d_{\sigma} d_{\lambda}}}{|G|^{3/2}} \sum_g \sum_{\gamma,\delta} \lambda(g)_{\gamma t}\overline{\rho(g)}_{\alpha \delta}  \overline{\rho(y^{-1})}_{\delta r}\overline{\sigma(g)}_{\beta s} e^{\lambda}_{\gamma} \,.
\end{split}
\end{equation}
To proceed, notice that for $\rho(k^{-1}) \otimes \sigma({k^{-1}})$ we have
    \begin{equation}
        \rho(g^{-1}) \otimes \sigma({g^{-1}}) = \sum_{\lambda',i}I^{\vee \rho,\sigma}_{\lambda';i} \circ I^{\lambda';i}_{\rho,\sigma}\circ \rho(g^{-1}) \otimes \sigma({g^{-1}}) = \sum_{\lambda',i}I^{\vee \rho,\sigma}_{\lambda';i} \circ \lambda'(g^{-1}) \circ I^{\lambda';i}_{\rho,\sigma}\,,
    \end{equation}
acting on $e^{\rho}_{\alpha}\otimes e^{\sigma}_{\beta}$, we have the coefficients
    \begin{equation}
        \rho(g^{-1})_{\delta \alpha} \sigma(g^{-1})_{s\beta} = \sum_{\lambda',i} \sum_{a,b}\mathcal{I}^{\vee (\rho,\delta),(\sigma,s)}_{(\lambda',a);i} \lambda'(g^{-1})_{ab} \mathcal{I}^{(\lambda',b);i}_{(\rho,\alpha),(\sigma,\beta)}\,,
    \end{equation}
and substitute that back we obtain
    \begin{equation}
        \frac{\sqrt{d_{\rho} d_{\sigma} d_{\lambda}}}{|G|^{3/2}} \sum_g \sum_{\gamma,\delta} \lambda(g)_{\gamma t}  \overline{\rho(y^{-1})}_{\delta r} \sum_{\lambda',i} \sum_{a,b}\mathcal{I}^{\vee (\rho,\delta),(\sigma,s)}_{(\lambda',a);i} \lambda'(g^{-1})_{ab} \mathcal{I}^{(\lambda',b);i}_{(\rho,\alpha),(\sigma,\beta)} e^{\lambda}_{\gamma}\,.
    \end{equation}
We then sum over $g$ and use the Schur orthogonality
    \begin{equation}
        \frac{1}{|G|}\sum_g \lambda(g)_{\gamma t} \overline{\lambda'}(g)_{ba} = \frac{\delta_{\lambda,\lambda'}\delta_{\gamma,b}\delta_{t,a}}{d_{\lambda}}\,,
    \end{equation}
and we have
    \begin{equation}
        \sqrt{\frac{d_{\rho}d_{\sigma}}{|G|d_{\lambda}}} \sum_i\sum_{\gamma,\delta} \overline{\rho(y^{-1})}_{\delta r} \mathcal{I}^{\vee (\rho,\delta),(\sigma,s)}_{(\lambda,t);i} \mathcal{I}^{(\lambda,\gamma);i}_{(\rho,\alpha),(\sigma,\beta)} e^{\lambda}_{\gamma}\,.
    \end{equation}
Compare to \eqref{eq:M-coefficient}, we have the $M$-coefficient in the chosen gauge
    \begin{equation}
        M_{(\rho,r),(\sigma,s)}^{(\lambda,t),i}(y)=\sqrt{\frac{d_{\rho}d_{\sigma}}{|G|d_{\lambda}}} \sum_{\delta} \overline{\rho(y^{-1})}_{\delta r} \mathcal{I}^{\vee (\rho,\delta),(\sigma,s)}_{(\lambda,t);i}\,.
    \end{equation}

Then consider the comultiplication
\begin{equation}
    m^{\vee(\rho,r),(\sigma,s)}_{(\lambda,t)}(x)
    =
    (J^{\vee}_{\rho,r}\otimes J^{\vee}_{\sigma,s})
    \circ
    (R_{x^{-1}}\otimes \text{id}_{\mathcal{A}})
    \circ
    m^{\vee}
    \circ
    J_{\lambda,t}\,,
\end{equation}
acting on the basis $e^{\lambda}_{\gamma}$, we have
    \begin{equation}
        \begin{split}
        &(J^{\vee}_{\rho,r}\otimes J^{\vee}_{\sigma,s})
    \circ
    (R_{x^{-1}}\otimes \text{id}_{\mathcal{A}})
    \circ
    m^{\vee}
    \circ
    J_{\lambda,t}\circ e^{\lambda}_{\gamma} \\
            =&(J^{\vee}_{\rho,r}\otimes J^{\vee}_{\sigma,s})
    \circ
    (R_{x^{-1}}\otimes \text{id}_{\mathcal{A}})
    \circ
    m^{\vee}\circ \sqrt{\frac{d_{\lambda}}{|G|}} \sum_g \overline{\lambda(g)}_{\gamma t}v_g\\
    =&(J^{\vee}_{\rho,r}\otimes J^{\vee}_{\sigma,s})
    \circ
    (R_{x^{-1}}\otimes \text{id}_{\mathcal{A}})
    \circ \sqrt{\frac{d_{\lambda}}{|G|}} \sum_g \overline{\lambda(g)}_{\gamma t}v_g\otimes v_g\\
    =&(J^{\vee}_{\rho,r}\otimes J^{\vee}_{\sigma,s})
    \circ \sqrt{\frac{d_{\lambda}}{|G|}} \sum_g \overline{\lambda(g)}_{\gamma t}v_{gx^{-1}}\otimes v_g\\
    =&\sqrt{\frac{d_{\rho}d_{\sigma}d_{\lambda}}{|G|^{3}}} \sum_g \sum_{\alpha,\beta}\rho(gx^{-1})_{\alpha r} \sigma(g)_{\beta s} \overline{\lambda(g)}_{\gamma t} e_{\alpha}^{\rho}\otimes e_{\beta}^{\sigma}\\
    =&\sqrt{\frac{d_{\rho}d_{\sigma}d_{\lambda}}{|G|^{3}}} \sum_g \sum_{\alpha,\beta,\delta}\rho(g)_{\alpha \delta} \rho(x^{-1})_{\delta r} \sigma(g)_{\beta s} \overline{\lambda(g)}_{\gamma t} e_{\alpha}^{\rho}\otimes e_{\beta}^{\sigma}\,.
        \end{split}
    \end{equation}
Using the same trick
    \begin{equation}
        \rho(g)_{\alpha \delta} \sigma(g)_{\beta s} = \sum_{\lambda',i} \sum_{a,b}\mathcal{I}^{\vee (\rho,\alpha),(\sigma,\beta)}_{(\lambda',a);i} \lambda'(g)_{ab} \mathcal{I}^{(\lambda',b);i}_{(\rho,\delta),(\sigma,s)}\,,
    \end{equation}
we have
    \begin{equation}
        \sqrt{\frac{d_{\rho}d_{\sigma}d_{\lambda}}{|G|^{3}}} \sum_g \sum_{\alpha,\beta,\delta}\sum_{\lambda',i} \sum_{a,b}\mathcal{I}^{\vee (\rho,\alpha),(\sigma,\beta)}_{(\lambda',a);i} \lambda'(g)_{ab} \mathcal{I}^{(\lambda',b);i}_{(\rho,\delta),(\sigma,s)} \rho(x^{-1})_{\delta r}\overline{\lambda(g)}_{\gamma t} e_{\alpha}^{\rho}\otimes e_{\beta}^{\sigma}\,.
    \end{equation}
Sum over $g$ and using the Schur orthogonality
    \begin{equation}
        \frac{1}{|G|}\sum_g \lambda'(g)_{ab} \overline{\lambda(g)}_{\gamma t} = \frac{\delta_{\lambda,\lambda'} \delta_{a,\gamma} \delta_{b,t}}{d_{\lambda}}\,,
    \end{equation}
we have
    \begin{equation}
        \sqrt{\frac{d_{\rho}d_{\sigma}}{|G| d_{\lambda}}} \sum_{\alpha,\beta,\delta} \sum_i \mathcal{I}^{\vee (\rho,\alpha),(\sigma,\beta)}_{(\lambda,\gamma);i}\mathcal{I}^{(\lambda,t);i}_{(\rho,\delta),(\sigma,s)} \rho(x^{-1})_{\delta r}e_{\alpha}^{\rho}\otimes e_{\beta}^{\sigma}\,.
\end{equation}
Compare to \eqref{eq:Mvee-coefficient}, we have the $M^{\vee}$ coefficient
    \begin{equation}
        M_{(\lambda,t),i}^{\vee (\rho,r),(\sigma,s)}(x) = \sqrt{\frac{d_{\rho}d_{\sigma}}{|G| d_{\lambda}}}\sum_{\delta}\mathcal{I}^{(\lambda,t);i}_{(\rho,\delta),(\sigma,s)}\rho(x^{-1})_{\delta r}\,.
    \end{equation}

The inverse Fourier kernel \eqref{eq:gauge_Rep_general_coefficient} can be evaluated directly as
    \begin{equation}
    \begin{split}
        C_{\rho,\sigma}^{\lambda;i,j}(x,y) =& \frac{d_{\rho}d_{\sigma}}{d_{\lambda}} \sum_{r,s,t}\sum_{\delta,\delta'} \sigma(x^{-1})_{\delta s}\mathcal{I}^{(\lambda,t);j}_{(\sigma,\delta),(\rho,r)} \rho(y)_{r \delta'} \mathcal{I}^{\vee (\rho,\delta'),(\sigma,s)}_{(\lambda,t);i}\\
        =&\frac{d_{\rho}d_{\sigma}}{d_{\lambda}}\text{Tr}_{E_{\sigma}\otimes E_{\rho}} \left[ S_{\rho,\sigma} \circ I^{\vee \rho,\sigma}_{\lambda;i} I_{\sigma,\rho}^{\lambda;j}\circ \sigma(x^{-1})\otimes \rho(y) \right]\\
        =&\frac{d_{\rho}d_{\sigma}}{d_{\lambda}}\text{Tr}_{E_{\rho}\otimes E_{\sigma}} \left[  I^{\vee \rho,\sigma}_{\lambda;i} I_{\sigma,\rho}^{\lambda;j}\circ S_{\rho,\sigma} \circ  \rho(y)\otimes \sigma(x^{-1}) \right]\,.
    \end{split}
    \end{equation}
Compare to the Fourier kernel $W$ in \eqref{eq:Rep(G)_G_kernel}
\begin{equation}
    W_{\rho,\sigma}^{\lambda;i,j}(g,h)
    =
    \textrm{Tr}_{E_{\rho}\otimes E_{\sigma}}
    \left[
        S_{\sigma,\rho}
        \circ
        I^{\vee \sigma,\rho}_{\lambda;j}
        I_{\rho,\sigma}^{\lambda;i}
        \circ
        \left(\rho(h^{-1})\otimes \sigma(g)\right)
    \right]\,,
\end{equation}
we have the particular simple relation
    \begin{equation}\label{eq:gauge-fixed-inverse-kernel}
        C^{\lambda;i,j}_{\rho,\sigma} (x,y) = \frac{d_{\rho}d_{\sigma}}{d_{\lambda}}W^{\lambda;j,i}_{\sigma,\rho}(y,x)\,.
    \end{equation}

Finally, we wish to examine
    \begin{equation}
        \sum_{\rho,\sigma}\sum_{\lambda,i,j} C_{\rho,\sigma}^{\lambda;i,j}(x,y)W_{\rho,\sigma}^{\lambda;i,j}(g,h)\,.
    \end{equation}
Introduce the notation
    \begin{equation}
        D_{\lambda;i,j}^{\rho,\sigma}=I^{\vee \rho,\sigma}_{\lambda;i} I^{\lambda;j}_{\sigma,\rho}\circ S_{\rho,\sigma}\,,\quad P^{\rho,\sigma}_{\lambda;i,j} = S_{\sigma,\rho} \circ I^{\vee \sigma,\rho}_{\lambda;j} \circ I^{\lambda;i}_{\rho,\sigma}\,,
    \end{equation}
which satisfies the trace orthogonality
    \begin{equation}
        \text{Tr}(D_{\lambda;i,j}^{\rho,\sigma}P^{\rho,\sigma}_{\lambda';i',j'})= d_{\lambda} \delta_{\lambda,\lambda'} \delta_{i,i'} \delta_{j,j'}\,.
    \end{equation}
and the $C,W$ coefficients are written as
    \begin{equation}
    \begin{gathered}
        C^{\lambda;i,j}_{\rho,\sigma}(x,y) = \frac{d_{\rho}d_{\sigma}}{d_{\lambda}} \text{Tr}_{E_{\rho}\otimes E_{\sigma}} \left[D_{\lambda;i,j}^{\rho,\sigma} \circ \rho(y)\otimes \sigma(x^{-1}) \right]\\
        W^{\lambda;i,j}_{\rho,\sigma}(g,h)=\text{Tr}_{E_{\rho}\otimes E_{\sigma}}  \left[P^{\rho,\sigma}_{\lambda;i,j} \circ \rho(h^{-1}) \otimes \sigma(g) \right]\,,
    \end{gathered}
    \end{equation}
    
To begin with, we need the following identity. Denote $R(g) = \rho(g)\otimes \sigma(g)$, for arbitrary $B\in \text{End}(E_{\rho}\otimes E_{\sigma})$ we have
    \begin{equation}
        \Lambda(B) \equiv \frac{1}{|G|} \sum_k R(k) BR(k^{-1}) = \sum_{\lambda;i,j}\frac{1}{d_{\lambda}}D^{\rho,\sigma}_{\lambda;i,j}\text{Tr} (P^{\rho,\sigma}_{\lambda;i,j}B)\,,
    \end{equation}
To prove that, it is easy to check $\Lambda(B)$ commute with any $R(g)$. Then insert the completeness for the two two different tensor order
    \begin{equation}
        \mathbf{1}_{\rho}\otimes \mathbf{1}_{\sigma} = \sum_{\lambda,i} I^{\vee \rho,\sigma}_{\lambda;i} \circ I^{\lambda;i}_{\rho,\sigma}= \sum_{\lambda,i} S_{\sigma,\rho} \circ I^{\vee \sigma,\rho}_{\lambda;i} \circ I^{\lambda;i}_{\sigma,\rho}\circ S_{\rho,\sigma}\,,
    \end{equation}
we have
    \begin{equation}
        \Lambda(B) = \sum_{\lambda,i} \sum_{\lambda',i'} I^{\vee \rho,\sigma}_{\lambda;i} \circ \left( I^{\lambda;i}_{\rho,\sigma} \circ \Lambda(B)\circ S_{\sigma,\rho} \circ I^{\vee \sigma,\rho}_{\lambda';i'}\right) \circ I^{\lambda';i'}_{\sigma,\rho}\circ S_{\rho,\sigma}\,.
    \end{equation}
Since $I^{\lambda;i}_{\rho,\sigma} \circ \Lambda(B)\circ S_{\sigma,\rho} \circ I^{\vee \sigma,\rho}_{\lambda';i'}$ is an intertwiner mapping $E_{\lambda'}\rightarrow E_{\lambda}$, and by Schur lemma, it is only nonzero when $\lambda=\lambda'$. Thus we can introduce the scalar $x_{\lambda;i,i'}$ such that
    \begin{equation}
        I^{\lambda;i}_{\rho,\sigma} \circ \Lambda(B)\circ S_{\sigma,\rho} \circ I^{\vee \sigma,\rho}_{\lambda';i'} = \delta_{\lambda,\lambda'} x_{\lambda;i,i'} \mathbf{1}_{\lambda}\,,
    \end{equation}
and expand
    \begin{equation}
        \Lambda(B) = \sum_{\lambda,i,j} x_{\lambda;i,j}I^{\vee \rho,\sigma}_{\lambda;i} I^{\lambda;j}_{\sigma,\rho}\circ S_{\rho,\sigma}\equiv \sum_{\lambda,i,j} x_{\lambda;i,j} D_{\lambda;i,j}^{\rho,\sigma}\,.
    \end{equation}
Multiply $P$ on both side and take the trace, we have
    \begin{equation}
        x_{\lambda;i,j} = \frac{1}{d_{\lambda}}\text{Tr} (P^{\rho,\sigma}_{\lambda;i,j}\Lambda(B)) = \frac{1}{d_{\lambda}}\text{Tr} (P^{\rho,\sigma}_{\lambda;i,j}B)\,,
    \end{equation}
where we use the fact that $P^{\rho,\sigma}_{\lambda;i,j}$ commute with all $R(g)$. Therefore we have proved the identity 
    \begin{equation}
        \Lambda(B) \equiv \frac{1}{|G|}\sum_k R(k)BR(k^{-1})= \sum_{\lambda;i,j} \frac{1}{d_{\lambda}}D^{\rho,\sigma}_{\lambda;i,j}\text{Tr} (P^{\rho,\sigma}_{\lambda;i,j}\Lambda(B))\,.
    \end{equation}

Multiply another endomorphism $A$ on both side and take the trace again, we have then
    \begin{equation}
         \frac{1}{|G|} \sum_k \text{Tr} \left[AR(k) BR(k^{-1})\right] = \sum_{\lambda;i,j}\frac{1}{d_{\lambda}} \text{Tr}(D^{\rho,\sigma}_{\lambda;i,j}A)\text{Tr} (P^{\rho,\sigma}_{\lambda;i,j}B)\,.
    \end{equation}
Choose
    \begin{equation}
        A=\rho(y)\otimes \sigma(x^{-1})\,,\quad B = \rho(h^{-1})\otimes \sigma(g)\,,
    \end{equation}
the LHS is
    \begin{equation}
        \frac{1}{|G|} \sum_k\text{Tr} \rho(y k h^{-1}k^{-1})\otimes \sigma(x^{-1}kgk^{-1}) = \frac{1}{|G|}\sum_k\chi_{\rho}(ykh^{-1}k^{-1}) \chi_{\sigma}(x^{-1}kgk^{-1})\,,
    \end{equation}
and the RHS is simply
    \begin{equation}
        \frac{1}{d_{\rho}d_{\sigma}} \sum_{\lambda;i,j}C^{\lambda;i,j}_{\rho,\sigma}(x,y) W^{\lambda;i,j}_{\rho,\sigma}(g,h)\,.
    \end{equation}
Multiply $d_{\rho} d_{\sigma}$ and sum over $\rho,\sigma$, one has
    \begin{equation}
    \begin{split}
        &\frac{1}{|G|^2}\sum_{\rho,\sigma,\lambda;i,j}C^{\lambda;i,j}_{\rho,\sigma}(x,y) W^{\lambda;i,j}_{\rho,\sigma}(g,h)\\
        =& \frac{1}{|G|^3}\sum_{\rho,\sigma}\sum_k d_{\rho} \chi_{\rho}(ykh^{-1}k^{-1}) d_{\sigma} \chi_{\sigma}(x^{-1}kgk^{-1})\\
        =&\frac{1}{|G|}\sum_k \delta_{x,kgk^{-1}} \delta_{y,khk^{-1}}\,,
    \end{split}
    \end{equation}
which is nonzero if $(x,y)$ is conjugate with $(g,h)$. Therefore one has
\begin{equation}
    \begin{split}
    &\frac{1}{|G|^2}\sum_{\rho,\sigma,\lambda;i,j}\sum_{\substack{g,h\in G\\ gh=hg}}C^{\lambda;i,j}_{\rho,\sigma}(x,y) W^{\lambda;i,j}_{\rho,\sigma}(g,h)Z_{g,h}\\
    =&\frac{1}{|G|}\sum_k \sum_{\substack{g,h\in G\\ gh=hg}} \delta_{x,kgk^{-1}} \delta_{y,khk^{-1}} Z_{g,h}\\
    =& \frac{|\text{Stab}(x,y)|}{|G|} \sum_{(g,h) \in \text{Conj}(x,y)}Z_{g,h} = \frac{1}{|\text{Conj}(x,y)|}\sum_{(g,h) \in \text{Conj}(x,y)}Z_{g,h}\,,
    \end{split}
\end{equation}
where the stabilizer group is defined by
    \begin{equation}
        \text{Stab}(g,h) = \{k\in G|kgk^{-1}=g,khk^{-1}=h \} \,,
    \end{equation}
and $|\text{Conj}(x,y)| = |G|/|\text{Stab}(g,h)|$ is the number of commuting pair conjugated with $(x,y)$. Since $Z_{g,h}$ are the same for all $(g,h)\in \text{Conj}(x,y)$, we simply have
\begin{equation}
    \frac{1}{|G|^2}\sum_{\rho,\sigma,\lambda;i,j}\sum_{\substack{g,h\in G\\ gh=hg}}C^{\lambda;i,j}_{\rho,\sigma}(x,y) W^{\lambda;i,j}_{\rho,\sigma}(g,h)Z_{g,h}=Z_{x,y}\,,
\end{equation}
as expected.

\section{Examples : Finite $\text{Rep}(G)$ category with multiplicity-free fusion}

In this section, we examine representation categories $\text{Rep}(G)$ with multiplicity-free fusion rules, namely those for which $N_{\rho\sigma}^{\lambda}=0,1$ for all $\rho,\sigma,\lambda\in\widehat{G}$. Our main examples are $\text{Rep}(G)$ for $G=S_3$, $D_4$, and $Q_8$, which were analyzed recently in \cite{Banerjee:2026mnx}. In later sections, we will turn to $\text{Rep}(A_4)$ as a nontrivial example with higher-multiplicity fusion, and to $\text{Rep}(SU(2))$ as an example of an infinite representation category.

For each example discussed in this section and in the following sections, we summarize the following data:
\begin{itemize}
    \item the conventions for the group structure and irreducible representations;
    \item the intertwiners $I$ and co-intertwiners $I^{\vee}$;
    \item the regular representation $\mathcal{A}$, viewed as a Frobenius algebra, together with the intertwiners $J$ and co-intertwiners $J^{\vee}$;
    \item the Fourier kernel;
    \item the inverse Fourier kernel.
\end{itemize}

To compare with the results of \cite{Banerjee:2026mnx}, we will adopt the same conventions for representation matrices, (co-)intertwiners, and (co-)multiplications as in that work, as well as in the earlier references \cite{Perez-Lona:2023djo,Perez-Lona:2024sds}. The group-twisted partition
functions in the two conventions are related by
    \begin{equation}
        Z^{\mathrm{BS}}_{g,h}
    = Z^{\mathrm{ours}}_{g,h^{-1}}\,,
    \end{equation}
and the two $\mathrm{Rep}(G)$ defect labels are exchanged
    \begin{equation}
        Z_{\rho,\sigma}^{\lambda;\mathrm{BS}}
    = \widetilde{Z}_{\sigma,\rho}^{\lambda;\mathrm{ours}}\,.
    \end{equation}
After these identifications, the forward kernels agree
exactly for $\mathrm{Rep}(S_3)$. For $\mathrm{Rep}(D_4)$ and
$\mathrm{Rep}(Q_8)$, they agree except for a phase in the
channels involving two distinct nontrivial invertible
objects. The invertible simple objects
$\{\mathbf{1},a,b,c\}$ form a fusion group
$\mathbb{Z}_2\times\mathbb{Z}_2$, and the corresponding
kernels satisfy
\begin{equation}
    W_{p,q}^{pq;\mathrm{ours}}(g,h)
    =
    \epsilon(p,q)\,
    W_{q,p}^{pq;\mathrm{BS}}(g,h^{-1})\,,
    \qquad p,q\in \mathbb{Z}_2\times\mathbb{Z}_2\,,
\end{equation}
where
\begin{equation}
    \epsilon(p,q)=
    \begin{cases}
        -1, & p,q\in\{a,b,c\},\quad p\neq q\,,\\
        +1, & \text{otherwise}\,.
    \end{cases}
\end{equation}
All other fusion channels agree under the same
identifications. The sign factor $\epsilon$ coincides with the genus-one discrete-torsion phase as commented in \cite{Perez-Lona:2023djo}.

\subsection{$\text{Rep}(S_3)$}

\paragraph{Group data}
We begin with the symmetric group $S_3$ and its representation category $\text{Rep}(S_3)$. The group $S_3$, namely the permutation group on three elements, is defined by
\begin{equation}
    S_3=\langle a,b \mid a^2=b^3=e\,,\ ab=b^2a\rangle\,,
\end{equation}
where $a$ and $b$ generate the $\mathbb{Z}_2$ and $\mathbb{Z}_3$ subgroups, respectively. Its six elements are
\begin{equation}
    S_3=\{e,b,b^2,a,ab,ab^2\}\,,
\end{equation}
and the three conjugacy classes may be represented by $e$, $a$, and $b$:
\begin{equation}
    [e]=\{e\}\,,\qquad [a]=\{a,ab,ab^2\}\,,\qquad [b]=\{b,b^2\}\,.
\end{equation}

There are three irreducible representations, denoted by $1$, $X$, and $Y$, with representation matrices
\begin{equation}
    \begin{gathered}
        1(a)=1(b)=1\,,\\
        X(a)=-1\,,\qquad X(b)=1\,,\\
        Y(a)=
        \begin{pmatrix}
            -1&0\\
            0&1
        \end{pmatrix},
        \qquad
        Y(b)=
        \begin{pmatrix}
            -\frac{1}{2}&-\frac{\sqrt{3}}{2}\\
            \frac{\sqrt{3}}{2}&-\frac{1}{2}
        \end{pmatrix}.
    \end{gathered}
\end{equation}
The non-trivial fusion rules are
\begin{equation}
    X\otimes X=1\,,\qquad
    X\otimes Y=Y\otimes X=Y\,,\qquad
    Y\otimes Y=1\oplus X\oplus Y\,,
\end{equation}
and the character table is
\begin{table}[!h]
    \centering
    \begin{tabular}{c|c|c|c}
         & $[e]$ & $[a]$ & $[b]$\\
         \hline
         $\chi_1$ & $1$ & $1$ & $1$\\
         \hline
         $\chi_X$ & $1$ & $-1$ & $1$\\
         \hline
         $\chi_Y$ & $2$ & $0$ & $-1$
    \end{tabular}
\end{table}

We fix the following ordering conventions for tensor-product bases. For the representation $Y$, we use the ordered basis
\begin{equation}
    \{e^Y_1,e^Y_2\}.
\end{equation}
For $Y\otimes Y$, we use the ordered tensor-product basis
\begin{equation}
    \{e^Y_1\otimes e^Y_1,\ e^Y_1\otimes e^Y_2,\ e^Y_2\otimes e^Y_1,\ e^Y_2\otimes e^Y_2\}.
\end{equation}
For $X\otimes Y$ and $Y\otimes X$, we use the natural ordered bases
\begin{equation}
    \{e^X\otimes e^Y_1,\ e^X\otimes e^Y_2\},
    \qquad
    \{e^Y_1\otimes e^X,\ e^Y_2\otimes e^X\}.
\end{equation}

\paragraph{Intertwiners and cointertwiners}

The intertwiner $I_{\rho,\sigma}^{\lambda}:E_\rho\otimes E_\sigma\to E_\lambda$ are represented, in the above bases, by the following matrices
\begin{equation}
\begin{gathered}
    \mathcal{I}_{1,1}^{1}=1,\qquad
\mathcal{I}_{1,X}^{X}=1,\qquad
\mathcal{I}_{X,1}^{X}=1,\qquad \mathcal{I}_{X,X}^{1}=\beta_1\,,\qquad
\mathcal{I}_{1,Y}^{Y}=\mathbf{1}_2,\qquad
\mathcal{I}_{Y,1}^{Y}=\mathbf{1}_2,
\end{gathered}
\end{equation}
\begin{equation}
\begin{gathered}
\mathcal{I}_{X,Y}^{Y}=
\begin{pmatrix}
0&-\beta_2\\
\beta_2&0
\end{pmatrix},
\qquad
\mathcal{I}_{Y,X}^{Y}=
\begin{pmatrix}
0&-\beta_3\\
\beta_3&0
\end{pmatrix}\,,
\end{gathered}
\end{equation}
\begin{equation}
\begin{gathered}
\mathcal{I}_{Y,Y}^{1}=
\begin{pmatrix}
\beta_4&0&0&\beta_4
\end{pmatrix},
\qquad
\mathcal{I}_{Y,Y}^{X}=
\begin{pmatrix}
0&\beta_6&-\beta_6&0
\end{pmatrix}\,,
\end{gathered}
\end{equation}
\begin{equation}
\begin{gathered}
\mathcal{I}_{Y,Y}^{Y}=
\begin{pmatrix}
0&\beta_5&\beta_5&0\\
\beta_5&0&0&-\beta_5
\end{pmatrix}\,.
\end{gathered}
\end{equation}
And the co-intertwiners $I^{\vee \rho,\sigma}_{\lambda}:E_{\lambda}\to E_{\rho}\otimes E_{\sigma}$ are given by
\begin{equation}
    \begin{gathered}
        (\mathcal{I}^\vee)_{1}^{1,1}=1,\qquad
(\mathcal{I}^\vee)_{X}^{1,X}=1,\qquad
(\mathcal{I}^\vee)_{X}^{X,1}=1,\qquad (\mathcal{I}^\vee)_{1}^{X,X}=
1/\beta_1,
\end{gathered}
\end{equation}
\begin{equation}
\begin{gathered}
(\mathcal{I}^\vee)_{Y}^{1,Y}=\mathbf{1}_2,\qquad
(\mathcal{I}^\vee)_{Y}^{Y,1}=\mathbf{1}_2\,,
\end{gathered}
\end{equation}
\begin{equation}
\begin{gathered}
(\mathcal{I}^\vee)_{1}^{Y,Y}=
\begin{pmatrix}
1/\beta_4\\
0\\
0\\
1/\beta_4
\end{pmatrix},
\qquad
(\mathcal{I}^\vee)_{X}^{Y,Y}=
\begin{pmatrix}
0\\
-\beta_2/\beta_4\\
\beta_2/\beta_4\\
0
\end{pmatrix},
\end{gathered}
\end{equation}
\begin{equation}
\begin{gathered}
(\mathcal{I}^\vee)_{Y}^{X,Y}=
\begin{pmatrix}
0&-\beta_6/\beta_4\\
\beta_6/\beta_4&0
\end{pmatrix},
\qquad
(\mathcal{I}^\vee)_{Y}^{Y,X}=
\begin{pmatrix}
0&-\beta_3/\beta_1\\
\beta_3/\beta_1&0
\end{pmatrix},
\end{gathered}
\end{equation}
\begin{equation}
\begin{gathered}
(\mathcal{I}^\vee)_{Y}^{Y,Y}=
\begin{pmatrix}
0&\beta_5/\beta_4\\
\beta_5/\beta_4&0\\
\beta_5/\beta_4&0\\
0&-\beta_5/\beta_4
\end{pmatrix}.
    \end{gathered}
\end{equation}
Here $\beta_1,\ldots,\beta_6$ are arbitrary parameters.

\paragraph{Frobenius algebra}
The Frobenius algebra relevant for gauging $\text{Rep}(S_3)$ is given by the regular representation, which decomposes as
\begin{equation}
\mathcal A \cong 1\oplus X\oplus Y_1\oplus Y_2,
\end{equation}
where $Y_1$ and $Y_2$ denote the two copies of the irreducible representation $Y$. The embedding maps $J_{\rho,r}:E_{\rho}\to \mathcal{A}$ are defined by
\begin{equation}\label{eq:RepS3_Frobenius_embeding}
\begin{split}
J_1(e^1)
&=v_e+v_b+v_{b^2}+v_a+v_{ab}+v_{ab^2},\\
J_X(e^X)
&=\frac{c_x}{\sqrt6}
\left(v_e+v_b+v_{b^2}-v_a-v_{ab}-v_{ab^2}\right),\\
J_{Y,1}(e^{Y}_1)
&=\frac{c_1}{2}
\left(v_b-v_{b^2}-v_{ab}+v_{ab^2}\right),\\
J_{Y,1}(e^{Y}_2)
&=\frac{c_1}{2\sqrt3}
\left(-2v_e+v_b+v_{b^2}-2v_a+v_{ab}+v_{ab^2}\right),\\
J_{Y,2}(e^{Y}_1)
&=\frac{c_2}{2\sqrt3}
\left(-2v_e+v_b+v_{b^2}+2v_a-v_{ab}-v_{ab^2}\right),\\
J_{Y,2}(e^{Y}_2)
&=\frac{c_2}{2}
\left(-v_b+v_{b^2}-v_{ab}+v_{ab^2}\right).
\end{split}
\end{equation}
In matrix form, these maps are collected as
    \begin{equation}
        \mathcal{J}^g_{\rho,r} = \left(\mathcal{J}^g_1,\mathcal{J}^g_X,\mathcal{J}^g_{(Y,1),1},\mathcal{J}^g_{(Y,2),1},\mathcal{J}^g_{(Y,1),2},\mathcal{J}^g_{(Y,2),2}  \right)\,,
    \end{equation}
where the column vectors are the coefficients on the RHS of \eqref{eq:RepS3_Frobenius_embeding}
    \begin{equation}
    \begin{gathered}
        \mathcal{J}^g_{1} = (1,1,1,1,1,1)^T\,,\quad \mathcal{J}_{X}^g = \frac{c_x}{\sqrt{6}}(1,1,1,-1,-1,-1)^T\,,\\
        \mathcal{J}^{g}_{(Y,1),1}=\frac{c_1}{2}(0,1,-1,0,-1,1)^T\,,\quad \mathcal{J}^g_{(Y,2),1}=\frac{c_1}{2\sqrt{3}}(-2,1,1,-2,1,1)^T\,,\\
        \mathcal{J}^{g}_{(Y,1),2}=\frac{c_2}{2\sqrt{3}}(-2,1,1,2,-1,-1)^T\,,\quad \mathcal{J}^g_{(Y,2),2}=\frac{c_2}{2}(0,-1,1,0,-1,1)^T\,.
    \end{gathered}
    \end{equation}
The projection maps $J^{\vee}_{\rho,r}:\mathcal{A}\to E_\rho$ are then defined as the inverse maps to $J_{a,r}$, satisfying 
\begin{equation}
    J^{\vee}_{\rho,r} \circ J_{\sigma,s} = \delta_{\rho,\sigma} \delta_{r,s} \mathbf{1}_{E_{\rho}}\,,
\end{equation}
or equivalently
    \begin{equation}
        \sum_g \mathcal{J}^{\vee g}_{(\sigma,\beta),s} \mathcal{J}^g_{(\rho,\alpha),r} = \delta_{\rho,\sigma} \delta_{\alpha,\beta} \delta_{r,s}\,.
    \end{equation}
Therefore $\mathcal{J}^{\vee g}_{(\sigma,\beta),s}$ is the inverse matrix of $\mathcal{J}^g_{(\rho,\alpha),r}$ and is organized as
    \begin{equation}
        \mathcal{J}^{\vee g}_{(\sigma,\beta),s} = \left(\mathcal{J}_{(\sigma,\beta),s}^{\vee e},\mathcal{J}_{(\sigma,\beta),s}^{\vee b},\mathcal{J}_{(\sigma,\beta),s}^{\vee b^2},\mathcal{J}_{(\sigma,\beta),s}^{\vee a},\mathcal{J}_{(\sigma,\beta),s}^{\vee ab},\mathcal{J}_{(\sigma,\beta),s}^{\vee ab^2} \right)\,.
    \end{equation}

\paragraph{Fourier kernels}

The coefficients for projectors $P_{\rho,\sigma}^{\lambda}=p_{\rho,\sigma}^{\lambda} \Pi_{\rho,\sigma}^{\lambda}$ can summarized as
\begin{equation}
\begin{gathered}
    p_{1,1}^1=p_{1,X}^X=p_{1,Y}^Y=p_{X,1}^X=p_{Y,1}^Y=p_{X,X}^1=1\,,\\
    p_{Y,Y}^1=2\,,p_{X,Y}^Y=-\frac{\beta_2 \beta_3}{\beta_1}\,,p_{Y,X}^Y=-\frac{\beta_3\beta_6}{\beta_4}\,,p_{Y,Y}^X=\frac{2\beta_2 \beta_6}{\beta_4}\,,p_{Y,Y}^Y=\frac{2\beta_5^2}{\beta_4}\,,
\end{gathered}
\end{equation}
and the dual partition function for the defect background is read from \eqref{eq:dual_partition_function_multiplicity_free}.

\paragraph{Inverse Fourier kernels}
The inverse kernels assigns the same expression to all members of the same conjugacy orbit of commuting pairs. We have
\begin{equation}
    \begin{split}
        Z_{e,e}
&=\frac{1}{6}\Big(
Z_{1,1}^{1}
+Z_{1,X}^{X}
+2Z_{1,Y}^{Y}
+Z_{X,1}^{X}
+Z_{X,X}^{1}\\
&\hspace{2em}
-\frac{2\beta_1}{\beta_2\beta_3}Z_{X,Y}^{Y}
+2Z_{Y,1}^{Y}
-\frac{2\beta_4}{\beta_3\beta_6}Z_{Y,X}^{Y}
+2Z_{Y,Y}^{1}\\
&\hspace{2em}
+\frac{2\beta_4}{\beta_2\beta_6}Z_{Y,Y}^{X}
+\frac{2\beta_4}{\beta_5^2}Z_{Y,Y}^{Y}
\Big),
    \end{split}
\end{equation}

\begin{equation}
    \begin{split}
        Z_{e,b}
=Z_{e,b^2}
&=\frac{1}{6}\Big(
Z_{1,1}^{1}
+Z_{1,X}^{X}
+2Z_{1,Y}^{Y}
+Z_{X,1}^{X}
+Z_{X,X}^{1}\\
&\hspace{2em}
-\frac{2\beta_1}{\beta_2\beta_3}Z_{X,Y}^{Y}
-Z_{Y,1}^{Y}
+\frac{\beta_4}{\beta_3\beta_6}Z_{Y,X}^{Y}
-Z_{Y,Y}^{1}\\
&\hspace{2em}
-\frac{\beta_4}{\beta_2\beta_6}Z_{Y,Y}^{X}
-\frac{\beta_4}{\beta_5^2}Z_{Y,Y}^{Y}
\Big),
    \end{split}
\end{equation}

\begin{equation}
    \begin{split}
        Z_{e,a}
=Z_{e,ab}
=Z_{e,ab^2}
&=\frac{1}{6}\Big(
Z_{1,1}^{1}
+Z_{1,X}^{X}
+2Z_{1,Y}^{Y}
-Z_{X,1}^{X}
-Z_{X,X}^{1}
+\frac{2\beta_1}{\beta_2\beta_3}Z_{X,Y}^{Y}
\Big),
    \end{split}
\end{equation}

\begin{equation}
    \begin{split}
        Z_{b,e}
=Z_{b^2,e}
&=\frac{1}{6}\Big(
Z_{1,1}^{1}
+Z_{1,X}^{X}
-Z_{1,Y}^{Y}
+Z_{X,1}^{X}
+Z_{X,X}^{1}\\
&\hspace{2em}
+\frac{\beta_1}{\beta_2\beta_3}Z_{X,Y}^{Y}
+2Z_{Y,1}^{Y}
-\frac{2\beta_4}{\beta_3\beta_6}Z_{Y,X}^{Y}
-Z_{Y,Y}^{1}\\
&\hspace{2em}
-\frac{\beta_4}{\beta_2\beta_6}Z_{Y,Y}^{X}
-\frac{\beta_4}{\beta_5^2}Z_{Y,Y}^{Y}
\Big),
    \end{split}
\end{equation}

\begin{equation}
    \begin{split}
        Z_{b,b}
=Z_{b^2,b^2}
&=\frac{1}{6}\Big(
Z_{1,1}^{1}
+Z_{1,X}^{X}
-Z_{1,Y}^{Y}
+Z_{X,1}^{X}
+Z_{X,X}^{1}\\
&\hspace{2em}
+\frac{\beta_1}{\beta_2\beta_3}Z_{X,Y}^{Y}
-Z_{Y,1}^{Y}
+\frac{\beta_4}{\beta_3\beta_6}Z_{Y,X}^{Y}
-Z_{Y,Y}^{1}\\
&\hspace{2em}
-\frac{\beta_4}{\beta_2\beta_6}Z_{Y,Y}^{X}
+\frac{2\beta_4}{\beta_5^2}Z_{Y,Y}^{Y}
\Big),
    \end{split}
\end{equation}

\begin{equation}
    \begin{split}
        Z_{b,b^2}
=Z_{b^2,b}
&=\frac{1}{6}\Big(
Z_{1,1}^{1}
+Z_{1,X}^{X}
-Z_{1,Y}^{Y}
+Z_{X,1}^{X}
+Z_{X,X}^{1}\\
&\hspace{2em}
+\frac{\beta_1}{\beta_2\beta_3}Z_{X,Y}^{Y}
-Z_{Y,1}^{Y}
+\frac{\beta_4}{\beta_3\beta_6}Z_{Y,X}^{Y}
+2Z_{Y,Y}^{1}\\
&\hspace{2em}
+\frac{2\beta_4}{\beta_2\beta_6}Z_{Y,Y}^{X}
-\frac{\beta_4}{\beta_5^2}Z_{Y,Y}^{Y}
\Big),
    \end{split}
\end{equation}

\begin{equation}
    \begin{split}
        Z_{a,e}
=Z_{ab,e}
=Z_{ab^2,e}
&=\frac{1}{6}\Big(
Z_{1,1}^{1}
-Z_{1,X}^{X}
+Z_{X,1}^{X}
-Z_{X,X}^{1}
+2Z_{Y,1}^{Y}
+\frac{2\beta_4}{\beta_3\beta_6}Z_{Y,X}^{Y}
\Big),
    \end{split}
\end{equation}

\begin{equation}
    \begin{split}
        Z_{a,a}
=Z_{ab,ab}
=Z_{ab^2,ab^2}
&=\frac{1}{6}\Big(
Z_{1,1}^{1}
-Z_{1,X}^{X}
-Z_{X,1}^{X}
+Z_{X,X}^{1}
+2Z_{Y,Y}^{1}
-\frac{2\beta_4}{\beta_2\beta_6}Z_{Y,Y}^{X}
\Big).
    \end{split}
\end{equation}

\subsection{$\text{Rep}(D_4)$}

\paragraph{Group data}
We then consider the $D_4$ group and its representation category $\text{Rep}(D_4)$. The group $D_4$ is the dihedral group with eight elements, which is defined by
    \begin{equation}
        D_4=\langle x,y|x^4=y^2=(xy)^2=e\rangle\,,
    \end{equation}
where $x$ and $y$ generate the $\mathbb{Z}_4$ and $\mathbb{Z}_2$ subgroups, respectively. Its eight elements represented by
    \begin{equation}
        D_4=\{ e,x,x^2,x^3,y,xy,x^2y,x^3y\}\,,
    \end{equation}
and the five conjugacy classes are represented by $e,x,x^2,y,xy$
    \begin{equation}
        [e]=\{e\}\,,\quad [x]=\{x,x^3\}\,,\quad [x^2]=\{x^2\}\,,\quad [y]=\{y,x^2y\}\,,\quad [xy]=\{xy,x^3y\}\,.
    \end{equation}
    
There are five irreducible representations $1,a,b,c,m$. The one-dimensional irreducible representations $1,a,b,c$ are
    \begin{equation}
    \begin{gathered}
        1(x)=1(y)=1\,\\
        a(x)=1\,,\quad a(y)=-1\,,\\
        b(x)=-1\,,\quad b(y)=1\,,\\
        c(x)=-1\,,\quad c(y)=-1\,,
    \end{gathered}
    \end{equation}
and the two-dimensional irreducible representation $m$ is given by
    \begin{equation}
        m(x)=\left(\begin{array}{cc}
            i & 0 \\
            0 & -i
        \end{array} \right)\,,\quad m(y)=\left(\begin{array}{cc}
            0 & 1 \\
            1 & 0
        \end{array} \right)\,.
    \end{equation}
The non-trivial fusion rules are
    \begin{equation}
    \begin{gathered}
        a\otimes a=b\otimes b=c\otimes c=1\,,\quad a\otimes b=c\,,\quad a\otimes c=b\,,\quad b\otimes c=a\,,\\
        a\otimes m=b\otimes m=c\otimes m = m\,,\quad m\otimes m=1\oplus a\oplus b\oplus c\,,
    \end{gathered}
    \end{equation}
and the character table is
    \begin{table}[!h]
        \centering
        \begin{tabular}{c|c|c|c|c|c}
            & $[1]$ & $[x^2]$ & $[x]$ & $[y]$ & $[xy]$\\
            \hline $\chi_1$ & 1 & 1 & 1 & 1 & 1\\
            \hline
            $\chi_a$ & 1 & 1 & 1 & -1 & -1\\
            \hline
            $\chi_b$ & 1 & 1 & -1 & 1 & -1\\
            \hline
            $\chi_c$ & 1 & 1 & -1 & -1 & 1\\
            \hline
            $\chi_m$ & 2 & -2 & 0 & 0 & 0
        \end{tabular}
    \end{table}

For the $m$-representation, we use the ordered basis $\{e^m_1,e^m_2\}$. And for $m\otimes m$, we use the ordered tensor-product basis
\begin{equation}
\{e^m_1\otimes e^m_1,\ e^m_1\otimes e^m_2,\ e^m_2\otimes e^m_1,\ e^m_2\otimes e^m_2\}.
\end{equation}
For $1$-dimensional objects tensored with $m$, and for $m$ tensored with a $1$-dimensional object, we use the obvious ordered bases.

\paragraph{Intertwiners and cointertwiners} The intertwiners are represented, in the above bases, by the following matrices
\begin{equation}
    \begin{gathered}
        \mathcal{I}_{1,1}^{1}=1,
\qquad
\mathcal{I}_{1,a}^{a}=\mathcal{I}_{a,1}^{a}=1,
\qquad
\mathcal{I}_{1,b}^{b}=\mathcal{I}_{b,1}^{b}=1,
\qquad
\mathcal{I}_{1,c}^{c}=\mathcal{I}_{c,1}^{c}=1,
\end{gathered}
\end{equation}
\begin{equation}
\begin{gathered}
\mathcal{I}_{1,m}^{m}=\mathcal{I}_{m,1}^{m}=\mathbf 1_2,
\qquad
\mathcal{I}_{a,a}^{1}=\beta_1,
\qquad
\mathcal{I}_{a,b}^{c}=\beta_2,
\qquad
\mathcal{I}_{a,c}^{b}=\beta_3,
\end{gathered}
\end{equation}
\begin{equation}
\begin{gathered}
\mathcal{I}_{a,m}^{m}=\begin{pmatrix}\beta_4&0\\0&-\beta_4\end{pmatrix},
\qquad
\mathcal{I}_{b,a}^{c}=\beta_5,
\qquad
\mathcal{I}_{b,b}^{1}=\beta_6,
\qquad
\mathcal{I}_{b,c}^{a}=\beta_7,
\end{gathered}
\end{equation}
\begin{equation}
\begin{gathered}
\mathcal{I}_{b,m}^{m}=\begin{pmatrix}0&\beta_8\\ \beta_8&0\end{pmatrix},
\qquad
\mathcal{I}_{c,a}^{b}=\beta_9,
\qquad
\mathcal{I}_{c,b}^{a}=\beta_{10},
\qquad
\mathcal{I}_{c,c}^{1}=\beta_{11},
\end{gathered}
\end{equation}
\begin{equation}
\begin{gathered}
\mathcal{I}_{c,m}^{m}=\begin{pmatrix}0&-\beta_{12}\\ \beta_{12}&0\end{pmatrix},
\qquad
\mathcal{I}_{m,a}^{m}=\begin{pmatrix}\beta_{13}&0\\0&-\beta_{13}\end{pmatrix},
\qquad
\mathcal{I}_{m,b}^{m}=\begin{pmatrix}0&\beta_{14}\\ \beta_{14}&0\end{pmatrix},\end{gathered}
\end{equation}
\begin{equation}
\begin{gathered}
\mathcal{I}_{m,c}^{m}=\begin{pmatrix}0&-\beta_{15}\\ \beta_{15}&0\end{pmatrix},
\qquad
\mathcal{I}_{m,m}^{1}=\begin{pmatrix}0&\beta_{18}&\beta_{18}&0\end{pmatrix},
\qquad
\mathcal{I}_{m,m}^{a}=\begin{pmatrix}0&\beta_{19}&-\beta_{19}&0\end{pmatrix},\end{gathered}
\end{equation}
\begin{equation}
\begin{gathered}
\mathcal{I}_{m,m}^{b}=\begin{pmatrix}\beta_{16}&0&0&\beta_{16}\end{pmatrix},
\qquad
\mathcal{I}_{m,m}^{c}=\begin{pmatrix}\beta_{17}&0&0&-\beta_{17}\end{pmatrix}.
    \end{gathered}
\end{equation}
And the cointertwiners are given by
\begin{equation}
    \begin{gathered}
        (\mathcal{I}^\vee)_{1}^{1,1}=1,
\qquad
(\mathcal{I}^\vee)_{a}^{1,a}=(\mathcal{I}^\vee)_{a}^{a,1}=1,
\qquad
(\mathcal{I}^\vee)_{b}^{1,b}=(\mathcal{I}^\vee)_{b}^{b,1}=1,
\qquad
(\mathcal{I}^\vee)_{c}^{1,c}=(\mathcal{I}^\vee)_{c}^{c,1}=1,
    \end{gathered}
\end{equation}
\begin{equation}
    \begin{gathered}
(\mathcal{I}^\vee)_{m}^{1,m}=(\mathcal{I}^\vee)_{m}^{m,1}=\mathbf 1_2,
\qquad
(\mathcal{I}^\vee)_{1}^{a,a}=\frac{1}{\beta_1},
\qquad
(\mathcal{I}^\vee)_{1}^{b,b}=\frac{1}{\beta_6},
\qquad
(\mathcal{I}^\vee)_{1}^{c,c}=\frac{1}{\beta_{11}},        
    \end{gathered}
\end{equation}
\begin{equation}
    \begin{gathered}
(\mathcal{I}^\vee)_{1}^{m,m}=\begin{pmatrix}0\\ 1/\beta_{18}\\ 1/\beta_{18}\\ 0\end{pmatrix},
\qquad
(\mathcal{I}^\vee)_{a}^{b,c}=\frac{\beta_3}{\beta_{11}},
\qquad
(\mathcal{I}^\vee)_{a}^{c,b}=\frac{\beta_2}{\beta_6},        
    \end{gathered}
\end{equation}
\begin{equation}
    \begin{gathered}
(\mathcal{I}^\vee)_{a}^{m,m}=\begin{pmatrix}0\\ \beta_4/\beta_{18}\\ -\beta_4/\beta_{18}\\ 0\end{pmatrix},
\qquad
(\mathcal{I}^\vee)_{b}^{a,c}=\frac{\beta_7}{\beta_{11}},
\qquad
(\mathcal{I}^\vee)_{b}^{c,a}=\frac{\beta_5}{\beta_1},        
    \end{gathered}
\end{equation}
\begin{equation}
    \begin{gathered}
(\mathcal{I}^\vee)_{b}^{m,m}=\begin{pmatrix}\beta_8/\beta_{18}\\ 0\\ 0\\ \beta_8/\beta_{18}\end{pmatrix},
\qquad
(\mathcal{I}^\vee)_{c}^{a,b}=\frac{\beta_{10}}{\beta_6},
\qquad
(\mathcal{I}^\vee)_{c}^{b,a}=\frac{\beta_9}{\beta_1},        
    \end{gathered}
\end{equation}
\begin{equation}
    \begin{gathered}
(\mathcal{I}^\vee)_{c}^{m,m}=\begin{pmatrix}-\beta_{12}/\beta_{18}\\ 0\\ 0\\ \beta_{12}/\beta_{18}\end{pmatrix},
\qquad
(\mathcal{I}^\vee)_{m}^{a,m}=\begin{pmatrix}\beta_{19}/\beta_{18}&0\\ 0&-\beta_{19}/\beta_{18}\end{pmatrix},
\qquad
(\mathcal{I}^\vee)_{m}^{m,a}=\begin{pmatrix}\beta_{13}/\beta_1&0\\ 0&-\beta_{13}/\beta_1\end{pmatrix},        
    \end{gathered}
\end{equation}
\begin{equation}
    \begin{gathered}
(\mathcal{I}^\vee)_{m}^{b,m}=\begin{pmatrix}0&\beta_{16}/\beta_{18}\\ \beta_{16}/\beta_{18}&0\end{pmatrix},
\qquad
(\mathcal{I}^\vee)_{m}^{m,b}=\begin{pmatrix}0&\beta_{14}/\beta_6\\ \beta_{14}/\beta_6&0\end{pmatrix},
    \end{gathered}
\end{equation}
\begin{equation}
    \begin{gathered}
(\mathcal{I}^\vee)_{m}^{c,m}=\begin{pmatrix}0&-\beta_{17}/\beta_{18}\\ \beta_{17}/\beta_{18}&0\end{pmatrix},
\qquad
(\mathcal{I}^\vee)_{m}^{m,c}=\begin{pmatrix}0&-\beta_{15}/\beta_{11}\\ \beta_{15}/\beta_{11}&0\end{pmatrix}.        
    \end{gathered}
\end{equation}
Here $\beta_1,\ldots,\beta_{19}$ are arbitrary parameters.

\paragraph{Frobenius algebra}
The Frobenius algebra relevant for gauging $\text{Rep}(D_4)$ is given by the regular representation, which decomposes as
\begin{equation}
\mathcal A\cong 1\oplus a\oplus b\oplus c\oplus m_1\oplus m_2.
\end{equation}
The embedding maps are fixed by
\begin{equation}\label{eq:RepD4_Frobenius_embeding}
\begin{split}
J_1(e^1)
&=v_e+v_x+v_{x^2}+v_{x^3}+v_y+v_{xy}+v_{x^2y}+v_{x^3y},\\
J_a(e^a)
&=v_e+v_x+v_{x^2}+v_{x^3}-v_y-v_{xy}-v_{x^2y}-v_{x^3y},\\
J_b(e^b)
&=v_e-v_x+v_{x^2}-v_{x^3}+v_y-v_{xy}+v_{x^2y}-v_{x^3y},\\
J_c(e^c)
&=v_e-v_x+v_{x^2}-v_{x^3}-v_y+v_{xy}-v_{x^2y}+v_{x^3y},\\
J_{m,1}(e^m_1)
&=v_e-i v_x-v_{x^2}+i v_{x^3}+v_y-i v_{xy}-v_{x^2y}+i v_{x^3y},\\
J_{m,1}(e^m_2)
&=v_e+i v_x-v_{x^2}-i v_{x^3}+v_y+i v_{xy}-v_{x^2y}-i v_{x^3y},\\
J_{m,2}(e^m_1)
&=-v_e+i v_x+v_{x^2}-i v_{x^3}+v_y-i v_{xy}-v_{x^2y}+i v_{x^3y},\\
J_{m,2}(e^m_2)
&=v_e+i v_x-v_{x^2}-i v_{x^3}-v_y-i v_{xy}+v_{x^2y}+i v_{x^3y}.
\end{split}
\end{equation}
In matrix form, these maps are collected as
\begin{equation}
    \mathcal{J}^g_{\rho,r} = \left( \mathcal{J}^g_{1},\mathcal{J}^g_a,\mathcal{J}^g_b,\mathcal{J}^g_c,\mathcal{J}^g_{(m,1),1},\mathcal{J}^g_{(m,2),1},\mathcal{J}^g_{(m,1),2},\mathcal{J}^g_{(m,2),2} \right)\,,
\end{equation}
where the column vectors are the coefficients on the RHS of \eqref{eq:RepD4_Frobenius_embeding}
\begin{equation}
    \begin{gathered}
        \mathcal{J}^g_{1}=(1,1,1,1,1,1,1,1)^T\,,
    \end{gathered}
\end{equation}
\begin{equation}
    \begin{gathered}
        \mathcal{J}^g_{a}=(1,1,1,1,-1,-1,-1,-1)^T\,,
    \end{gathered}
\end{equation}
\begin{equation}
    \begin{gathered}
        \mathcal{J}^g_{b}=(1,-1,1,-1,1,-1,1,-1)^T\,,
    \end{gathered}
\end{equation}
\begin{equation}
    \begin{gathered}
        \mathcal{J}^g_c=(1,-1,1,-1,-1,1,-1,1)^T\,,
    \end{gathered}
\end{equation}
\begin{equation}
    \begin{gathered}
        \mathcal{J}^g_{(m,1),1}=(1,-i,-1,i,1,-i,-1,i)^T\,,
    \end{gathered}
\end{equation}
\begin{equation}
    \begin{gathered}
        \mathcal{J}^g_{(m,2),1}=(1,i,-1,-i,1,i,-1,-i)^T\,,
    \end{gathered}
\end{equation}
\begin{equation}
    \begin{gathered}
        \mathcal{J}^g_{(m,1),2}=(-1,i,1,-i,1,-i,-1,i)^T\,,
    \end{gathered}
\end{equation}
\begin{equation}
    \begin{gathered}
        \mathcal{J}^g_{(m,2),2}=(1,i,-1,-i,-1,-i,1,i)^T\,.
    \end{gathered}
\end{equation}
Similarly, the projection map $J^{\vee}_{\rho,r}$ is the inverse map to $J_{\rho,r}$ and $\mathcal{J}^g_{(\rho,\alpha),r}$ is the inverse matrix of $\mathcal{J}^{\vee g}_{(\rho,\alpha),r}$.

\paragraph{Fourier kernel} The coefficients for projectors $P_{\rho,\sigma}^{\lambda}=p_{\rho,\sigma}^{\lambda} \Pi_{\rho,\sigma}^{\lambda}$ can summarized as

\begin{equation}
    \begin{gathered}
        p_{11}^1=p_{1a}^a=p_{1b}^b=p_{1c}^c=p_{1m}^m=p_{a1}^a=p_{b1}^b=p_{c1}^c=p_{m1}^m=p_{aa}^1=p_{bb}^1=p_{cc}^1=1
    \end{gathered}
\end{equation}
\begin{equation}
    \begin{gathered}
        p_{ab}^c=\frac{\beta_2 \beta_9}{\beta_1}\,,\quad p_{ba}^c = \frac{\beta_{10}\beta_5}{\beta_6}\,,\quad p_{ac}^b=\frac{\beta_3 \beta_5}{\beta_1}\,,\quad p_{ca}^b=\frac{\beta_7 \beta_9}{\beta_{11}}\,,\quad p_{bc}^a=\frac{\beta_2 \beta_7}{\beta_6}\,,\quad p_{cb}^a=\frac{\beta_{10}\beta_3}{\beta_{11}}\,,
    \end{gathered}
\end{equation}
\begin{equation}
    \begin{gathered}
        p_{am}^m= \frac{\beta_{13}\beta_4}{\beta_1}\,,\quad p_{bm}^m=\frac{\beta_{14}\beta_8}{\beta_6}\,,\quad p_{cm}^m=-\frac{\beta_{12}\beta_{15}}{\beta_{11}}\,,
    \end{gathered}
\end{equation}
\begin{equation}
    \begin{gathered}
        p_{ma}^m= \frac{\beta_{13}\beta_{19}}{\beta_{18}}\,,\quad p_{mb}^m=\frac{\beta_{14}\beta_{16}}{\beta_{18}}\,,\quad p_{mc}^m=-\frac{\beta_{15}\beta_{17}}{\beta_{18}}\,, 
    \end{gathered}
\end{equation}
\begin{equation}
    \begin{gathered}
        p_{mm}^1=2\,,\quad p_{mm}^a=-\frac{2\beta_{19}\beta_4}{\beta_{18}}\,,\quad p_{mm}^b=\frac{2\beta_8\beta_{16}}{\beta_{18}}\,,\quad p_{mm}^c=-\frac{2\beta_{12}\beta_{17}}{\beta_{18}}\,.
    \end{gathered}
\end{equation}
and the dual partition function for the defect background is read from \eqref{eq:dual_partition_function_multiplicity_free}.

\paragraph{Inverse Fourier kernels}
The inverse kernels assigns the same expression to all members of the same conjugacy orbit of commuting pairs. We have
\begin{equation}
    \begin{split}
Z_{e,e}
&= \frac{1}{8}\Big(Z_{1,1}^{1} + Z_{1,a}^{a} + Z_{1,b}^{b} + Z_{1,c}^{c} + 2 Z_{1,m}^{m} + Z_{a,1}^{a} +\\
&\hspace{2em}Z_{a,a}^{1} + \frac{\beta_{1}}{\beta_{2} \beta_{9}} Z_{a,b}^{c} + \frac{\beta_{1}}{\beta_{3} \beta_{5}} Z_{a,c}^{b} + \frac{2 \beta_{1}}{\beta_{13} \beta_{4}} Z_{a,m}^{m} + Z_{b,1}^{b} + \frac{\beta_{6}}{\beta_{10} \beta_{5}} Z_{b,a}^{c} +\\
&\hspace{2em}Z_{b,b}^{1} + \frac{\beta_{6}}{\beta_{2} \beta_{7}} Z_{b,c}^{a} + \frac{2 \beta_{6}}{\beta_{14} \beta_{8}} Z_{b,m}^{m} + Z_{c,1}^{c} + \frac{\beta_{11}}{\beta_{7} \beta_{9}} Z_{c,a}^{b} + \frac{\beta_{11}}{\beta_{10} \beta_{3}} Z_{c,b}^{a} +\\
&\hspace{2em}Z_{c,c}^{1} - \frac{2 \beta_{11}}{\beta_{12} \beta_{15}} Z_{c,m}^{m} + 2 Z_{m,1}^{m} + \frac{2 \beta_{18}}{\beta_{13} \beta_{19}} Z_{m,a}^{m} + \frac{2 \beta_{18}}{\beta_{14} \beta_{16}} Z_{m,b}^{m} - \frac{2 \beta_{18}}{\beta_{15} \beta_{17}} Z_{m,c}^{m} +\\
&\hspace{2em}2 Z_{m,m}^{1} - \frac{2 \beta_{18}}{\beta_{19} \beta_{4}} Z_{m,m}^{a} + \frac{2 \beta_{18}}{\beta_{16} \beta_{8}} Z_{m,m}^{b} - \frac{2 \beta_{18}}{\beta_{12} \beta_{17}} Z_{m,m}^{c}\Big)
    \end{split}
\end{equation}
\begin{equation}
    \begin{split}
Z_{e,x} = Z_{e,x^3}
&= \frac{1}{8}\Big(Z_{1,1}^{1} + Z_{1,a}^{a} + Z_{1,b}^{b} + Z_{1,c}^{c} + 2 Z_{1,m}^{m} + Z_{a,1}^{a} +\\
&\hspace{2em}Z_{a,a}^{1} + \frac{\beta_{1}}{\beta_{2} \beta_{9}} Z_{a,b}^{c} + \frac{\beta_{1}}{\beta_{3} \beta_{5}} Z_{a,c}^{b} + \frac{2 \beta_{1}}{\beta_{13} \beta_{4}} Z_{a,m}^{m} - Z_{b,1}^{b} - \frac{\beta_{6}}{\beta_{10} \beta_{5}} Z_{b,a}^{c} +\\
&\hspace{2em}-Z_{b,b}^{1} - \frac{\beta_{6}}{\beta_{2} \beta_{7}} Z_{b,c}^{a} - \frac{2 \beta_{6}}{\beta_{14} \beta_{8}} Z_{b,m}^{m} - Z_{c,1}^{c} - \frac{\beta_{11}}{\beta_{7} \beta_{9}} Z_{c,a}^{b} - \frac{\beta_{11}}{\beta_{10} \beta_{3}} Z_{c,b}^{a} +\\
&\hspace{2em}-Z_{c,c}^{1} + \frac{2 \beta_{11}}{\beta_{12} \beta_{15}} Z_{c,m}^{m}\Big)
    \end{split}
\end{equation}
\begin{equation}
    \begin{split}
Z_{e,x^2}
&= \frac{1}{8}\Big(Z_{1,1}^{1} + Z_{1,a}^{a} + Z_{1,b}^{b} + Z_{1,c}^{c} + 2 Z_{1,m}^{m} + Z_{a,1}^{a} +\\
&\hspace{2em}Z_{a,a}^{1} + \frac{\beta_{1}}{\beta_{2} \beta_{9}} Z_{a,b}^{c} + \frac{\beta_{1}}{\beta_{3} \beta_{5}} Z_{a,c}^{b} + \frac{2 \beta_{1}}{\beta_{13} \beta_{4}} Z_{a,m}^{m} + Z_{b,1}^{b} + \frac{\beta_{6}}{\beta_{10} \beta_{5}} Z_{b,a}^{c} +\\
&\hspace{2em}Z_{b,b}^{1} + \frac{\beta_{6}}{\beta_{2} \beta_{7}} Z_{b,c}^{a} + \frac{2 \beta_{6}}{\beta_{14} \beta_{8}} Z_{b,m}^{m} + Z_{c,1}^{c} + \frac{\beta_{11}}{\beta_{7} \beta_{9}} Z_{c,a}^{b} + \frac{\beta_{11}}{\beta_{10} \beta_{3}} Z_{c,b}^{a} +\\
&\hspace{2em}Z_{c,c}^{1} - \frac{2 \beta_{11}}{\beta_{12} \beta_{15}} Z_{c,m}^{m} - 2 Z_{m,1}^{m} - \frac{2 \beta_{18}}{\beta_{13} \beta_{19}} Z_{m,a}^{m} - \frac{2 \beta_{18}}{\beta_{14} \beta_{16}} Z_{m,b}^{m} + \frac{2 \beta_{18}}{\beta_{15} \beta_{17}} Z_{m,c}^{m} +\\
&\hspace{2em}-2 Z_{m,m}^{1} + \frac{2 \beta_{18}}{\beta_{19} \beta_{4}} Z_{m,m}^{a} - \frac{2 \beta_{18}}{\beta_{16} \beta_{8}} Z_{m,m}^{b} + \frac{2 \beta_{18}}{\beta_{12} \beta_{17}} Z_{m,m}^{c}\Big)
    \end{split}
\end{equation}
\begin{equation}
    \begin{split}
Z_{e,y} = Z_{e,x^2y}
&= \frac{1}{8}\Big(Z_{1,1}^{1} + Z_{1,a}^{a} + Z_{1,b}^{b} + Z_{1,c}^{c} + 2 Z_{1,m}^{m} - Z_{a,1}^{a} +\\
&\hspace{2em}-Z_{a,a}^{1} - \frac{\beta_{1}}{\beta_{2} \beta_{9}} Z_{a,b}^{c} - \frac{\beta_{1}}{\beta_{3} \beta_{5}} Z_{a,c}^{b} - \frac{2 \beta_{1}}{\beta_{13} \beta_{4}} Z_{a,m}^{m} + Z_{b,1}^{b} + \frac{\beta_{6}}{\beta_{10} \beta_{5}} Z_{b,a}^{c} +\\
&\hspace{2em}Z_{b,b}^{1} + \frac{\beta_{6}}{\beta_{2} \beta_{7}} Z_{b,c}^{a} + \frac{2 \beta_{6}}{\beta_{14} \beta_{8}} Z_{b,m}^{m} - Z_{c,1}^{c} - \frac{\beta_{11}}{\beta_{7} \beta_{9}} Z_{c,a}^{b} - \frac{\beta_{11}}{\beta_{10} \beta_{3}} Z_{c,b}^{a} +\\
&\hspace{2em}-Z_{c,c}^{1} + \frac{2 \beta_{11}}{\beta_{12} \beta_{15}} Z_{c,m}^{m}\Big)
    \end{split}
\end{equation}
\begin{equation}
    \begin{split}
Z_{e,xy} = Z_{e,x^3y}
&= \frac{1}{8}\Big(Z_{1,1}^{1} + Z_{1,a}^{a} + Z_{1,b}^{b} + Z_{1,c}^{c} + 2 Z_{1,m}^{m} - Z_{a,1}^{a} +\\
&\hspace{2em}-Z_{a,a}^{1} - \frac{\beta_{1}}{\beta_{2} \beta_{9}} Z_{a,b}^{c} - \frac{\beta_{1}}{\beta_{3} \beta_{5}} Z_{a,c}^{b} - \frac{2 \beta_{1}}{\beta_{13} \beta_{4}} Z_{a,m}^{m} - Z_{b,1}^{b} - \frac{\beta_{6}}{\beta_{10} \beta_{5}} Z_{b,a}^{c} +\\
&\hspace{2em}-Z_{b,b}^{1} - \frac{\beta_{6}}{\beta_{2} \beta_{7}} Z_{b,c}^{a} - \frac{2 \beta_{6}}{\beta_{14} \beta_{8}} Z_{b,m}^{m} + Z_{c,1}^{c} + \frac{\beta_{11}}{\beta_{7} \beta_{9}} Z_{c,a}^{b} + \frac{\beta_{11}}{\beta_{10} \beta_{3}} Z_{c,b}^{a} +\\
&\hspace{2em}Z_{c,c}^{1} - \frac{2 \beta_{11}}{\beta_{12} \beta_{15}} Z_{c,m}^{m}\Big)
    \end{split}
\end{equation}
\begin{equation}
    \begin{split}
Z_{x,e} = Z_{x^3,e}
&= \frac{1}{8}\Big(Z_{1,1}^{1} + Z_{1,a}^{a} - Z_{1,b}^{b} - Z_{1,c}^{c} + Z_{a,1}^{a} + Z_{a,a}^{1} +\\
&\hspace{2em}-\frac{\beta_{1}}{\beta_{2} \beta_{9}} Z_{a,b}^{c} - \frac{\beta_{1}}{\beta_{3} \beta_{5}} Z_{a,c}^{b} + Z_{b,1}^{b} + \frac{\beta_{6}}{\beta_{10} \beta_{5}} Z_{b,a}^{c} - Z_{b,b}^{1} - \frac{\beta_{6}}{\beta_{2} \beta_{7}} Z_{b,c}^{a} +\\
&\hspace{2em}Z_{c,1}^{c} + \frac{\beta_{11}}{\beta_{7} \beta_{9}} Z_{c,a}^{b} - \frac{\beta_{11}}{\beta_{10} \beta_{3}} Z_{c,b}^{a} - Z_{c,c}^{1} + 2 Z_{m,1}^{m} + \frac{2 \beta_{18}}{\beta_{13} \beta_{19}} Z_{m,a}^{m} +\\
&\hspace{2em}-\frac{2 \beta_{18}}{\beta_{14} \beta_{16}} Z_{m,b}^{m} + \frac{2 \beta_{18}}{\beta_{15} \beta_{17}} Z_{m,c}^{m}\Big)
    \end{split}
\end{equation}
\begin{equation}
    \begin{split}
Z_{x,x} = Z_{x^3,x^3}
&= \frac{1}{8}\Big(Z_{1,1}^{1} + Z_{1,a}^{a} - Z_{1,b}^{b} - Z_{1,c}^{c} + Z_{a,1}^{a} + Z_{a,a}^{1} +\\
&\hspace{2em}-\frac{\beta_{1}}{\beta_{2} \beta_{9}} Z_{a,b}^{c} - \frac{\beta_{1}}{\beta_{3} \beta_{5}} Z_{a,c}^{b} - Z_{b,1}^{b} - \frac{\beta_{6}}{\beta_{10} \beta_{5}} Z_{b,a}^{c} + Z_{b,b}^{1} + \frac{\beta_{6}}{\beta_{2} \beta_{7}} Z_{b,c}^{a} +\\
&\hspace{2em}-Z_{c,1}^{c} - \frac{\beta_{11}}{\beta_{7} \beta_{9}} Z_{c,a}^{b} + \frac{\beta_{11}}{\beta_{10} \beta_{3}} Z_{c,b}^{a} + Z_{c,c}^{1} - 2 Z_{m,m}^{1} + \frac{2 \beta_{18}}{\beta_{19} \beta_{4}} Z_{m,m}^{a} +\\
&\hspace{2em}\frac{2 \beta_{18}}{\beta_{16} \beta_{8}} Z_{m,m}^{b} - \frac{2 \beta_{18}}{\beta_{12} \beta_{17}} Z_{m,m}^{c}\Big)
    \end{split}
\end{equation}
\begin{equation}
    \begin{split}
Z_{x,x^2} = Z_{x^3,x^2}
&= \frac{1}{8}\Big(Z_{1,1}^{1} + Z_{1,a}^{a} - Z_{1,b}^{b} - Z_{1,c}^{c} + Z_{a,1}^{a} + Z_{a,a}^{1} +\\
&\hspace{2em}-\frac{\beta_{1}}{\beta_{2} \beta_{9}} Z_{a,b}^{c} - \frac{\beta_{1}}{\beta_{3} \beta_{5}} Z_{a,c}^{b} + Z_{b,1}^{b} + \frac{\beta_{6}}{\beta_{10} \beta_{5}} Z_{b,a}^{c} - Z_{b,b}^{1} - \frac{\beta_{6}}{\beta_{2} \beta_{7}} Z_{b,c}^{a} +\\
&\hspace{2em}Z_{c,1}^{c} + \frac{\beta_{11}}{\beta_{7} \beta_{9}} Z_{c,a}^{b} - \frac{\beta_{11}}{\beta_{10} \beta_{3}} Z_{c,b}^{a} - Z_{c,c}^{1} - 2 Z_{m,1}^{m} - \frac{2 \beta_{18}}{\beta_{13} \beta_{19}} Z_{m,a}^{m} +\\
&\hspace{2em}\frac{2 \beta_{18}}{\beta_{14} \beta_{16}} Z_{m,b}^{m} - \frac{2 \beta_{18}}{\beta_{15} \beta_{17}} Z_{m,c}^{m}\Big)
    \end{split}
\end{equation}
\begin{equation}
    \begin{split}
Z_{x,x^3} = Z_{x^3,x}
&= \frac{1}{8}\Big(Z_{1,1}^{1} + Z_{1,a}^{a} - Z_{1,b}^{b} - Z_{1,c}^{c} + Z_{a,1}^{a} + Z_{a,a}^{1} +\\
&\hspace{2em}-\frac{\beta_{1}}{\beta_{2} \beta_{9}} Z_{a,b}^{c} - \frac{\beta_{1}}{\beta_{3} \beta_{5}} Z_{a,c}^{b} - Z_{b,1}^{b} - \frac{\beta_{6}}{\beta_{10} \beta_{5}} Z_{b,a}^{c} + Z_{b,b}^{1} + \frac{\beta_{6}}{\beta_{2} \beta_{7}} Z_{b,c}^{a} +\\
&\hspace{2em}-Z_{c,1}^{c} - \frac{\beta_{11}}{\beta_{7} \beta_{9}} Z_{c,a}^{b} + \frac{\beta_{11}}{\beta_{10} \beta_{3}} Z_{c,b}^{a} + Z_{c,c}^{1} + 2 Z_{m,m}^{1} - \frac{2 \beta_{18}}{\beta_{19} \beta_{4}} Z_{m,m}^{a} +\\
&\hspace{2em}-\frac{2 \beta_{18}}{\beta_{16} \beta_{8}} Z_{m,m}^{b} + \frac{2 \beta_{18}}{\beta_{12} \beta_{17}} Z_{m,m}^{c}\Big)
    \end{split}
\end{equation}
\begin{equation}
    \begin{split}
Z_{x^2,e}
&= \frac{1}{8}\Big(Z_{1,1}^{1} + Z_{1,a}^{a} + Z_{1,b}^{b} + Z_{1,c}^{c} - 2 Z_{1,m}^{m} + Z_{a,1}^{a} +\\
&\hspace{2em}Z_{a,a}^{1} + \frac{\beta_{1}}{\beta_{2} \beta_{9}} Z_{a,b}^{c} + \frac{\beta_{1}}{\beta_{3} \beta_{5}} Z_{a,c}^{b} - \frac{2 \beta_{1}}{\beta_{13} \beta_{4}} Z_{a,m}^{m} + Z_{b,1}^{b} + \frac{\beta_{6}}{\beta_{10} \beta_{5}} Z_{b,a}^{c} +\\
&\hspace{2em}Z_{b,b}^{1} + \frac{\beta_{6}}{\beta_{2} \beta_{7}} Z_{b,c}^{a} - \frac{2 \beta_{6}}{\beta_{14} \beta_{8}} Z_{b,m}^{m} + Z_{c,1}^{c} + \frac{\beta_{11}}{\beta_{7} \beta_{9}} Z_{c,a}^{b} + \frac{\beta_{11}}{\beta_{10} \beta_{3}} Z_{c,b}^{a} +\\
&\hspace{2em}Z_{c,c}^{1} + \frac{2 \beta_{11}}{\beta_{12} \beta_{15}} Z_{c,m}^{m} + 2 Z_{m,1}^{m} + \frac{2 \beta_{18}}{\beta_{13} \beta_{19}} Z_{m,a}^{m} + \frac{2 \beta_{18}}{\beta_{14} \beta_{16}} Z_{m,b}^{m} - \frac{2 \beta_{18}}{\beta_{15} \beta_{17}} Z_{m,c}^{m} +\\
&\hspace{2em}-2 Z_{m,m}^{1} + \frac{2 \beta_{18}}{\beta_{19} \beta_{4}} Z_{m,m}^{a} - \frac{2 \beta_{18}}{\beta_{16} \beta_{8}} Z_{m,m}^{b} + \frac{2 \beta_{18}}{\beta_{12} \beta_{17}} Z_{m,m}^{c}\Big)
    \end{split}
\end{equation}
\begin{equation}
    \begin{split}
Z_{x^2,x} = Z_{x^2,x^3}
&= \frac{1}{8}\Big(Z_{1,1}^{1} + Z_{1,a}^{a} + Z_{1,b}^{b} + Z_{1,c}^{c} - 2 Z_{1,m}^{m} + Z_{a,1}^{a} +\\
&\hspace{2em}Z_{a,a}^{1} + \frac{\beta_{1}}{\beta_{2} \beta_{9}} Z_{a,b}^{c} + \frac{\beta_{1}}{\beta_{3} \beta_{5}} Z_{a,c}^{b} - \frac{2 \beta_{1}}{\beta_{13} \beta_{4}} Z_{a,m}^{m} - Z_{b,1}^{b} - \frac{\beta_{6}}{\beta_{10} \beta_{5}} Z_{b,a}^{c} +\\
&\hspace{2em}-Z_{b,b}^{1} - \frac{\beta_{6}}{\beta_{2} \beta_{7}} Z_{b,c}^{a} + \frac{2 \beta_{6}}{\beta_{14} \beta_{8}} Z_{b,m}^{m} - Z_{c,1}^{c} - \frac{\beta_{11}}{\beta_{7} \beta_{9}} Z_{c,a}^{b} - \frac{\beta_{11}}{\beta_{10} \beta_{3}} Z_{c,b}^{a} +\\
&\hspace{2em}-Z_{c,c}^{1} - \frac{2 \beta_{11}}{\beta_{12} \beta_{15}} Z_{c,m}^{m}\Big)
    \end{split}
\end{equation}
\begin{equation}
    \begin{split}
Z_{x^2,x^2}
&= \frac{1}{8}\Big(Z_{1,1}^{1} + Z_{1,a}^{a} + Z_{1,b}^{b} + Z_{1,c}^{c} - 2 Z_{1,m}^{m} + Z_{a,1}^{a} +\\
&\hspace{2em}Z_{a,a}^{1} + \frac{\beta_{1}}{\beta_{2} \beta_{9}} Z_{a,b}^{c} + \frac{\beta_{1}}{\beta_{3} \beta_{5}} Z_{a,c}^{b} - \frac{2 \beta_{1}}{\beta_{13} \beta_{4}} Z_{a,m}^{m} + Z_{b,1}^{b} + \frac{\beta_{6}}{\beta_{10} \beta_{5}} Z_{b,a}^{c} +\\
&\hspace{2em}Z_{b,b}^{1} + \frac{\beta_{6}}{\beta_{2} \beta_{7}} Z_{b,c}^{a} - \frac{2 \beta_{6}}{\beta_{14} \beta_{8}} Z_{b,m}^{m} + Z_{c,1}^{c} + \frac{\beta_{11}}{\beta_{7} \beta_{9}} Z_{c,a}^{b} + \frac{\beta_{11}}{\beta_{10} \beta_{3}} Z_{c,b}^{a} +\\
&\hspace{2em}Z_{c,c}^{1} + \frac{2 \beta_{11}}{\beta_{12} \beta_{15}} Z_{c,m}^{m} - 2 Z_{m,1}^{m} - \frac{2 \beta_{18}}{\beta_{13} \beta_{19}} Z_{m,a}^{m} - \frac{2 \beta_{18}}{\beta_{14} \beta_{16}} Z_{m,b}^{m} + \frac{2 \beta_{18}}{\beta_{15} \beta_{17}} Z_{m,c}^{m} +\\
&\hspace{2em}2 Z_{m,m}^{1} - \frac{2 \beta_{18}}{\beta_{19} \beta_{4}} Z_{m,m}^{a} + \frac{2 \beta_{18}}{\beta_{16} \beta_{8}} Z_{m,m}^{b} - \frac{2 \beta_{18}}{\beta_{12} \beta_{17}} Z_{m,m}^{c}\Big)
    \end{split}
\end{equation}
\begin{equation}
    \begin{split}
Z_{x^2,y} = Z_{x^2,x^2y}
&= \frac{1}{8}\Big(Z_{1,1}^{1} + Z_{1,a}^{a} + Z_{1,b}^{b} + Z_{1,c}^{c} - 2 Z_{1,m}^{m} - Z_{a,1}^{a} +\\
&\hspace{2em}-Z_{a,a}^{1} - \frac{\beta_{1}}{\beta_{2} \beta_{9}} Z_{a,b}^{c} - \frac{\beta_{1}}{\beta_{3} \beta_{5}} Z_{a,c}^{b} + \frac{2 \beta_{1}}{\beta_{13} \beta_{4}} Z_{a,m}^{m} + Z_{b,1}^{b} + \frac{\beta_{6}}{\beta_{10} \beta_{5}} Z_{b,a}^{c} +\\
&\hspace{2em}Z_{b,b}^{1} + \frac{\beta_{6}}{\beta_{2} \beta_{7}} Z_{b,c}^{a} - \frac{2 \beta_{6}}{\beta_{14} \beta_{8}} Z_{b,m}^{m} - Z_{c,1}^{c} - \frac{\beta_{11}}{\beta_{7} \beta_{9}} Z_{c,a}^{b} - \frac{\beta_{11}}{\beta_{10} \beta_{3}} Z_{c,b}^{a} +\\
&\hspace{2em}-Z_{c,c}^{1} - \frac{2 \beta_{11}}{\beta_{12} \beta_{15}} Z_{c,m}^{m}\Big)
    \end{split}
\end{equation}
\begin{equation}
    \begin{split}
Z_{x^2,xy} = Z_{x^2,x^3y}
&= \frac{1}{8}\Big(Z_{1,1}^{1} + Z_{1,a}^{a} + Z_{1,b}^{b} + Z_{1,c}^{c} - 2 Z_{1,m}^{m} - Z_{a,1}^{a} +\\
&\hspace{2em}-Z_{a,a}^{1} - \frac{\beta_{1}}{\beta_{2} \beta_{9}} Z_{a,b}^{c} - \frac{\beta_{1}}{\beta_{3} \beta_{5}} Z_{a,c}^{b} + \frac{2 \beta_{1}}{\beta_{13} \beta_{4}} Z_{a,m}^{m} - Z_{b,1}^{b} - \frac{\beta_{6}}{\beta_{10} \beta_{5}} Z_{b,a}^{c} +\\
&\hspace{2em}-Z_{b,b}^{1} - \frac{\beta_{6}}{\beta_{2} \beta_{7}} Z_{b,c}^{a} + \frac{2 \beta_{6}}{\beta_{14} \beta_{8}} Z_{b,m}^{m} + Z_{c,1}^{c} + \frac{\beta_{11}}{\beta_{7} \beta_{9}} Z_{c,a}^{b} + \frac{\beta_{11}}{\beta_{10} \beta_{3}} Z_{c,b}^{a} +\\
&\hspace{2em}Z_{c,c}^{1} + \frac{2 \beta_{11}}{\beta_{12} \beta_{15}} Z_{c,m}^{m}\Big)
    \end{split}
\end{equation}
\begin{equation}
    \begin{split}
Z_{y,e} = Z_{x^2y,e}
&= \frac{1}{8}\Big(Z_{1,1}^{1} - Z_{1,a}^{a} + Z_{1,b}^{b} - Z_{1,c}^{c} + Z_{a,1}^{a} - Z_{a,a}^{1} +\\
&\hspace{2em}\frac{\beta_{1}}{\beta_{2} \beta_{9}} Z_{a,b}^{c} - \frac{\beta_{1}}{\beta_{3} \beta_{5}} Z_{a,c}^{b} + Z_{b,1}^{b} - \frac{\beta_{6}}{\beta_{10} \beta_{5}} Z_{b,a}^{c} + Z_{b,b}^{1} - \frac{\beta_{6}}{\beta_{2} \beta_{7}} Z_{b,c}^{a} +\\
&\hspace{2em}Z_{c,1}^{c} - \frac{\beta_{11}}{\beta_{7} \beta_{9}} Z_{c,a}^{b} + \frac{\beta_{11}}{\beta_{10} \beta_{3}} Z_{c,b}^{a} - Z_{c,c}^{1} + 2 Z_{m,1}^{m} - \frac{2 \beta_{18}}{\beta_{13} \beta_{19}} Z_{m,a}^{m} +\\
&\hspace{2em}\frac{2 \beta_{18}}{\beta_{14} \beta_{16}} Z_{m,b}^{m} + \frac{2 \beta_{18}}{\beta_{15} \beta_{17}} Z_{m,c}^{m}\Big)
    \end{split}
\end{equation}
\begin{equation}
    \begin{split}
Z_{y,x^2} = Z_{x^2y,x^2}
&= \frac{1}{8}\Big(Z_{1,1}^{1} - Z_{1,a}^{a} + Z_{1,b}^{b} - Z_{1,c}^{c} + Z_{a,1}^{a} - Z_{a,a}^{1} +\\
&\hspace{2em}\frac{\beta_{1}}{\beta_{2} \beta_{9}} Z_{a,b}^{c} - \frac{\beta_{1}}{\beta_{3} \beta_{5}} Z_{a,c}^{b} + Z_{b,1}^{b} - \frac{\beta_{6}}{\beta_{10} \beta_{5}} Z_{b,a}^{c} + Z_{b,b}^{1} - \frac{\beta_{6}}{\beta_{2} \beta_{7}} Z_{b,c}^{a} +\\
&\hspace{2em}Z_{c,1}^{c} - \frac{\beta_{11}}{\beta_{7} \beta_{9}} Z_{c,a}^{b} + \frac{\beta_{11}}{\beta_{10} \beta_{3}} Z_{c,b}^{a} - Z_{c,c}^{1} - 2 Z_{m,1}^{m} + \frac{2 \beta_{18}}{\beta_{13} \beta_{19}} Z_{m,a}^{m} +\\
&\hspace{2em}-\frac{2 \beta_{18}}{\beta_{14} \beta_{16}} Z_{m,b}^{m} - \frac{2 \beta_{18}}{\beta_{15} \beta_{17}} Z_{m,c}^{m}\Big)
    \end{split}
\end{equation}
\begin{equation}
    \begin{split}
Z_{y,y} = Z_{x^2y,x^2y}
&= \frac{1}{8}\Big(Z_{1,1}^{1} - Z_{1,a}^{a} + Z_{1,b}^{b} - Z_{1,c}^{c} - Z_{a,1}^{a} + Z_{a,a}^{1} +\\
&\hspace{2em}-\frac{\beta_{1}}{\beta_{2} \beta_{9}} Z_{a,b}^{c} + \frac{\beta_{1}}{\beta_{3} \beta_{5}} Z_{a,c}^{b} + Z_{b,1}^{b} - \frac{\beta_{6}}{\beta_{10} \beta_{5}} Z_{b,a}^{c} + Z_{b,b}^{1} - \frac{\beta_{6}}{\beta_{2} \beta_{7}} Z_{b,c}^{a} +\\
&\hspace{2em}-Z_{c,1}^{c} + \frac{\beta_{11}}{\beta_{7} \beta_{9}} Z_{c,a}^{b} - \frac{\beta_{11}}{\beta_{10} \beta_{3}} Z_{c,b}^{a} + Z_{c,c}^{1} + 2 Z_{m,m}^{1} + \frac{2 \beta_{18}}{\beta_{19} \beta_{4}} Z_{m,m}^{a} +\\
&\hspace{2em}\frac{2 \beta_{18}}{\beta_{16} \beta_{8}} Z_{m,m}^{b} + \frac{2 \beta_{18}}{\beta_{12} \beta_{17}} Z_{m,m}^{c}\Big)
    \end{split}
\end{equation}
\begin{equation}
    \begin{split}
Z_{y,x^2y} = Z_{x^2y,y}
&= \frac{1}{8}\Big(Z_{1,1}^{1} - Z_{1,a}^{a} + Z_{1,b}^{b} - Z_{1,c}^{c} - Z_{a,1}^{a} + Z_{a,a}^{1} +\\
&\hspace{2em}-\frac{\beta_{1}}{\beta_{2} \beta_{9}} Z_{a,b}^{c} + \frac{\beta_{1}}{\beta_{3} \beta_{5}} Z_{a,c}^{b} + Z_{b,1}^{b} - \frac{\beta_{6}}{\beta_{10} \beta_{5}} Z_{b,a}^{c} + Z_{b,b}^{1} - \frac{\beta_{6}}{\beta_{2} \beta_{7}} Z_{b,c}^{a} +\\
&\hspace{2em}-Z_{c,1}^{c} + \frac{\beta_{11}}{\beta_{7} \beta_{9}} Z_{c,a}^{b} - \frac{\beta_{11}}{\beta_{10} \beta_{3}} Z_{c,b}^{a} + Z_{c,c}^{1} - 2 Z_{m,m}^{1} - \frac{2 \beta_{18}}{\beta_{19} \beta_{4}} Z_{m,m}^{a} +\\
&\hspace{2em}-\frac{2 \beta_{18}}{\beta_{16} \beta_{8}} Z_{m,m}^{b} - \frac{2 \beta_{18}}{\beta_{12} \beta_{17}} Z_{m,m}^{c}\Big)
    \end{split}
\end{equation}
\begin{equation}
    \begin{split}
Z_{xy,e} = Z_{x^3y,e}
&= \frac{1}{8}\Big(Z_{1,1}^{1} - Z_{1,a}^{a} - Z_{1,b}^{b} + Z_{1,c}^{c} + Z_{a,1}^{a} - Z_{a,a}^{1} +\\
&\hspace{2em}-\frac{\beta_{1}}{\beta_{2} \beta_{9}} Z_{a,b}^{c} + \frac{\beta_{1}}{\beta_{3} \beta_{5}} Z_{a,c}^{b} + Z_{b,1}^{b} - \frac{\beta_{6}}{\beta_{10} \beta_{5}} Z_{b,a}^{c} - Z_{b,b}^{1} + \frac{\beta_{6}}{\beta_{2} \beta_{7}} Z_{b,c}^{a} +\\
&\hspace{2em}Z_{c,1}^{c} - \frac{\beta_{11}}{\beta_{7} \beta_{9}} Z_{c,a}^{b} - \frac{\beta_{11}}{\beta_{10} \beta_{3}} Z_{c,b}^{a} + Z_{c,c}^{1} + 2 Z_{m,1}^{m} - \frac{2 \beta_{18}}{\beta_{13} \beta_{19}} Z_{m,a}^{m} +\\
&\hspace{2em}-\frac{2 \beta_{18}}{\beta_{14} \beta_{16}} Z_{m,b}^{m} - \frac{2 \beta_{18}}{\beta_{15} \beta_{17}} Z_{m,c}^{m}\Big)
    \end{split}
\end{equation}
\begin{equation}
    \begin{split}
Z_{xy,x^2} = Z_{x^3y,x^2}
&= \frac{1}{8}\Big(Z_{1,1}^{1} - Z_{1,a}^{a} - Z_{1,b}^{b} + Z_{1,c}^{c} + Z_{a,1}^{a} - Z_{a,a}^{1} +\\
&\hspace{2em}-\frac{\beta_{1}}{\beta_{2} \beta_{9}} Z_{a,b}^{c} + \frac{\beta_{1}}{\beta_{3} \beta_{5}} Z_{a,c}^{b} + Z_{b,1}^{b} - \frac{\beta_{6}}{\beta_{10} \beta_{5}} Z_{b,a}^{c} - Z_{b,b}^{1} + \frac{\beta_{6}}{\beta_{2} \beta_{7}} Z_{b,c}^{a} +\\
&\hspace{2em}Z_{c,1}^{c} - \frac{\beta_{11}}{\beta_{7} \beta_{9}} Z_{c,a}^{b} - \frac{\beta_{11}}{\beta_{10} \beta_{3}} Z_{c,b}^{a} + Z_{c,c}^{1} - 2 Z_{m,1}^{m} + \frac{2 \beta_{18}}{\beta_{13} \beta_{19}} Z_{m,a}^{m} +\\
&\hspace{2em}\frac{2 \beta_{18}}{\beta_{14} \beta_{16}} Z_{m,b}^{m} + \frac{2 \beta_{18}}{\beta_{15} \beta_{17}} Z_{m,c}^{m}\Big)
    \end{split}
\end{equation}
\begin{equation}
    \begin{split}
Z_{xy,xy} = Z_{x^3y,x^3y}
&= \frac{1}{8}\Big(Z_{1,1}^{1} - Z_{1,a}^{a} - Z_{1,b}^{b} + Z_{1,c}^{c} - Z_{a,1}^{a} + Z_{a,a}^{1} +\\
&\hspace{2em}\frac{\beta_{1}}{\beta_{2} \beta_{9}} Z_{a,b}^{c} - \frac{\beta_{1}}{\beta_{3} \beta_{5}} Z_{a,c}^{b} - Z_{b,1}^{b} + \frac{\beta_{6}}{\beta_{10} \beta_{5}} Z_{b,a}^{c} + Z_{b,b}^{1} - \frac{\beta_{6}}{\beta_{2} \beta_{7}} Z_{b,c}^{a} +\\
&\hspace{2em}Z_{c,1}^{c} - \frac{\beta_{11}}{\beta_{7} \beta_{9}} Z_{c,a}^{b} - \frac{\beta_{11}}{\beta_{10} \beta_{3}} Z_{c,b}^{a} + Z_{c,c}^{1} + 2 Z_{m,m}^{1} + \frac{2 \beta_{18}}{\beta_{19} \beta_{4}} Z_{m,m}^{a} +\\
&\hspace{2em}-\frac{2 \beta_{18}}{\beta_{16} \beta_{8}} Z_{m,m}^{b} - \frac{2 \beta_{18}}{\beta_{12} \beta_{17}} Z_{m,m}^{c}\Big)
    \end{split}
\end{equation}
\begin{equation}
    \begin{split}
Z_{xy,x^3y} = Z_{x^3y,xy}
&= \frac{1}{8}\Big(Z_{1,1}^{1} - Z_{1,a}^{a} - Z_{1,b}^{b} + Z_{1,c}^{c} - Z_{a,1}^{a} + Z_{a,a}^{1} +\\
&\hspace{2em}\frac{\beta_{1}}{\beta_{2} \beta_{9}} Z_{a,b}^{c} - \frac{\beta_{1}}{\beta_{3} \beta_{5}} Z_{a,c}^{b} - Z_{b,1}^{b} + \frac{\beta_{6}}{\beta_{10} \beta_{5}} Z_{b,a}^{c} + Z_{b,b}^{1} - \frac{\beta_{6}}{\beta_{2} \beta_{7}} Z_{b,c}^{a} +\\
&\hspace{2em}Z_{c,1}^{c} - \frac{\beta_{11}}{\beta_{7} \beta_{9}} Z_{c,a}^{b} - \frac{\beta_{11}}{\beta_{10} \beta_{3}} Z_{c,b}^{a} + Z_{c,c}^{1} - 2 Z_{m,m}^{1} - \frac{2 \beta_{18}}{\beta_{19} \beta_{4}} Z_{m,m}^{a} +\\
&\hspace{2em}\frac{2 \beta_{18}}{\beta_{16} \beta_{8}} Z_{m,m}^{b} + \frac{2 \beta_{18}}{\beta_{12} \beta_{17}} Z_{m,m}^{c}\Big)        
    \end{split}
\end{equation}

\subsection{$\text{Rep}(Q_8)$}

\paragraph{Group data} The $Q_8$ group is similar to $D_4$ group and is defined by
\begin{equation}
    Q_8=\langle x,y\mid x^2=y^2=(xy)^2,\ x^4=1\rangle\,,
\end{equation}
and it has eight elements represented by
    \begin{equation}
        Q_8 = \{e,x,x^2,x^3,y,xy,x^2y,x^3y\}\,,
    \end{equation}
and the five conjugacy classes are represented by $e,x,x^2,y,xy$
\begin{equation}
    [e]=\{e\}\,,\quad [x]=\{x,x^3\}\,,\quad [x^2]=\{x^2\}\,,\quad [y]=\{y,y^3\}\,,\quad [xy]=\{xy,x^3y\}\,.
\end{equation}

There are five irreducible representations $I,a,b,c,m$. The one-dimensional irreducible representations $1,a,b,c$ are the same to $D_4$
    \begin{equation}
    \begin{gathered}
        1(x)=1(y)=1\,\\
        a(x)=1\,,\quad a(y)=-1\,,\\
        b(x)=-1\,,\quad b(y)=1\,,\\
        c(x)=-1\,,\quad c(y)=-1\,,
    \end{gathered}
    \end{equation}
and the two-dimensional irreducible representation $m$ is different and is given by
    \begin{equation}
        m(x)=\left(\begin{array}{cc}
            i & 0 \\
            0 & -i
        \end{array} \right)\,,\quad m(y)=\left(\begin{array}{cc}
            0 & 1 \\
            -1 & 0
        \end{array} \right)\,.
    \end{equation}
The fusion rules are also the same to thoes of $\text{Rep}(D_4)$
    \begin{equation}
    \begin{gathered}
        a\otimes a=b\otimes b=c\otimes c=1\,,\quad a\otimes b=c\,,\quad a\otimes c=b\,,\quad b\otimes c=a\,,\\
        a\otimes m=b\otimes m=c\otimes m = m\,,\quad m\otimes m=1\oplus a\oplus b\oplus c\,,
    \end{gathered}
    \end{equation}
and the character table is also the same
    \begin{table}[!h]
        \centering
        \begin{tabular}{c|c|c|c|c|c}
            & $[1]$ & $[x^2]$ & $[x]$ & $[y]$ & $[xy]$\\
            \hline $\chi_1$ & 1 & 1 & 1 & 1 & 1\\
            \hline
            $\chi_a$ & 1 & 1 & 1 & -1 & -1\\
            \hline
            $\chi_b$ & 1 & 1 & -1 & 1 & -1\\
            \hline
            $\chi_c$ & 1 & 1 & -1 & -1 & 1\\
            \hline
            $\chi_m$ & 2 & -2 & 0 & 0 & 0
        \end{tabular}
    \end{table}

We will use the same ordered basis for $m$-representation and also the tensor product representation $m\otimes m$ as in the $\text{Rep}(D_4)$ case.

\paragraph{Intertwiners and cointertwiners}

The intertwiners are also similar to the $\text{Rep}(D_4)$ case, and are represented by the following matrices
\begin{equation}
    \begin{gathered}
        \mathcal{I}_{1,1}^{1}=1,
\qquad
\mathcal{I}_{1,a}^{a}=\mathcal{I}_{a,1}^{a}=1,
\qquad
\mathcal{I}_{1,b}^{b}=\mathcal{I}_{b,1}^{b}=1,
\qquad
\mathcal{I}_{1,c}^{c}=\mathcal{I}_{c,1}^{c}=1,
\end{gathered}
\end{equation}
\begin{equation}
\begin{gathered}
\mathcal{I}_{1,m}^{m}=\mathcal{I}_{m,1}^{m}=\mathbf 1_2,
\qquad
\mathcal{I}_{a,a}^{1}=\beta_1,
\qquad
\mathcal{I}_{a,b}^{c}=\beta_2,
\qquad
\mathcal{I}_{a,c}^{b}=\beta_3,
\end{gathered}
\end{equation}
\begin{equation}
\begin{gathered}
\mathcal{I}_{a,m}^{m}=\begin{pmatrix}\beta_4&0\\0&-\beta_4\end{pmatrix},
\qquad
\mathcal{I}_{b,a}^{c}=\beta_5,
\qquad
\mathcal{I}_{b,b}^{1}=\beta_6,
\qquad
\mathcal{I}_{b,c}^{a}=\beta_7,
\end{gathered}
\end{equation}
\begin{equation}
\begin{gathered}
\mathcal{I}_{b,m}^{m}=\begin{pmatrix}0&{ {\color{red}{-\beta_8}}}\\ \beta_8&0\end{pmatrix},
\qquad
\mathcal{I}_{c,a}^{b}=\beta_9,
\qquad
\mathcal{I}_{c,b}^{a}=\beta_{10},
\qquad
\mathcal{I}_{c,c}^{1}=\beta_{11},
\end{gathered}
\end{equation}
\begin{equation}
\begin{gathered}
\mathcal{I}_{c,m}^{m}=\begin{pmatrix}0& {\color{red}{\beta_{12}}}\\ \beta_{12}&0\end{pmatrix},
\qquad
\mathcal{I}_{m,a}^{m}=\begin{pmatrix}\beta_{13}&0\\0&-\beta_{13}\end{pmatrix},
\qquad
\mathcal{I}_{m,b}^{m}=\begin{pmatrix}0&{\color{red}{-\beta_{14}}}\\ \beta_{14}&0\end{pmatrix},\end{gathered}
\end{equation}
\begin{equation}
\begin{gathered}
\mathcal{I}_{m,c}^{m}=\begin{pmatrix}0&{\color{red}{\beta_{15}}}\\ \beta_{15}&0\end{pmatrix},
\qquad
\mathcal{I}_{m,m}^{1}=\begin{pmatrix}0&\beta_{18}&{\color{red}{-\beta_{18}}}&0\end{pmatrix},
\qquad
\mathcal{I}_{m,m}^{a}=\begin{pmatrix}0&\beta_{19}&{\color{red}{\beta_{19}}}&0\end{pmatrix},\end{gathered}
\end{equation}
\begin{equation}
\begin{gathered}
\mathcal{I}_{m,m}^{b}=\begin{pmatrix}\beta_{16}&0&0&\beta_{16}\end{pmatrix},
\qquad
\mathcal{I}_{m,m}^{c}=\begin{pmatrix}\beta_{17}&0&0&-\beta_{17}\end{pmatrix}.
    \end{gathered}
\end{equation}
which differs by some sign factors from the $\text{Rep}(D_4)$ case, and we highlight them in red here and also in the follows. The cointertwiners are given by
\begin{equation}
    \begin{gathered}
        (\mathcal{I}^\vee)_{1}^{1,1}=1,
\qquad
(\mathcal{I}^\vee)_{a}^{1,a}=(\mathcal{I}^\vee)_{a}^{a,1}=1,
\qquad
(\mathcal{I}^\vee)_{b}^{1,b}=(\mathcal{I}^\vee)_{b}^{b,1}=1,
\qquad
(\mathcal{I}^\vee)_{c}^{1,c}=(\mathcal{I}^\vee)_{c}^{c,1}=1,
    \end{gathered}
\end{equation}
\begin{equation}
    \begin{gathered}
(\mathcal{I}^\vee)_{m}^{1,m}=(\mathcal{I}^\vee)_{m}^{m,1}=\mathbf 1_2,
\qquad
(\mathcal{I}^\vee)_{1}^{a,a}=\frac{1}{\beta_1},
\qquad
(\mathcal{I}^\vee)_{1}^{b,b}=\frac{1}{\beta_6},
\qquad
(\mathcal{I}^\vee)_{1}^{c,c}=\frac{1}{\beta_{11}},        
    \end{gathered}
\end{equation}
\begin{equation}
    \begin{gathered}
(\mathcal{I}^\vee)_{1}^{m,m}=\begin{pmatrix}0\\ {\color{red}{-1/\beta_{18}}}\\ 1/\beta_{18}\\ 0\end{pmatrix},
\qquad
(\mathcal{I}^\vee)_{a}^{b,c}=\frac{\beta_3}{\beta_{11}},
\qquad
(\mathcal{I}^\vee)_{a}^{c,b}=\frac{\beta_2}{\beta_6},        
    \end{gathered}
\end{equation}
\begin{equation}
    \begin{gathered}
(\mathcal{I}^\vee)_{a}^{m,m}=\begin{pmatrix}0\\ {\color{red}{-\beta_4/\beta_{18}}}\\ -\beta_4/\beta_{18}\\ 0\end{pmatrix},
\qquad
(\mathcal{I}^\vee)_{b}^{a,c}=\frac{\beta_7}{\beta_{11}},
\qquad
(\mathcal{I}^\vee)_{b}^{c,a}=\frac{\beta_5}{\beta_1},        
    \end{gathered}
\end{equation}
\begin{equation}
    \begin{gathered}
(\mathcal{I}^\vee)_{b}^{m,m}=\begin{pmatrix}{\color{red}{-\beta_8/\beta_{18}}}\\ 0\\ 0\\ {\color{red}{-\beta_8/\beta_{18}}}\end{pmatrix},
\qquad
(\mathcal{I}^\vee)_{c}^{a,b}=\frac{\beta_{10}}{\beta_6},
\qquad
(\mathcal{I}^\vee)_{c}^{b,a}=\frac{\beta_9}{\beta_1},        
    \end{gathered}
\end{equation}
\begin{equation}
    \begin{gathered}
(\mathcal{I}^\vee)_{c}^{m,m}=\begin{pmatrix}{\color{red}{\beta_{12}/\beta_{18}}}\\ 0\\ 0\\ {\color{red}{-\beta_{12}/\beta_{18}}}\end{pmatrix},
\qquad
(\mathcal{I}^\vee)_{m}^{a,m}=\begin{pmatrix}\beta_{19}/\beta_{18}&0\\ 0&-\beta_{19}/\beta_{18}\end{pmatrix},
\qquad
(\mathcal{I}^\vee)_{m}^{m,a}=\begin{pmatrix}\beta_{13}/\beta_1&0\\ 0&-\beta_{13}/\beta_1\end{pmatrix},        
    \end{gathered}
\end{equation}
\begin{equation}
    \begin{gathered}
(\mathcal{I}^\vee)_{m}^{b,m}=\begin{pmatrix}0&\beta_{16}/\beta_{18}\\ {\color{red}{-\beta_{16}/\beta_{18}}}&0\end{pmatrix},
\qquad
(\mathcal{I}^\vee)_{m}^{m,b}=\begin{pmatrix}0&{\color{red}{-\beta_{14}/\beta_6}}\\ \beta_{14}/\beta_6&0\end{pmatrix},
    \end{gathered}
\end{equation}
\begin{equation}
    \begin{gathered}
(\mathcal{I}^\vee)_{m}^{c,m}=\begin{pmatrix}0&-\beta_{17}/\beta_{18}\\ {\color{red}{-\beta_{17}/\beta_{18}}}&0\end{pmatrix},
\qquad
(\mathcal{I}^\vee)_{m}^{m,c}=\begin{pmatrix}0&{\color{red}{\beta_{15}/\beta_{11}}}\\ \beta_{15}/\beta_{11}&0\end{pmatrix}.        
    \end{gathered}
\end{equation}
where also color the different terms in red.

\paragraph{Frobenius algebra}
The Frobenius algebra relevant for gauging $\text{Rep}(Q8)$ is given by the regular representation, which decomposes as
\begin{equation}
\mathcal A\cong 1\oplus a\oplus b\oplus c\oplus m_1\oplus m_2.
\end{equation}
The embedding maps are fixed by
\begin{equation}\label{eq:RepQ8_Frobenius_embeding}
\begin{split}
J_1(e^1)
&=v_e+v_x+v_{x^2}+v_{x^3}+v_y+v_{xy}+v_{x^2y}+v_{x^3y},\\
J_a(e^a)
&=v_e+v_x+v_{x^2}+v_{x^3}-v_y-v_{xy}-v_{x^2y}-v_{x^3y},\\
J_b(e^b)
&=v_e-v_x+v_{x^2}-v_{x^3}+v_y-v_{xy}+v_{x^2y}-v_{x^3y},\\
J_c(e^c)
&=v_e-v_x+v_{x^2}-v_{x^3}-v_y+v_{xy}-v_{x^2y}+v_{x^3y},\\
J_{m,1}(e^m_1)
&={\color{red}{ -v_e+i v_x+v_{x^2}-i v_{x^3}}}+ v_y-i v_{xy}-v_{x^2y}+i v_{x^3y},\\
J_{m,1}(e^m_2)
&=v_e+i v_x-v_{x^2}-i v_{x^3}+v_y+i v_{xy}-v_{x^2y}-i v_{x^3y},\\
J_{m,2}(e^m_1)
&={\color{red}{v_e-i v_x-v_{x^2}+i v_{x^3}}}+v_y-i v_{xy}-v_{x^2y}+i v_{x^3y},\\
J_{m,2}(e^m_2)
&=v_e+i v_x-v_{x^2}-i v_{x^3}-v_y-i v_{xy}+v_{x^2y}+i v_{x^3y}.
\end{split}
\end{equation}
In matrix form, these maps are collected as
\begin{equation}
    \mathcal{J}^g_{\rho,r} = \left( \mathcal{J}^g_{1},\mathcal{J}^g_a,\mathcal{J}^g_b,\mathcal{J}^g_c,\mathcal{J}^g_{(m,1),1},\mathcal{J}^g_{(m,2),1},\mathcal{J}^g_{(m,1),2},\mathcal{J}^g_{(m,2),2} \right)\,,
\end{equation}
where the column vectors are the coefficients on the RHS of \eqref{eq:RepQ8_Frobenius_embeding}
\begin{equation}
    \begin{gathered}
        \mathcal{J}^g_{1}=(1,1,1,1,1,1,1,1)^T\,,
    \end{gathered}
\end{equation}
\begin{equation}
    \begin{gathered}
        \mathcal{J}^g_{a}=(1,1,1,1,-1,-1,-1,-1)^T\,,
    \end{gathered}
\end{equation}
\begin{equation}
    \begin{gathered}
        \mathcal{J}^g_{b}=(1,-1,1,-1,1,-1,1,-1)^T\,,
    \end{gathered}
\end{equation}
\begin{equation}
    \begin{gathered}
        \mathcal{J}^g_c=(1,-1,1,-1,-1,1,-1,1)^T\,,
    \end{gathered}
\end{equation}
\begin{equation}
    \begin{gathered}
        \mathcal{J}^g_{(m,1),1}=({\color{red}{-1}},{\color{red}{i}},{\color{red}{1}},{\color{red}{-i}},1,-i,-1,i)^T\,,
    \end{gathered}
\end{equation}
\begin{equation}
    \begin{gathered}
        \mathcal{J}^g_{(m,2),1}=(1,i,-1,-i,1,i,-1,-i)^T\,,
    \end{gathered}
\end{equation}
\begin{equation}
    \begin{gathered}
        \mathcal{J}^g_{(m,1),2}=({\color{red}{1}},{\color{red}{-i}},{\color{red}{-1}},{\color{red}{i}},1,-i,-1,i)^T\,,
    \end{gathered}
\end{equation}
\begin{equation}
    \begin{gathered}
        \mathcal{J}^g_{(m,2),2}=(1,i,-1,-i,-1,-i,1,i)^T\,.
    \end{gathered}
\end{equation}
Similarly, the projection map $J^{\vee}_{\rho,r}$ is the inverse map to $J_{\rho,r}$ and $\mathcal{J}^g_{(\rho,\alpha),r}$ is the inverse matrix of $\mathcal{J}^g_{(\rho,\alpha),r}$.

\paragraph{Fourier kernel} The coefficients for projectors $P_{\rho,\sigma}^{\lambda}=p_{\rho,\sigma}^{\lambda} \Pi_{\rho,\sigma}^{\lambda}$ can summarized as

\begin{equation}
    \begin{gathered}
        p_{11}^1=p_{1a}^a=p_{1b}^b=p_{1c}^c=p_{1m}^m=p_{a1}^a=p_{b1}^b=p_{c1}^c=p_{m1}^m=p_{aa}^1=p_{bb}^1=p_{cc}^1=1
    \end{gathered}
\end{equation}
\begin{equation}
    \begin{gathered}
        p_{ab}^c=\frac{\beta_2 \beta_9}{\beta_1}\,,\quad p_{ba}^c = \frac{\beta_{10}\beta_5}{\beta_6}\,,\quad p_{ac}^b=\frac{\beta_3 \beta_5}{\beta_1}\,,\quad p_{ca}^b=\frac{\beta_7 \beta_9}{\beta_{11}}\,,\quad p_{bc}^a=\frac{\beta_2 \beta_7}{\beta_6}\,,\quad p_{cb}^a=\frac{\beta_{10}\beta_3}{\beta_{11}}\,,
    \end{gathered}
\end{equation}
\begin{equation}
    \begin{gathered}
        p_{am}^m= \frac{\beta_{13}\beta_4}{\beta_1}\,,\quad p_{bm}^m={\color{red}{-\frac{\beta_{14}\beta_8}{\beta_6}}}\,,\quad p_{cm}^m=\color{red}{\frac{\beta_{12}\beta_{15}}{\beta_{11}}}\,,
    \end{gathered}
\end{equation}
\begin{equation}
    \begin{gathered}
        p_{ma}^m= \frac{\beta_{13}\beta_{19}}{\beta_{18}}\,,\quad p_{mb}^m=\frac{\beta_{14}\beta_{16}}{\beta_{18}}\,,\quad p_{mc}^m=-\frac{\beta_{15}\beta_{17}}{\beta_{18}}\,, 
    \end{gathered}
\end{equation}
\begin{equation}
    \begin{gathered}
        p_{mm}^1=2\,,\quad p_{mm}^a=-\frac{2\beta_{19}\beta_4}{\beta_{18}}\,,\quad p_{mm}^b={\color{red}{-\frac{2\beta_8\beta_{16}}{\beta_{18}}}}\,,\quad p_{mm}^c={\color{red}{\frac{2\beta_{12}\beta_{17}}{\beta_{18}}}}\,.
    \end{gathered}
\end{equation}
and the dual partition function for the defect background is read from \eqref{eq:dual_partition_function_multiplicity_free}.

\paragraph{Inverse Fourier kernels}
The inverse kernels assigns the same expression to all members of the same conjugacy orbit of commuting pairs. We have
\begin{equation}
\begin{split}
Z_{e,e}
&= \frac{1}{8}\Big(Z_{1,1}^{1} + Z_{1,a}^{a} + Z_{1,b}^{b} + Z_{1,c}^{c} + 2 Z_{1,m}^{m} + Z_{a,1}^{a} +\\
&\hspace{2em}Z_{a,a}^{1} + \frac{\beta_{1}}{\beta_{2} \beta_{9}} Z_{a,b}^{c} + \frac{\beta_{1}}{\beta_{3} \beta_{5}} Z_{a,c}^{b} + \frac{2 \beta_{1}}{\beta_{13} \beta_{4}} Z_{a,m}^{m} + Z_{b,1}^{b} + \frac{\beta_{6}}{\beta_{10} \beta_{5}} Z_{b,a}^{c} +\\
&\hspace{2em}Z_{b,b}^{1} + \frac{\beta_{6}}{\beta_{2} \beta_{7}} Z_{b,c}^{a} {\color{red}{-\frac{2 \beta_{6}}{\beta_{14} \beta_{8}}}} Z_{b,m}^{m} + Z_{c,1}^{c} + \frac{\beta_{11}}{\beta_{7} \beta_{9}} Z_{c,a}^{b} + \frac{\beta_{11}}{\beta_{10} \beta_{3}} Z_{c,b}^{a} +\\
&\hspace{2em}Z_{c,c}^{1} {\color{red}{+\frac{2 \beta_{11}}{\beta_{12} \beta_{15}}}} Z_{c,m}^{m} + 2 Z_{m,1}^{m} + \frac{2 \beta_{18}}{\beta_{13} \beta_{19}} Z_{m,a}^{m} + \frac{2 \beta_{18}}{\beta_{14} \beta_{16}} Z_{m,b}^{m} - \frac{2 \beta_{18}}{\beta_{15} \beta_{17}} Z_{m,c}^{m} +\\
&\hspace{2em}2 Z_{m,m}^{1} - \frac{2 \beta_{18}}{\beta_{19} \beta_{4}} Z_{m,m}^{a} {\color{red}{-\frac{2 \beta_{18}}{\beta_{16} \beta_{8}}}} Z_{m,m}^{b} {\color{red}{+\frac{2 \beta_{18}}{\beta_{12} \beta_{17}}}} Z_{m,m}^{c}\Big)
\end{split}
\end{equation}

\begin{equation}
\begin{split}
Z_{e,x} = Z_{e,x^3}
&= \frac{1}{8}\Big(Z_{1,1}^{1} + Z_{1,a}^{a} + Z_{1,b}^{b} + Z_{1,c}^{c} + 2 Z_{1,m}^{m} + Z_{a,1}^{a} +\\
&\hspace{2em}Z_{a,a}^{1} + \frac{\beta_{1}}{\beta_{2} \beta_{9}} Z_{a,b}^{c} + \frac{\beta_{1}}{\beta_{3} \beta_{5}} Z_{a,c}^{b} + \frac{2 \beta_{1}}{\beta_{13} \beta_{4}} Z_{a,m}^{m} - Z_{b,1}^{b} - \frac{\beta_{6}}{\beta_{10} \beta_{5}} Z_{b,a}^{c} +\\
&\hspace{2em}-Z_{b,b}^{1} - \frac{\beta_{6}}{\beta_{2} \beta_{7}} Z_{b,c}^{a} {\color{red}{+\frac{2 \beta_{6}}{\beta_{14} \beta_{8}}}} Z_{b,m}^{m} - Z_{c,1}^{c} - \frac{\beta_{11}}{\beta_{7} \beta_{9}} Z_{c,a}^{b} - \frac{\beta_{11}}{\beta_{10} \beta_{3}} Z_{c,b}^{a} +\\
&\hspace{2em}-Z_{c,c}^{1} {\color{red}{-\frac{2 \beta_{11}}{\beta_{12} \beta_{15}}}} Z_{c,m}^{m}\Big)
\end{split}
\end{equation}

\begin{equation}
\begin{split}
Z_{e,x^2}
&= \frac{1}{8}\Big(Z_{1,1}^{1} + Z_{1,a}^{a} + Z_{1,b}^{b} + Z_{1,c}^{c} + 2 Z_{1,m}^{m} + Z_{a,1}^{a} +\\
&\hspace{2em}Z_{a,a}^{1} + \frac{\beta_{1}}{\beta_{2} \beta_{9}} Z_{a,b}^{c} + \frac{\beta_{1}}{\beta_{3} \beta_{5}} Z_{a,c}^{b} + \frac{2 \beta_{1}}{\beta_{13} \beta_{4}} Z_{a,m}^{m} + Z_{b,1}^{b} + \frac{\beta_{6}}{\beta_{10} \beta_{5}} Z_{b,a}^{c} +\\
&\hspace{2em}Z_{b,b}^{1} + \frac{\beta_{6}}{\beta_{2} \beta_{7}} Z_{b,c}^{a} {\color{red}{-\frac{2 \beta_{6}}{\beta_{14} \beta_{8}}}} Z_{b,m}^{m} + Z_{c,1}^{c} + \frac{\beta_{11}}{\beta_{7} \beta_{9}} Z_{c,a}^{b} + \frac{\beta_{11}}{\beta_{10} \beta_{3}} Z_{c,b}^{a} +\\
&\hspace{2em}Z_{c,c}^{1} {\color{red}{+\frac{2 \beta_{11}}{\beta_{12} \beta_{15}}}} Z_{c,m}^{m} - 2 Z_{m,1}^{m} - \frac{2 \beta_{18}}{\beta_{13} \beta_{19}} Z_{m,a}^{m} - \frac{2 \beta_{18}}{\beta_{14} \beta_{16}} Z_{m,b}^{m} + \frac{2 \beta_{18}}{\beta_{15} \beta_{17}} Z_{m,c}^{m} +\\
&\hspace{2em}-2 Z_{m,m}^{1} + \frac{2 \beta_{18}}{\beta_{19} \beta_{4}} Z_{m,m}^{a} {\color{red}{+\frac{2 \beta_{18}}{\beta_{16} \beta_{8}}}} Z_{m,m}^{b} {\color{red}{-\frac{2 \beta_{18}}{\beta_{12} \beta_{17}}}} Z_{m,m}^{c}\Big)
\end{split}
\end{equation}

\begin{equation}
\begin{split}
Z_{e,y} = Z_{e,x^2y}
&= \frac{1}{8}\Big(Z_{1,1}^{1} + Z_{1,a}^{a} + Z_{1,b}^{b} + Z_{1,c}^{c} + 2 Z_{1,m}^{m} - Z_{a,1}^{a} +\\
&\hspace{2em}-Z_{a,a}^{1} - \frac{\beta_{1}}{\beta_{2} \beta_{9}} Z_{a,b}^{c} - \frac{\beta_{1}}{\beta_{3} \beta_{5}} Z_{a,c}^{b} - \frac{2 \beta_{1}}{\beta_{13} \beta_{4}} Z_{a,m}^{m} + Z_{b,1}^{b} + \frac{\beta_{6}}{\beta_{10} \beta_{5}} Z_{b,a}^{c} +\\
&\hspace{2em}Z_{b,b}^{1} + \frac{\beta_{6}}{\beta_{2} \beta_{7}} Z_{b,c}^{a} {\color{red}{-\frac{2 \beta_{6}}{\beta_{14} \beta_{8}}}} Z_{b,m}^{m} - Z_{c,1}^{c} - \frac{\beta_{11}}{\beta_{7} \beta_{9}} Z_{c,a}^{b} - \frac{\beta_{11}}{\beta_{10} \beta_{3}} Z_{c,b}^{a} +\\
&\hspace{2em}-Z_{c,c}^{1} {\color{red}{-\frac{2 \beta_{11}}{\beta_{12} \beta_{15}}}} Z_{c,m}^{m}\Big)
\end{split}
\end{equation}

\begin{equation}
\begin{split}
Z_{e,xy} = Z_{e,x^3y}
&= \frac{1}{8}\Big(Z_{1,1}^{1} + Z_{1,a}^{a} + Z_{1,b}^{b} + Z_{1,c}^{c} + 2 Z_{1,m}^{m} - Z_{a,1}^{a} +\\
&\hspace{2em}-Z_{a,a}^{1} - \frac{\beta_{1}}{\beta_{2} \beta_{9}} Z_{a,b}^{c} - \frac{\beta_{1}}{\beta_{3} \beta_{5}} Z_{a,c}^{b} - \frac{2 \beta_{1}}{\beta_{13} \beta_{4}} Z_{a,m}^{m} - Z_{b,1}^{b} - \frac{\beta_{6}}{\beta_{10} \beta_{5}} Z_{b,a}^{c} +\\
&\hspace{2em}-Z_{b,b}^{1} - \frac{\beta_{6}}{\beta_{2} \beta_{7}} Z_{b,c}^{a} {\color{red}{+\frac{2 \beta_{6}}{\beta_{14} \beta_{8}}}} Z_{b,m}^{m} + Z_{c,1}^{c} + \frac{\beta_{11}}{\beta_{7} \beta_{9}} Z_{c,a}^{b} + \frac{\beta_{11}}{\beta_{10} \beta_{3}} Z_{c,b}^{a} +\\
&\hspace{2em}Z_{c,c}^{1} {\color{red}{+\frac{2 \beta_{11}}{\beta_{12} \beta_{15}}}} Z_{c,m}^{m}\Big)
\end{split}
\end{equation}

\begin{equation}
\begin{split}
Z_{x,e} = Z_{x^3,e}
&= \frac{1}{8}\Big(Z_{1,1}^{1} + Z_{1,a}^{a} - Z_{1,b}^{b} - Z_{1,c}^{c} + Z_{a,1}^{a} + Z_{a,a}^{1} +\\
&\hspace{2em}-\frac{\beta_{1}}{\beta_{2} \beta_{9}} Z_{a,b}^{c} - \frac{\beta_{1}}{\beta_{3} \beta_{5}} Z_{a,c}^{b} + Z_{b,1}^{b} + \frac{\beta_{6}}{\beta_{10} \beta_{5}} Z_{b,a}^{c} - Z_{b,b}^{1} - \frac{\beta_{6}}{\beta_{2} \beta_{7}} Z_{b,c}^{a} +\\
&\hspace{2em}Z_{c,1}^{c} + \frac{\beta_{11}}{\beta_{7} \beta_{9}} Z_{c,a}^{b} - \frac{\beta_{11}}{\beta_{10} \beta_{3}} Z_{c,b}^{a} - Z_{c,c}^{1} + 2 Z_{m,1}^{m} + \frac{2 \beta_{18}}{\beta_{13} \beta_{19}} Z_{m,a}^{m} +\\
&\hspace{2em}-\frac{2 \beta_{18}}{\beta_{14} \beta_{16}} Z_{m,b}^{m} + \frac{2 \beta_{18}}{\beta_{15} \beta_{17}} Z_{m,c}^{m}\Big)
\end{split}
\end{equation}

\begin{equation}
\begin{split}
Z_{x,x} = Z_{x^3,x^3}
&= \frac{1}{8}\Big(Z_{1,1}^{1} + Z_{1,a}^{a} - Z_{1,b}^{b} - Z_{1,c}^{c} + Z_{a,1}^{a} + Z_{a,a}^{1} +\\
&\hspace{2em}-\frac{\beta_{1}}{\beta_{2} \beta_{9}} Z_{a,b}^{c} - \frac{\beta_{1}}{\beta_{3} \beta_{5}} Z_{a,c}^{b} - Z_{b,1}^{b} - \frac{\beta_{6}}{\beta_{10} \beta_{5}} Z_{b,a}^{c} + Z_{b,b}^{1} + \frac{\beta_{6}}{\beta_{2} \beta_{7}} Z_{b,c}^{a} +\\
&\hspace{2em}-Z_{c,1}^{c} - \frac{\beta_{11}}{\beta_{7} \beta_{9}} Z_{c,a}^{b} + \frac{\beta_{11}}{\beta_{10} \beta_{3}} Z_{c,b}^{a} + Z_{c,c}^{1} - 2 Z_{m,m}^{1} + \frac{2 \beta_{18}}{\beta_{19} \beta_{4}} Z_{m,m}^{a} +\\
&\hspace{2em}{\color{red}{-\frac{2 \beta_{18}}{\beta_{16} \beta_{8}}}} Z_{m,m}^{b} {\color{red}{+\frac{2 \beta_{18}}{\beta_{12} \beta_{17}}}} Z_{m,m}^{c}\Big)
\end{split}
\end{equation}

\begin{equation}
\begin{split}
Z_{x,x^2} = Z_{x^3,x^2}
&= \frac{1}{8}\Big(Z_{1,1}^{1} + Z_{1,a}^{a} - Z_{1,b}^{b} - Z_{1,c}^{c} + Z_{a,1}^{a} + Z_{a,a}^{1} +\\
&\hspace{2em}-\frac{\beta_{1}}{\beta_{2} \beta_{9}} Z_{a,b}^{c} - \frac{\beta_{1}}{\beta_{3} \beta_{5}} Z_{a,c}^{b} + Z_{b,1}^{b} + \frac{\beta_{6}}{\beta_{10} \beta_{5}} Z_{b,a}^{c} - Z_{b,b}^{1} - \frac{\beta_{6}}{\beta_{2} \beta_{7}} Z_{b,c}^{a} +\\
&\hspace{2em}Z_{c,1}^{c} + \frac{\beta_{11}}{\beta_{7} \beta_{9}} Z_{c,a}^{b} - \frac{\beta_{11}}{\beta_{10} \beta_{3}} Z_{c,b}^{a} - Z_{c,c}^{1} - 2 Z_{m,1}^{m} - \frac{2 \beta_{18}}{\beta_{13} \beta_{19}} Z_{m,a}^{m} +\\
&\hspace{2em}\frac{2 \beta_{18}}{\beta_{14} \beta_{16}} Z_{m,b}^{m} - \frac{2 \beta_{18}}{\beta_{15} \beta_{17}} Z_{m,c}^{m}\Big)
\end{split}
\end{equation}

\begin{equation}
\begin{split}
Z_{x,x^3} = Z_{x^3,x}
&= \frac{1}{8}\Big(Z_{1,1}^{1} + Z_{1,a}^{a} - Z_{1,b}^{b} - Z_{1,c}^{c} + Z_{a,1}^{a} + Z_{a,a}^{1} +\\
&\hspace{2em}-\frac{\beta_{1}}{\beta_{2} \beta_{9}} Z_{a,b}^{c} - \frac{\beta_{1}}{\beta_{3} \beta_{5}} Z_{a,c}^{b} - Z_{b,1}^{b} - \frac{\beta_{6}}{\beta_{10} \beta_{5}} Z_{b,a}^{c} + Z_{b,b}^{1} + \frac{\beta_{6}}{\beta_{2} \beta_{7}} Z_{b,c}^{a} +\\
&\hspace{2em}-Z_{c,1}^{c} - \frac{\beta_{11}}{\beta_{7} \beta_{9}} Z_{c,a}^{b} + \frac{\beta_{11}}{\beta_{10} \beta_{3}} Z_{c,b}^{a} + Z_{c,c}^{1} + 2 Z_{m,m}^{1} - \frac{2 \beta_{18}}{\beta_{19} \beta_{4}} Z_{m,m}^{a} +\\
&\hspace{2em}{\color{red}{\frac{2 \beta_{18}}{\beta_{16} \beta_{8}}}} Z_{m,m}^{b} {\color{red}{-\frac{2 \beta_{18}}{\beta_{12} \beta_{17}}}} Z_{m,m}^{c}\Big)
\end{split}
\end{equation}

\begin{equation}
\begin{split}
Z_{x^2,e}
&= \frac{1}{8}\Big(Z_{1,1}^{1} + Z_{1,a}^{a} + Z_{1,b}^{b} + Z_{1,c}^{c} - 2 Z_{1,m}^{m} + Z_{a,1}^{a} +\\
&\hspace{2em}Z_{a,a}^{1} + \frac{\beta_{1}}{\beta_{2} \beta_{9}} Z_{a,b}^{c} + \frac{\beta_{1}}{\beta_{3} \beta_{5}} Z_{a,c}^{b} - \frac{2 \beta_{1}}{\beta_{13} \beta_{4}} Z_{a,m}^{m} + Z_{b,1}^{b} + \frac{\beta_{6}}{\beta_{10} \beta_{5}} Z_{b,a}^{c} +\\
&\hspace{2em}Z_{b,b}^{1} + \frac{\beta_{6}}{\beta_{2} \beta_{7}} Z_{b,c}^{a} {\color{red}{+\frac{2 \beta_{6}}{\beta_{14} \beta_{8}}}} Z_{b,m}^{m} + Z_{c,1}^{c} + \frac{\beta_{11}}{\beta_{7} \beta_{9}} Z_{c,a}^{b} + \frac{\beta_{11}}{\beta_{10} \beta_{3}} Z_{c,b}^{a} +\\
&\hspace{2em}Z_{c,c}^{1} {\color{red}{-\frac{2 \beta_{11}}{\beta_{12} \beta_{15}}}} Z_{c,m}^{m} + 2 Z_{m,1}^{m} + \frac{2 \beta_{18}}{\beta_{13} \beta_{19}} Z_{m,a}^{m} + \frac{2 \beta_{18}}{\beta_{14} \beta_{16}} Z_{m,b}^{m} - \frac{2 \beta_{18}}{\beta_{15} \beta_{17}} Z_{m,c}^{m} +\\
&\hspace{2em}-2 Z_{m,m}^{1} + \frac{2 \beta_{18}}{\beta_{19} \beta_{4}} Z_{m,m}^{a} {\color{red}{+\frac{2 \beta_{18}}{\beta_{16} \beta_{8}}}} Z_{m,m}^{b} {\color{red}{-\frac{2 \beta_{18}}{\beta_{12} \beta_{17}}}} Z_{m,m}^{c}\Big)
\end{split}
\end{equation}

\begin{equation}
\begin{split}
Z_{x^2,x} = Z_{x^2,x^3}
&= \frac{1}{8}\Big(Z_{1,1}^{1} + Z_{1,a}^{a} + Z_{1,b}^{b} + Z_{1,c}^{c} - 2 Z_{1,m}^{m} + Z_{a,1}^{a} +\\
&\hspace{2em}Z_{a,a}^{1} + \frac{\beta_{1}}{\beta_{2} \beta_{9}} Z_{a,b}^{c} + \frac{\beta_{1}}{\beta_{3} \beta_{5}} Z_{a,c}^{b} - \frac{2 \beta_{1}}{\beta_{13} \beta_{4}} Z_{a,m}^{m} - Z_{b,1}^{b} - \frac{\beta_{6}}{\beta_{10} \beta_{5}} Z_{b,a}^{c} +\\
&\hspace{2em}-Z_{b,b}^{1} - \frac{\beta_{6}}{\beta_{2} \beta_{7}} Z_{b,c}^{a} {\color{red}{-\frac{2 \beta_{6}}{\beta_{14} \beta_{8}}}} Z_{b,m}^{m} - Z_{c,1}^{c} - \frac{\beta_{11}}{\beta_{7} \beta_{9}} Z_{c,a}^{b} - \frac{\beta_{11}}{\beta_{10} \beta_{3}} Z_{c,b}^{a} +\\
&\hspace{2em}-Z_{c,c}^{1} {\color{red}{+\frac{2 \beta_{11}}{\beta_{12} \beta_{15}}}} Z_{c,m}^{m}\Big)
\end{split}
\end{equation}

\begin{equation}
\begin{split}
Z_{x^2,x^2}
&= \frac{1}{8}\Big(Z_{1,1}^{1} + Z_{1,a}^{a} + Z_{1,b}^{b} + Z_{1,c}^{c} - 2 Z_{1,m}^{m} + Z_{a,1}^{a} +\\
&\hspace{2em}Z_{a,a}^{1} + \frac{\beta_{1}}{\beta_{2} \beta_{9}} Z_{a,b}^{c} + \frac{\beta_{1}}{\beta_{3} \beta_{5}} Z_{a,c}^{b} - \frac{2 \beta_{1}}{\beta_{13} \beta_{4}} Z_{a,m}^{m} + Z_{b,1}^{b} + \frac{\beta_{6}}{\beta_{10} \beta_{5}} Z_{b,a}^{c} +\\
&\hspace{2em}Z_{b,b}^{1} + \frac{\beta_{6}}{\beta_{2} \beta_{7}} Z_{b,c}^{a} {\color{red}{+\frac{2 \beta_{6}}{\beta_{14} \beta_{8}}}} Z_{b,m}^{m} + Z_{c,1}^{c} + \frac{\beta_{11}}{\beta_{7} \beta_{9}} Z_{c,a}^{b} + \frac{\beta_{11}}{\beta_{10} \beta_{3}} Z_{c,b}^{a} +\\
&\hspace{2em}Z_{c,c}^{1} {\color{red}{-\frac{2 \beta_{11}}{\beta_{12} \beta_{15}}}} Z_{c,m}^{m} - 2 Z_{m,1}^{m} - \frac{2 \beta_{18}}{\beta_{13} \beta_{19}} Z_{m,a}^{m} - \frac{2 \beta_{18}}{\beta_{14} \beta_{16}} Z_{m,b}^{m} + \frac{2 \beta_{18}}{\beta_{15} \beta_{17}} Z_{m,c}^{m} +\\
&\hspace{2em}2 Z_{m,m}^{1} - \frac{2 \beta_{18}}{\beta_{19} \beta_{4}} Z_{m,m}^{a} {\color{red}{-\frac{2 \beta_{18}}{\beta_{16} \beta_{8}}}} Z_{m,m}^{b} {\color{red}{+\frac{2 \beta_{18}}{\beta_{12} \beta_{17}}}} Z_{m,m}^{c}\Big)
\end{split}
\end{equation}

\begin{equation}
\begin{split}
Z_{x^2,y} = Z_{x^2,x^2y}
&= \frac{1}{8}\Big(Z_{1,1}^{1} + Z_{1,a}^{a} + Z_{1,b}^{b} + Z_{1,c}^{c} - 2 Z_{1,m}^{m} - Z_{a,1}^{a} +\\
&\hspace{2em}-Z_{a,a}^{1} - \frac{\beta_{1}}{\beta_{2} \beta_{9}} Z_{a,b}^{c} - \frac{\beta_{1}}{\beta_{3} \beta_{5}} Z_{a,c}^{b} + \frac{2 \beta_{1}}{\beta_{13} \beta_{4}} Z_{a,m}^{m} + Z_{b,1}^{b} + \frac{\beta_{6}}{\beta_{10} \beta_{5}} Z_{b,a}^{c} +\\
&\hspace{2em}Z_{b,b}^{1} + \frac{\beta_{6}}{\beta_{2} \beta_{7}} Z_{b,c}^{a} {\color{red}{+\frac{2 \beta_{6}}{\beta_{14} \beta_{8}}}} Z_{b,m}^{m} - Z_{c,1}^{c} - \frac{\beta_{11}}{\beta_{7} \beta_{9}} Z_{c,a}^{b} - \frac{\beta_{11}}{\beta_{10} \beta_{3}} Z_{c,b}^{a} +\\
&\hspace{2em}-Z_{c,c}^{1} {\color{red}{+\frac{2 \beta_{11}}{\beta_{12} \beta_{15}}}} Z_{c,m}^{m}\Big)
\end{split}
\end{equation}

\begin{equation}
\begin{split}
Z_{x^2,xy} = Z_{x^2,x^3y}
&= \frac{1}{8}\Big(Z_{1,1}^{1} + Z_{1,a}^{a} + Z_{1,b}^{b} + Z_{1,c}^{c} - 2 Z_{1,m}^{m} - Z_{a,1}^{a} +\\
&\hspace{2em}-Z_{a,a}^{1} - \frac{\beta_{1}}{\beta_{2} \beta_{9}} Z_{a,b}^{c} - \frac{\beta_{1}}{\beta_{3} \beta_{5}} Z_{a,c}^{b} + \frac{2 \beta_{1}}{\beta_{13} \beta_{4}} Z_{a,m}^{m} - Z_{b,1}^{b} - \frac{\beta_{6}}{\beta_{10} \beta_{5}} Z_{b,a}^{c} +\\
&\hspace{2em}-Z_{b,b}^{1} - \frac{\beta_{6}}{\beta_{2} \beta_{7}} Z_{b,c}^{a} {\color{red}{-\frac{2 \beta_{6}}{\beta_{14} \beta_{8}}}} Z_{b,m}^{m} + Z_{c,1}^{c} + \frac{\beta_{11}}{\beta_{7} \beta_{9}} Z_{c,a}^{b} + \frac{\beta_{11}}{\beta_{10} \beta_{3}} Z_{c,b}^{a} +\\
&\hspace{2em}Z_{c,c}^{1} {\color{red}{-\frac{2 \beta_{11}}{\beta_{12} \beta_{15}}}} Z_{c,m}^{m}\Big)
\end{split}
\end{equation}

\begin{equation}
\begin{split}
Z_{y,e} = Z_{x^2y,e}
&= \frac{1}{8}\Big(Z_{1,1}^{1} - Z_{1,a}^{a} + Z_{1,b}^{b} - Z_{1,c}^{c} + Z_{a,1}^{a} - Z_{a,a}^{1} +\\
&\hspace{2em}\frac{\beta_{1}}{\beta_{2} \beta_{9}} Z_{a,b}^{c} - \frac{\beta_{1}}{\beta_{3} \beta_{5}} Z_{a,c}^{b} + Z_{b,1}^{b} - \frac{\beta_{6}}{\beta_{10} \beta_{5}} Z_{b,a}^{c} + Z_{b,b}^{1} - \frac{\beta_{6}}{\beta_{2} \beta_{7}} Z_{b,c}^{a} +\\
&\hspace{2em}Z_{c,1}^{c} - \frac{\beta_{11}}{\beta_{7} \beta_{9}} Z_{c,a}^{b} + \frac{\beta_{11}}{\beta_{10} \beta_{3}} Z_{c,b}^{a} - Z_{c,c}^{1} + 2 Z_{m,1}^{m} - \frac{2 \beta_{18}}{\beta_{13} \beta_{19}} Z_{m,a}^{m} +\\
&\hspace{2em}\frac{2 \beta_{18}}{\beta_{14} \beta_{16}} Z_{m,b}^{m} + \frac{2 \beta_{18}}{\beta_{15} \beta_{17}} Z_{m,c}^{m}\Big)
\end{split}
\end{equation}

\begin{equation}
\begin{split}
Z_{y,x^2} = Z_{x^2y,x^2}
&= \frac{1}{8}\Big(Z_{1,1}^{1} - Z_{1,a}^{a} + Z_{1,b}^{b} - Z_{1,c}^{c} + Z_{a,1}^{a} - Z_{a,a}^{1} +\\
&\hspace{2em}\frac{\beta_{1}}{\beta_{2} \beta_{9}} Z_{a,b}^{c} - \frac{\beta_{1}}{\beta_{3} \beta_{5}} Z_{a,c}^{b} + Z_{b,1}^{b} - \frac{\beta_{6}}{\beta_{10} \beta_{5}} Z_{b,a}^{c} + Z_{b,b}^{1} - \frac{\beta_{6}}{\beta_{2} \beta_{7}} Z_{b,c}^{a} +\\
&\hspace{2em}Z_{c,1}^{c} - \frac{\beta_{11}}{\beta_{7} \beta_{9}} Z_{c,a}^{b} + \frac{\beta_{11}}{\beta_{10} \beta_{3}} Z_{c,b}^{a} - Z_{c,c}^{1} - 2 Z_{m,1}^{m} + \frac{2 \beta_{18}}{\beta_{13} \beta_{19}} Z_{m,a}^{m} +\\
&\hspace{2em}-\frac{2 \beta_{18}}{\beta_{14} \beta_{16}} Z_{m,b}^{m} - \frac{2 \beta_{18}}{\beta_{15} \beta_{17}} Z_{m,c}^{m}\Big)
\end{split}
\end{equation}

\begin{equation}
\begin{split}
Z_{y,y} = Z_{x^2y,x^2y}
&= \frac{1}{8}\Big(Z_{1,1}^{1} - Z_{1,a}^{a} + Z_{1,b}^{b} - Z_{1,c}^{c} - Z_{a,1}^{a} + Z_{a,a}^{1} +\\
&\hspace{2em}-\frac{\beta_{1}}{\beta_{2} \beta_{9}} Z_{a,b}^{c} + \frac{\beta_{1}}{\beta_{3} \beta_{5}} Z_{a,c}^{b} + Z_{b,1}^{b} - \frac{\beta_{6}}{\beta_{10} \beta_{5}} Z_{b,a}^{c} + Z_{b,b}^{1} - \frac{\beta_{6}}{\beta_{2} \beta_{7}} Z_{b,c}^{a} +\\
&\hspace{2em}-Z_{c,1}^{c} + \frac{\beta_{11}}{\beta_{7} \beta_{9}} Z_{c,a}^{b} - \frac{\beta_{11}}{\beta_{10} \beta_{3}} Z_{c,b}^{a} + Z_{c,c}^{1} {\color{red}{-2}} Z_{m,m}^{1} {\color{red}{-\frac{2 \beta_{18}}{\beta_{19} \beta_{4}}}} Z_{m,m}^{a} +\\
&\hspace{2em}\frac{2 \beta_{18}}{\beta_{16} \beta_{8}} Z_{m,m}^{b} + \frac{2 \beta_{18}}{\beta_{12} \beta_{17}} Z_{m,m}^{c}\Big)
\end{split}
\end{equation}

\begin{equation}
\begin{split}
Z_{y,x^2y} = Z_{x^2y,y}
&= \frac{1}{8}\Big(Z_{1,1}^{1} - Z_{1,a}^{a} + Z_{1,b}^{b} - Z_{1,c}^{c} - Z_{a,1}^{a} + Z_{a,a}^{1} +\\
&\hspace{2em}-\frac{\beta_{1}}{\beta_{2} \beta_{9}} Z_{a,b}^{c} + \frac{\beta_{1}}{\beta_{3} \beta_{5}} Z_{a,c}^{b} + Z_{b,1}^{b} - \frac{\beta_{6}}{\beta_{10} \beta_{5}} Z_{b,a}^{c} + Z_{b,b}^{1} - \frac{\beta_{6}}{\beta_{2} \beta_{7}} Z_{b,c}^{a} +\\
&\hspace{2em}-Z_{c,1}^{c} + \frac{\beta_{11}}{\beta_{7} \beta_{9}} Z_{c,a}^{b} - \frac{\beta_{11}}{\beta_{10} \beta_{3}} Z_{c,b}^{a} + Z_{c,c}^{1} {\color{red}{+2}} Z_{m,m}^{1} {\color{red}{+\frac{2 \beta_{18}}{\beta_{19} \beta_{4}}}} Z_{m,m}^{a} +\\
&\hspace{2em}-\frac{2 \beta_{18}}{\beta_{16} \beta_{8}} Z_{m,m}^{b} - \frac{2 \beta_{18}}{\beta_{12} \beta_{17}} Z_{m,m}^{c}\Big)
\end{split}
\end{equation}

\begin{equation}
\begin{split}
Z_{xy,e} = Z_{x^3y,e}
&= \frac{1}{8}\Big(Z_{1,1}^{1} - Z_{1,a}^{a} - Z_{1,b}^{b} + Z_{1,c}^{c} + Z_{a,1}^{a} - Z_{a,a}^{1} +\\
&\hspace{2em}-\frac{\beta_{1}}{\beta_{2} \beta_{9}} Z_{a,b}^{c} + \frac{\beta_{1}}{\beta_{3} \beta_{5}} Z_{a,c}^{b} + Z_{b,1}^{b} - \frac{\beta_{6}}{\beta_{10} \beta_{5}} Z_{b,a}^{c} - Z_{b,b}^{1} + \frac{\beta_{6}}{\beta_{2} \beta_{7}} Z_{b,c}^{a} +\\
&\hspace{2em}Z_{c,1}^{c} - \frac{\beta_{11}}{\beta_{7} \beta_{9}} Z_{c,a}^{b} - \frac{\beta_{11}}{\beta_{10} \beta_{3}} Z_{c,b}^{a} + Z_{c,c}^{1} + 2 Z_{m,1}^{m} - \frac{2 \beta_{18}}{\beta_{13} \beta_{19}} Z_{m,a}^{m} +\\
&\hspace{2em}-\frac{2 \beta_{18}}{\beta_{14} \beta_{16}} Z_{m,b}^{m} - \frac{2 \beta_{18}}{\beta_{15} \beta_{17}} Z_{m,c}^{m}\Big)
\end{split}
\end{equation}

\begin{equation}
\begin{split}
Z_{xy,x^2} = Z_{x^3y,x^2}
&= \frac{1}{8}\Big(Z_{1,1}^{1} - Z_{1,a}^{a} - Z_{1,b}^{b} + Z_{1,c}^{c} + Z_{a,1}^{a} - Z_{a,a}^{1} +\\
&\hspace{2em}-\frac{\beta_{1}}{\beta_{2} \beta_{9}} Z_{a,b}^{c} + \frac{\beta_{1}}{\beta_{3} \beta_{5}} Z_{a,c}^{b} + Z_{b,1}^{b} - \frac{\beta_{6}}{\beta_{10} \beta_{5}} Z_{b,a}^{c} - Z_{b,b}^{1} + \frac{\beta_{6}}{\beta_{2} \beta_{7}} Z_{b,c}^{a} +\\
&\hspace{2em}Z_{c,1}^{c} - \frac{\beta_{11}}{\beta_{7} \beta_{9}} Z_{c,a}^{b} - \frac{\beta_{11}}{\beta_{10} \beta_{3}} Z_{c,b}^{a} + Z_{c,c}^{1} - 2 Z_{m,1}^{m} + \frac{2 \beta_{18}}{\beta_{13} \beta_{19}} Z_{m,a}^{m} +\\
&\hspace{2em}\frac{2 \beta_{18}}{\beta_{14} \beta_{16}} Z_{m,b}^{m} + \frac{2 \beta_{18}}{\beta_{15} \beta_{17}} Z_{m,c}^{m}\Big)
\end{split}
\end{equation}

\begin{equation}
\begin{split}
Z_{xy,xy} = Z_{x^3y,x^3y}
&= \frac{1}{8}\Big(Z_{1,1}^{1} - Z_{1,a}^{a} - Z_{1,b}^{b} + Z_{1,c}^{c} - Z_{a,1}^{a} + Z_{a,a}^{1} +\\
&\hspace{2em}\frac{\beta_{1}}{\beta_{2} \beta_{9}} Z_{a,b}^{c} - \frac{\beta_{1}}{\beta_{3} \beta_{5}} Z_{a,c}^{b} - Z_{b,1}^{b} + \frac{\beta_{6}}{\beta_{10} \beta_{5}} Z_{b,a}^{c} + Z_{b,b}^{1} - \frac{\beta_{6}}{\beta_{2} \beta_{7}} Z_{b,c}^{a} +\\
&\hspace{2em}Z_{c,1}^{c} - \frac{\beta_{11}}{\beta_{7} \beta_{9}} Z_{c,a}^{b} - \frac{\beta_{11}}{\beta_{10} \beta_{3}} Z_{c,b}^{a} + Z_{c,c}^{1} {\color{red}{-2}} Z_{m,m}^{1} {\color{red}{-\frac{2 \beta_{18}}{\beta_{19} \beta_{4}}}} Z_{m,m}^{a} +\\
&\hspace{2em}-\frac{2 \beta_{18}}{\beta_{16} \beta_{8}} Z_{m,m}^{b} - \frac{2 \beta_{18}}{\beta_{12} \beta_{17}} Z_{m,m}^{c}\Big)
\end{split}
\end{equation}

\begin{equation}
\begin{split}
Z_{xy,x^3y} = Z_{x^3y,xy}
&= \frac{1}{8}\Big(Z_{1,1}^{1} - Z_{1,a}^{a} - Z_{1,b}^{b} + Z_{1,c}^{c} - Z_{a,1}^{a} + Z_{a,a}^{1} +\\
&\hspace{2em}\frac{\beta_{1}}{\beta_{2} \beta_{9}} Z_{a,b}^{c} - \frac{\beta_{1}}{\beta_{3} \beta_{5}} Z_{a,c}^{b} - Z_{b,1}^{b} + \frac{\beta_{6}}{\beta_{10} \beta_{5}} Z_{b,a}^{c} + Z_{b,b}^{1} - \frac{\beta_{6}}{\beta_{2} \beta_{7}} Z_{b,c}^{a} +\\
&\hspace{2em}Z_{c,1}^{c} - \frac{\beta_{11}}{\beta_{7} \beta_{9}} Z_{c,a}^{b} - \frac{\beta_{11}}{\beta_{10} \beta_{3}} Z_{c,b}^{a} + Z_{c,c}^{1} {\color{red}{+2}} Z_{m,m}^{1} {\color{red}{+\frac{2 \beta_{18}}{\beta_{19} \beta_{4}}}} Z_{m,m}^{a} +\\
&\hspace{2em}\frac{2 \beta_{18}}{\beta_{16} \beta_{8}} Z_{m,m}^{b} + \frac{2 \beta_{18}}{\beta_{12} \beta_{17}} Z_{m,m}^{c}\Big)
\end{split}
\end{equation}

\section{Examples : Finite $\text{Rep}(G)$ category with higher-multiplicity fusion}

In this section, we will turn to the representation category $\text{Rep}(G)$ with higher multiplicity fusion rule. Our main example will be the $\text{Rep}(A_4)$, where $A_4$ is the alternating group of four elements. We will first summarize the group data, and give the intertwiners and cointertwiners, emphasizing the higher multiplicity of the tensor product. Then we review the Frobenius algebra as regular representation. After that, we give the Fourier kernel. For the multiplicity-free channel, the analyses is the same to the previous section, and we will focus on the higher multiplicity channel and give the explicit expression of the dual defect partition function. Then we will also give the inverse Fourier transformations.

\paragraph{Group data}

The $A_4$ group is defined by
\begin{equation}
    A_4=\langle a,b\mid a^3=e,\ b^2=e,\ (ab)^3=e\rangle\,,
\end{equation}
where $a,b$ are generators of $\mathbb{Z}_3$ and $\mathbb{Z}_2$. There are 12 elements listed as
    \begin{equation}
        A_4=\{e,a,ab,ba,bab,a^2,a^2b,aba,ba^2,b,a^2ba,aba^2 \}\,,
    \end{equation}
and the four conjugacy classes are represented by $e,a,a^2,b$
    \begin{equation}
        [e]=\{e\}\,,\quad [a]=\{a,ab,ba,bab\}\,,\quad [a^2]=\{a^2,a^2b,aba,ba^2\}\,,\quad [b]=\{b,a^2ba,aba^2\}\,.
    \end{equation}

There are four irreducible representations, denoted by $1,X,Y,Z$. The three one-dimensional irreducible representations
    \begin{equation}
        \begin{gathered}
            1(a)=1\,,\quad 1(b)=1\,,\\
            X(a)=\zeta\,,\quad X(b)=1\,,\\
            Y(a)=\zeta^2\,,\quad Y(b)=1\,,
        \end{gathered}
    \end{equation}
with $\zeta = e^{2\pi i /3}$. And the three-dimensional irreducible representation $Z$ is
    \begin{equation}
        Z(a)=\left(\begin{array}{ccc}
            -\frac{1}{2}&-\frac{\sqrt{3}}{2}&0\\
            \frac{\sqrt{3}}{2}&-\frac{1}{2}&0\\0&0&1
        \end{array}\right)\,,\quad Z(b)=\left(\begin{array}{ccc}
            -1&0&0\\
            0&\frac{1}{3}&\frac{2\sqrt{2}}{3}\\0&\frac{2\sqrt{2}}{3}&-\frac{1}{3}
        \end{array} \right)\,.
    \end{equation}
The non-trivial fusion rules are
\begin{equation}
\begin{gathered}
X\otimes X=Y,\qquad X\otimes Y=1,\qquad Y\otimes X=1,\qquad Y\otimes Y=X,\\
X\otimes Z=Z,\qquad Y\otimes Z=Z,\qquad Z\otimes X=Z,\qquad Z\otimes Y=Z,\\
Z\otimes Z=1\oplus X\oplus Y\oplus 2Z.
\end{gathered}
\end{equation}
Thus $\mathrm{Rep}(A_4)$ is multiplicity-free except for the channel $Z\otimes Z\to Z$, which has multiplicity $2$. The character table is
    \begin{table}[!h]
        \centering
        \begin{tabular}{c|c|c|c|c}
            &$[e]$ & $[a]$ & $[a^2]$ & $[b]$\\
            \hline
            $\chi_1$ & 1 & 1 & 1 & 1\\
            \hline
            $\chi_X$ & 1 & $\zeta$ & $\zeta^2$ & 1\\
            \hline
            $\chi_Y$ & 1 & $\zeta^2$ & $\zeta$ & 1\\
            \hline
            $\chi_Z$ & 3 & 0 & 0 & -1
        \end{tabular}
    \end{table}

For $Z$-representation, we use the ordered basis $\{e^Z_1,e^Z_2,e^Z_3 \}$. For $Z\otimes Z$ we use the ordered tensor-product basis
\begin{equation}
\{e^Z_1\otimes e^Z_1,\ e^Z_1\otimes e^Z_2,\ e^Z_1\otimes e^Z_3,\ e^Z_2\otimes e^Z_1,\ e^Z_2\otimes e^Z_2,\ e^Z_2\otimes e^Z_3,\ e^Z_3\otimes e^Z_1,\ e^Z_3\otimes e^Z_2,\ e^Z_3\otimes e^Z_3\}.
\end{equation}
For $1$-dimensional objects tensored with $Z$, and for $Z$ tensored with a $1$-dimensional object, we use the obvious ordered bases.

\paragraph{Intertwiners and cointertwiners}

To describe the (co)intertwiners, it is convenient to introduce
\begin{equation}
A=
\begin{pmatrix}
\tfrac12 & \tfrac{i}{2} & -\tfrac{i}{\sqrt2}\\
\tfrac{i}{2} & -\tfrac12 & -\tfrac{1}{\sqrt2}\\
-\tfrac{i}{\sqrt2} & -\tfrac{1}{\sqrt2} & 0
\end{pmatrix},
\qquad
B=
\begin{pmatrix}
-\tfrac{i}{2} & -\tfrac12 & \tfrac{1}{\sqrt2}\\
-\tfrac12 & \tfrac{i}{2} & \tfrac{i}{\sqrt2}\\
\tfrac{1}{\sqrt2} & \tfrac{i}{\sqrt2} & 0
\end{pmatrix},
\end{equation}
\begin{equation}
\begin{gathered}
\nu_1=(1,0,0,0,1,0,0,0,1),\\
\nu_X=
\left(-\tfrac{i}{2},-\tfrac12,\tfrac{1}{\sqrt2},-\tfrac12,\tfrac{i}{2},\tfrac{i}{\sqrt2},\tfrac{1}{\sqrt2},\tfrac{i}{\sqrt2},0\right),\\
\nu_Y=
\left(\tfrac12,\tfrac{i}{2},-\tfrac{i}{\sqrt2},\tfrac{i}{2},-\tfrac12,-\tfrac{1}{\sqrt2},-\tfrac{i}{\sqrt2},-\tfrac{1}{\sqrt2},0\right).
\end{gathered}
\end{equation}
For the multiplicity-two channel we also define
\begin{equation}
\mathcal C(\alpha,\beta)=
\begin{pmatrix}
0 & \tfrac{i\alpha}{\sqrt2} & \tfrac{i\alpha}{2} & \tfrac{i\alpha}{\sqrt2} & 0 & -\tfrac{\alpha}{2}-i\beta & \tfrac{i\alpha}{2} & \tfrac{\alpha}{2}+i\beta & 0\\
\tfrac{i\alpha}{\sqrt2} & 0 & \tfrac{\alpha}{2}+i\beta & 0 & -\tfrac{i\alpha}{\sqrt2} & \tfrac{i\alpha}{2} & -\tfrac{\alpha}{2}-i\beta & \tfrac{i\alpha}{2} & 0\\
\tfrac{i\alpha}{2} & -\tfrac{\alpha}{2}-i\beta & 0 & \tfrac{\alpha}{2}+i\beta & \tfrac{i\alpha}{2} & 0 & 0 & 0 & -i\alpha
\end{pmatrix}.
\end{equation}

Then the intertwiners are
\begin{equation}
I_{1,1}^{1}=1,
\qquad
I_{1,X}^{X}=I_{X,1}^{X}=1,
\qquad
I_{1,Y}^{Y}=I_{Y,1}^{Y}=1,\qquad 
I_{1,Z}^{Z}=I_{Z,1}^{Z}=\mathbf 1_3.
\end{equation}
\begin{equation}
I_{X,X}^{Y}=\beta_1,
\qquad
I_{X,Y}^{1}=\beta_2,
\qquad
I_{Y,X}^{1}=\beta_3,
\qquad
I_{Y,Y}^{X}=\beta_4,
\end{equation}
\begin{equation}
I_{X,Z}^{Z}=\beta_5 A,
\qquad
I_{Z,X}^{Z}=\beta_6 A,
\qquad
I_{Y,Z}^{Z}=\beta_7 B,
\qquad
I_{Z,Y}^{Z}=\beta_8 B,    
\end{equation}
\begin{equation}
I_{Z,Z}^{1}=\beta_9\,\nu_1,
\qquad
I_{Z,Z}^{X}=\beta_{10}\,\nu_X,
\qquad
I_{Z,Z}^{Y}=\beta_{11}\,\nu_Y,
\end{equation}
\begin{equation}
I_{Z,Z}^{Z;1}=\mathcal C(\beta_{12},\beta_{13}),
\qquad
I_{Z,Z}^{Z;2}=\mathcal C(\beta_{14},\beta_{15}).    
\end{equation}
And the cointertwiners are
\begin{equation}
    (I^\vee)_{1}^{1,1}=1,
\qquad
(I^\vee)_{X}^{1,X}=(I^\vee)_{X}^{X,1}=1,
\qquad
(I^\vee)_{Y}^{1,Y}=(I^\vee)_{Y}^{Y,1}=1, 
\end{equation}
\begin{equation}
    (I^\vee)_{Z}^{1,Z}=(I^\vee)_{Z}^{Z,1}=\mathbf 1_3.
\end{equation}
\begin{equation}
(I^\vee)_{1}^{X,Y}=\beta_3^{-1},
\qquad
(I^\vee)_{1}^{Y,X}=\beta_2^{-1},
\qquad
(I^\vee)_{X}^{Y,Y}=\frac{\beta_1}{\beta_3},
\qquad
(I^\vee)_{Y}^{X,X}=\frac{\beta_4}{\beta_2},
\end{equation}
\begin{equation}
(I^\vee)_{1}^{Z,Z}=\beta_9^{-1}\,\nu_1^{\mathrm T},
\qquad
(I^\vee)_{X}^{Z,Z}=\frac{\beta_5}{\beta_9}\,\nu_Y^{\mathrm T},
\qquad
(I^\vee)_{Y}^{Z,Z}=\frac{\beta_7}{\beta_9}\,\nu_X^{\mathrm T},
\end{equation}
\begin{equation}
(I^\vee)_{Z}^{X,Z}=\frac{\beta_{10}}{\beta_9} B,
\qquad
(I^\vee)_{Z}^{Y,Z}=\frac{\beta_{11}}{\beta_9} A,
\end{equation}
\begin{equation}
(I^\vee)_{Z}^{Z,X}=\frac{\beta_8}{\beta_2} B,
\qquad
(I^\vee)_{Z}^{Z,Y}=\frac{\beta_6}{\beta_3} A,
\end{equation}
\begin{equation}
(I^\vee)_{Z;1}^{Z,Z}=\beta_9^{-1}\,\mathcal C(\beta_{12},\beta_{13})^{\mathrm T},
\qquad
(I^\vee)_{Z;2}^{Z,Z}=\beta_9^{-1}\,\mathcal C(\beta_{14},\beta_{15})^{\mathrm T}.
\end{equation}

\paragraph{Frobenius Algebra}

The regular representation is decomposed as
\begin{equation}
    \mathcal A\cong 1\oplus X\oplus Y\oplus Z_1\oplus Z_2\oplus Z_3,
\end{equation}
and the embedding map $J_{\rho,r}$ reads
\begin{align*}
J_1(1)
&=v_e+v_a+v_{ab}+v_{ba}+v_{bab}+v_{a^2}+v_{a^2b}+v_{aba}+v_{ba^2}+v_b+v_{a^2ba}+v_{aba^2},\\
J_X(e^X)
&=v_e+v_b+v_{a^2ba}+v_{aba^2}
+\zeta^2\,(v_a+v_{ab}+v_{ba}+v_{bab})
+\zeta\,(v_{a^2}+v_{a^2b}+v_{aba}+v_{ba^2}),\\
J_Y(e^Y)
&=v_e+v_b+v_{a^2ba}+v_{aba^2}
+\zeta\,(v_a+v_{ab}+v_{ba}+v_{bab})
+\zeta^2\,(v_{a^2}+v_{a^2b}+v_{aba}+v_{ba^2}),\\
J_{Z,1}(e^Z_1)
&=\sqrt{\tfrac32}\,(-v_a-v_{ab}+v_{ba}+v_{bab}+v_{a^2}+v_{a^2b}-v_{aba}-v_{ba^2}),\\
J_{Z,1}(e^Z_2)
&=\tfrac{1}{\sqrt2}\,(2v_e-v_a-v_{ab}+v_{ba}+v_{bab}-v_{a^2}-v_{a^2b}+v_{aba}+v_{ba^2}+2v_b-2v_{a^2ba}-2v_{aba^2}),\\
J_{Z,1}(e^Z_3)
&=v_e+v_a+v_{ab}-v_{ba}-v_{bab}+v_{a^2}+v_{a^2b}-v_{aba}-v_{ba^2}+v_b-v_{a^2ba}-v_{aba^2},\\
J_{Z,2}(e^Z_1)
&=\sqrt{\tfrac32}\,(v_e-v_{a^2}+v_{a^2b}-v_{aba}+v_{ba^2}-v_b+v_{a^2ba}-v_{aba^2}),\\
J_{Z,2}(e^Z_2)
&=\tfrac{1}{\sqrt2}\,(-v_e+2v_a-2v_{ab}+2v_{ba}-2v_{bab}-v_{a^2}+v_{a^2b}-v_{aba}+v_{ba^2}+v_b-v_{a^2ba}+v_{aba^2}),\\
J_{Z,2}(e^Z_3)
&=v_e+v_a-v_{ab}+v_{ba}-v_{bab}+v_{a^2}-v_{a^2b}+v_{aba}-v_{ba^2}-v_b+v_{a^2ba}-v_{aba^2},\\
J_{Z,3}(e^Z_1)
&=\sqrt{\tfrac32}\,(-v_e+v_a-v_{ab}-v_{ba}+v_{bab}+v_b+v_{a^2ba}-v_{aba^2}),\\
J_{Z,3}(e^Z_2)
&=\tfrac{1}{\sqrt2}\,(-v_e-v_a+v_{ab}+v_{ba}-v_{bab}+2v_{a^2}-2v_{a^2b}-2v_{aba}+2v_{ba^2}+v_b+v_{a^2ba}-v_{aba^2}),\\
J_{Z,3}(e^Z_3)
&=v_e+v_a-v_{ab}-v_{ba}+v_{bab}+v_{a^2}-v_{a^2b}-v_{aba}+v_{ba^2}-v_b-v_{a^2ba}+v_{aba^2},
\end{align*}
where the we can read the coefficients $\mathcal{J}^g_{\rho,r}$. 

\paragraph{Fourier kernel}
We first consider the multiplicity-free channel. 

\begin{equation}
    p_{11}^1=p_{1X}^X=p_{1Y}^Y=p_{1Z}^Z=p_{X1}^X=p_{Y1}^Y=p_{Z1}^Z=p_{XY}^1=p_{YX}^1=1\,,
\end{equation}
\begin{equation}
    p_{XX}^Y=\frac{\beta_1 \beta_4}{\beta_2}\,,\quad p_{YY}^X=\frac{\beta_1 \beta_4}{\beta_3}\,,\quad p_{XZ}^Z=-i\frac{\beta_5 \beta_8}{\beta_2}\,,\quad p_{ZX}^Z=-i\frac{\beta_6 \beta_{10}}{\beta_9}\,,
\end{equation}
\begin{equation}
    p_{YZ}^Z=-i\frac{\beta_6 \beta_7}{\beta_3}\,,\quad p_{ZY}^Z=-i \frac{\beta_8 \beta_{11}}{\beta_9}\,,\quad
\end{equation}
\begin{equation}
    p_{ZZ}^1=3\,,\quad p_{ZZ}^X=-i\frac{3\beta_5 \beta_{10}}{\beta_9}\,,\quad p_{ZZ}^Y=-i \frac{3\beta_7 \beta_{11}}{\beta_9}\,.
\end{equation}

For the $ZZ\rightarrow Z$ channel, we introduce the orbit of the conjugated commuting pair as
\begin{equation}
\begin{split}
\mathcal O_{e,b}
&=Z_{e,b}+Z_{e,a^2ba}+Z_{e,aba^2},\\
\mathcal O_{b,e}
&=Z_{b,e}+Z_{a^2ba,e}+Z_{aba^2,e},\\
\mathcal O_{b,b}
&=Z_{b,b}+Z_{a^2ba,a^2ba}+Z_{aba^2,aba^2},\\
\mathcal O_{b,a^2ba}
&=Z_{b,a^2ba}+Z_{a^2ba,aba^2}+Z_{aba^2,b},\\
\mathcal O_{b,aba^2}
&=Z_{b,aba^2}+Z_{aba^2,a^2ba}+Z_{a^2ba,b},\\
\mathcal O_{a,a}
&=Z_{a,a}+Z_{ab,ab}+Z_{ba,ba}+Z_{bab,bab},\\
\mathcal O_{a^2,a^2}
&=Z_{a^2,a^2}+Z_{a^2b,a^2b}+Z_{aba,aba}+Z_{ba^2,ba^2}.
\end{split}
\end{equation}
Then the dual partition functions for the $ZZ\rightarrow Z$ channel become
\begin{equation}
\begin{split}
Z_{Z,Z}^{Z;1,1}
&=\frac{1}{12\beta_9}\Bigl[
-3\bigl(\beta_{12}^{2}-2i\beta_{12}\beta_{13}+2\beta_{13}^{2}\bigr) Z_{e,e}\\
&\qquad
+\bigl(\beta_{12}^{2}-2i\beta_{12}\beta_{13}+2\beta_{13}^{2}\bigr)
\bigl(\mathcal O_{e,b}+\mathcal O_{b,e}+\mathcal O_{b,b}\bigr)\\
&\qquad
+\bigl((-1+2i\sqrt3)\beta_{12}^{2}+(2i-4\sqrt3)\beta_{12}\beta_{13}-2\beta_{13}^{2}\bigr)
\mathcal O_{b,a^2ba}\\
&\qquad
+\bigl((-1-2i\sqrt3)\beta_{12}^{2}+2(i+2\sqrt3)\beta_{12}\beta_{13}-2\beta_{13}^{2}\bigr)
\mathcal O_{b,aba^2}\\
&\qquad
-3\bigl(\beta_{12}^{2}+i\beta_{12}\beta_{13}-\beta_{13}^{2}\bigr)
\bigl(\mathcal O_{a,a}+\mathcal O_{a^2,a^2}\bigr)
\Bigr].
\end{split}
\end{equation}
\begin{equation}
\begin{split}
Z_{Z,Z}^{Z;1,2}
&=\frac{1}{24\beta_9}\Bigl[
-6\bigl(\beta_{12}\beta_{14}-i\beta_{13}\beta_{14}-i\beta_{12}\beta_{15}+2\beta_{13}\beta_{15}\bigr) Z_{e,e}\\
&\qquad
+2\bigl(\beta_{12}\beta_{14}-i\beta_{13}\beta_{14}-i\beta_{12}\beta_{15}+2\beta_{13}\beta_{15}\bigr)
\bigl(\mathcal O_{e,b}+\mathcal O_{b,e}+\mathcal O_{b,b}\bigr)\\
&\qquad
+\bigl((-2+4i\sqrt3)\beta_{12}\beta_{14}+(2i-4\sqrt3)\beta_{12}\beta_{15}
+(2i-4\sqrt3)\beta_{13}\beta_{14}-4\beta_{13}\beta_{15}\bigr)
\mathcal O_{b,a^2ba}\\
&\qquad
+\bigl((-2-4i\sqrt3)\beta_{12}\beta_{14}+(2i+4\sqrt3)\beta_{12}\beta_{15}
+(2i+4\sqrt3)\beta_{13}\beta_{14}-4\beta_{13}\beta_{15}\bigr)
\mathcal O_{b,aba^2}\\
&\qquad
+\bigl(-6\beta_{12}\beta_{14}-3(i+\sqrt3)\beta_{13}\beta_{14}
+3(-i+\sqrt3)\beta_{12}\beta_{15}+6\beta_{13}\beta_{15}\bigr)
\mathcal O_{a,a}\\
&\qquad
+\bigl(-6\beta_{12}\beta_{14}+3(-i+\sqrt3)\beta_{13}\beta_{14}
-3(i+\sqrt3)\beta_{12}\beta_{15}+6\beta_{13}\beta_{15}\bigr)
\mathcal O_{a^2,a^2}
\Bigr].
\end{split}
\end{equation}
\begin{equation}
\begin{split}
Z_{Z,Z}^{Z;2,1}
&=\frac{1}{24\beta_9}\Bigl[
-6\bigl(\beta_{12}\beta_{14}-i\beta_{13}\beta_{14}-i\beta_{12}\beta_{15}+2\beta_{13}\beta_{15}\bigr) Z_{e,e}\\
&\qquad
+2\bigl(\beta_{12}\beta_{14}-i\beta_{13}\beta_{14}-i\beta_{12}\beta_{15}+2\beta_{13}\beta_{15}\bigr)
\bigl(\mathcal O_{e,b}+\mathcal O_{b,e}+\mathcal O_{b,b}\bigr)\\
&\qquad
+\bigl((-2+4i\sqrt3)\beta_{12}\beta_{14}+(2i-4\sqrt3)\beta_{12}\beta_{15}
+(2i-4\sqrt3)\beta_{13}\beta_{14}-4\beta_{13}\beta_{15}\bigr)
\mathcal O_{b,a^2ba}\\
&\qquad
+\bigl((-2-4i\sqrt3)\beta_{12}\beta_{14}+(2i+4\sqrt3)\beta_{12}\beta_{15}
+(2i+4\sqrt3)\beta_{13}\beta_{14}-4\beta_{13}\beta_{15}\bigr)
\mathcal O_{b,aba^2}\\
&\qquad
+\bigl(-6\beta_{12}\beta_{14}+3(-i+\sqrt3)\beta_{13}\beta_{14}
-3(i+\sqrt3)\beta_{12}\beta_{15}+6\beta_{13}\beta_{15}\bigr)
\mathcal O_{a,a}\\
&\qquad
+\bigl(-6\beta_{12}\beta_{14}-3(i+\sqrt3)\beta_{13}\beta_{14}
+3(-i+\sqrt3)\beta_{12}\beta_{15}+6\beta_{13}\beta_{15}\bigr)
\mathcal O_{a^2,a^2}
\Bigr].
\end{split}
\end{equation}
\begin{equation}
\begin{split}
Z_{Z,Z}^{Z;2,2}
&=\frac{1}{12\beta_9}\Bigl[
-3\bigl(\beta_{14}^{2}-2i\beta_{14}\beta_{15}+2\beta_{15}^{2}\bigr) Z_{e,e}\\
&\qquad
+\bigl(\beta_{14}^{2}-2i\beta_{14}\beta_{15}+2\beta_{15}^{2}\bigr)
\bigl(\mathcal O_{e,b}+\mathcal O_{b,e}+\mathcal O_{b,b}\bigr)\\
&\qquad
+\bigl((-1+2i\sqrt3)\beta_{14}^{2}+(2i-4\sqrt3)\beta_{14}\beta_{15}-2\beta_{15}^{2}\bigr)
\mathcal O_{b,a^2ba}\\
&\qquad
+\bigl((-1-2i\sqrt3)\beta_{14}^{2}+2(i+2\sqrt3)\beta_{14}\beta_{15}-2\beta_{15}^{2}\bigr)
\mathcal O_{b,aba^2}\\
&\qquad
-3\bigl(\beta_{14}^{2}+i\beta_{14}\beta_{15}-\beta_{15}^{2}\bigr)
\bigl(\mathcal O_{a,a}+\mathcal O_{a^2,a^2}\bigr)
\Bigr].
\end{split}
\end{equation}

As mentioned before, the character projector cannot distinguish the four different channels. Actually, one can show that the character projector produce the linear combination of the four channels as
\begin{equation}
\begin{split}
    &c_{1,1}Z_{ZZ}^{Z;1,1} + c_{1,2}Z_{ZZ}^{Z;1,2}+c_{2,1}Z_{ZZ}^{Z;2,1}+c_{2,2}Z_{ZZ}^{Z;2,2}\\
    =&\frac{d_{Z}}{|A_4|^2} \sum_{\substack{g,h\in A_4\\ gh=hg}}\sum_{k\in A_4} \chi_{Z}(k^{-1}) \chi_{Z}(kh^{-1})\chi_{Z}(kg) Z_{g,h}\,.
\end{split}
\end{equation}
with
    \begin{equation}
        c_{1,1}=-\frac{\beta_9 \left(\beta_{14}^2-2 i \beta_{14} \beta_{15}+2 \beta_{15}^2\right)}{3 \Omega^2}\,,\quad c_{2,2}=-\frac{\beta_{9} \left(\beta_{12}^2-2 i \beta_{12} \beta_{13}+2 \beta_{13}^2\right)}{3 \Omega^2}\,,
    \end{equation}
and
    \begin{equation}
        c_{1,2}=c_{2,1}=\frac{\beta_{9} (\beta_{12} \beta_{14}-i \beta_{12} \beta_{15}-i \beta_{13} \beta_{14}+2 \beta_{13} \beta_{15})}{3 \Omega^2}\,,
    \end{equation}
where $\Omega=\beta_{12}\beta_{15}-\beta_{13}\beta_{14}\neq 0$.

\paragraph{Inverse Fourier kernel}

For the inverse Fourier kennel, we choose the representative for each conjugacy orbits of commuting pair
\begin{equation}
\begin{split}
[Z_{e,e}]&=\{Z_{e,e}\},\\
[Z_{e,a}]&=\{Z_{e,a}, Z_{e,ab}, Z_{e,ba}, Z_{e,bab}\},\\
[Z_{e,a^2}]&=\{Z_{e,a^2}, Z_{e,a^2b}, Z_{e,aba}, Z_{e,ba^2}\},\\
[Z_{e,b}]&=\{Z_{e,b}, Z_{e,a^2ba}, Z_{e,aba^2}\},\\
[Z_{a,e}]&=\{Z_{a,e}, Z_{ab,e}, Z_{ba,e}, Z_{bab,e}\},\\
[Z_{a,a}]&=\{Z_{a,a}, Z_{ab,ab}, Z_{ba,ba}, Z_{bab,bab}\},\\
[Z_{a,a^2}]&=\{Z_{a,a^2}, Z_{ab,ba^2}, Z_{ba,a^2b}, Z_{bab,aba}\},\\
[Z_{a^2,e}]&=\{Z_{a^2,e}, Z_{a^2b,e}, Z_{aba,e}, Z_{ba^2,e}\},\\
[Z_{a^2,a}]&=\{Z_{a^2,a}, Z_{a^2b,ba}, Z_{aba,bab}, Z_{ba^2,ab}\},\\
[Z_{a^2,a^2}]&=\{Z_{a^2,a^2}, Z_{a^2b,a^2b}, Z_{aba,aba}, Z_{ba^2,ba^2}\},\\
[Z_{b,e}]&=\{Z_{b,e}, Z_{a^2ba,e}, Z_{aba^2,e}\},\\
[Z_{b,b}]&=\{Z_{b,b}, Z_{a^2ba,a^2ba}, Z_{aba^2,aba^2}\},\\
[Z_{b,a^2ba}]&=\{Z_{b,a^2ba}, Z_{a^2ba,aba^2}, Z_{aba^2,b}\},\\
[Z_{b,aba^2}]&=\{Z_{b,aba^2}, Z_{a^2ba,b}, Z_{aba^2,a^2ba}\},
\end{split}
\end{equation}

\begin{equation}
\begin{split}
Z_{e,e}&=\frac{1}{12}\Bigl[Z_{1,1}^{1}+Z_{1,X}^{X}+Z_{1,Y}^{Y}+3 Z_{1,Z}^{Z}+Z_{X,1}^{X}+\frac{\beta_{2}}{\beta_{1} \beta_{4}} Z_{X,X}^{Y}+Z_{X,Y}^{1}+\frac{3 i \beta_{2}}{\beta_{5} \beta_{8}} Z_{X,Z}^{Z}\\
&\qquad +Z_{Y,1}^{Y}+Z_{Y,X}^{1}+\frac{\beta_{3}}{\beta_{1} \beta_{4}} Z_{Y,Y}^{X}+\frac{3 i \beta_{3}}{\beta_{6} \beta_{7}} Z_{Y,Z}^{Z}+3 Z_{Z,1}^{Z}+\frac{3 i \beta_{9}}{\beta_{10} \beta_{6}} Z_{Z,X}^{Z}+\frac{3 i \beta_{9}}{\beta_{11} \beta_{8}} Z_{Z,Y}^{Z}+3 Z_{Z,Z}^{1}\\
&\qquad +\frac{3 i \beta_{9}}{\beta_{10} \beta_{5}} Z_{Z,Z}^{X}+\frac{3 i \beta_{9}}{\beta_{11} \beta_{7}} Z_{Z,Z}^{Y}-\frac{3 ({\beta_{14}}^{2}-2 i \beta_{14} \beta_{15}+2 {\beta_{15}}^{2}) \beta_{9}}{\Omega^{2}} Z_{Z,Z}^{Z;1,1}\\
&\qquad +\frac{3 (\beta_{12} \beta_{14}-i \beta_{13} \beta_{14}-i \beta_{12} \beta_{15}+2 \beta_{13} \beta_{15}) \beta_{9}}{\Omega^{2}} (Z_{Z,Z}^{Z;1,2}+Z_{Z,Z}^{Z;2,1})\\
&\qquad -\frac{3 ({\beta_{12}}^{2}-2 i \beta_{12} \beta_{13}+2 {\beta_{13}}^{2}) \beta_{9}}{\Omega^{2}} Z_{Z,Z}^{Z;2,2}
\Bigr].
\end{split}
\end{equation}

\begin{equation}
\begin{split}
Z_{e,a}&=\frac{1}{12}\Bigl[Z_{1,1}^{1}+Z_{1,X}^{X}+Z_{1,Y}^{Y}+3 Z_{1,Z}^{Z}\\
&\qquad +\frac{1}{2} (-1+i\sqrt{3}) Z_{X,1}^{X}+\frac{(-1+i\sqrt{3}) \beta_{2}}{2 \beta_{1} \beta_{4}} Z_{X,X}^{Y}+{(-1)}^{2/3} Z_{X,Y}^{1}+\frac{6 \beta_{2}}{i \beta_{5} \beta_{8}-\sqrt{3} \beta_{5} \beta_{8}} Z_{X,Z}^{Z}\\
&\qquad -\frac{1}{2} (1+i\sqrt{3}) Z_{Y,1}^{Y}-{(-1)}^{1/3} Z_{Y,X}^{1}-\frac{{(-1)}^{1/3} \beta_{3}}{\beta_{1} \beta_{4}} Z_{Y,Y}^{X}+\frac{6 \beta_{3}}{i \beta_{6} \beta_{7}+\sqrt{3} \beta_{6} \beta_{7}} Z_{Y,Z}^{Z}
\Bigr].
\end{split}
\end{equation}

\begin{equation}
\begin{split}
Z_{e,a^2}&=\frac{1}{12}\Bigl[Z_{1,1}^{1}+Z_{1,X}^{X}+Z_{1,Y}^{Y}+3 Z_{1,Z}^{Z}\\
&\qquad -{(-1)}^{1/3} Z_{X,1}^{X}-\frac{(\beta_{2}+i \sqrt{3} \beta_{2})}{2 \beta_{1} \beta_{4}} Z_{X,X}^{Y}-{(-1)}^{1/3} Z_{X,Y}^{1}+\frac{6 \beta_{2}}{i \beta_{5} \beta_{8}+\sqrt{3} \beta_{5} \beta_{8}} Z_{X,Z}^{Z}\\
&\qquad +\frac{1}{2} (-1+i\sqrt{3}) Z_{Y,1}^{Y}+{(-1)}^{2/3} Z_{Y,X}^{1}+\frac{(-1+i\sqrt{3}) \beta_{3}}{2 \beta_{1} \beta_{4}} Z_{Y,Y}^{X}-\frac{3 (i+\sqrt{3}) \beta_{3}}{2 \beta_{6} \beta_{7}} Z_{Y,Z}^{Z}
\Bigr].
\end{split}
\end{equation}

\begin{equation}
\begin{split}
Z_{e,b}&=\frac{1}{12}\Bigl[Z_{1,1}^{1}+Z_{1,X}^{X}+Z_{1,Y}^{Y}+3 Z_{1,Z}^{Z} +Z_{X,1}^{X}+\frac{\beta_{2}}{\beta_{1} \beta_{4}} Z_{X,X}^{Y}+Z_{X,Y}^{1}+\frac{3 i \beta_{2}}{\beta_{5} \beta_{8}} Z_{X,Z}^{Z}\\
&\qquad +Z_{Y,1}^{Y}+Z_{Y,X}^{1}+\frac{\beta_{3}}{\beta_{1} \beta_{4}} Z_{Y,Y}^{X}+\frac{3 i \beta_{3}}{\beta_{6} \beta_{7}} Z_{Y,Z}^{Z} -Z_{Z,1}^{Z}-\frac{i \beta_{9}}{\beta_{10} \beta_{6}} Z_{Z,X}^{Z}-\frac{i \beta_{9}}{\beta_{11} \beta_{8}} Z_{Z,Y}^{Z}-Z_{Z,Z}^{1}\\
&\qquad -\frac{i \beta_{9}}{\beta_{10} \beta_{5}} Z_{Z,Z}^{X}-\frac{i \beta_{9}}{\beta_{11} \beta_{7}} Z_{Z,Z}^{Y}+\frac{({\beta_{14}}^{2}-2 i \beta_{14} \beta_{15}+2 {\beta_{15}}^{2}) \beta_{9}}{\Omega^{2}} Z_{Z,Z}^{Z;1,1}\\
&\qquad +\frac{i (\beta_{13} (\beta_{14}+2 i \beta_{15})+\beta_{12} (i \beta_{14}+\beta_{15})) \beta_{9}}{\Omega^{2}} (Z_{Z,Z}^{Z;1,2}+Z_{Z,Z}^{Z;2,1})+\\
&\qquad \frac{({\beta_{12}}^{2}-2 i \beta_{12} \beta_{13}+2 {\beta_{13}}^{2}) \beta_{9}}{\Omega^{2}} Z_{Z,Z}^{Z;2,2}
\Bigr].
\end{split}
\end{equation}

\begin{equation}
\begin{split}
Z_{a,e}&=\frac{1}{12}\Bigl[ Z_{1,1}^{1}-{(-1)}^{1/3} Z_{1,X}^{X}+\frac{1}{2} (-1+i\sqrt{3}) Z_{1,Y}^{Y}+Z_{X,1}^{X}\\
&\qquad -\frac{{(-1)}^{1/3} \beta_{2}}{\beta_{1} \beta_{4}} Z_{X,X}^{Y}+\frac{1}{2} (-1+i\sqrt{3}) Z_{X,Y}^{1}+Z_{Y,1}^{Y}-\frac{1}{2} (1+i\sqrt{3}) Z_{Y,X}^{1}\\
&\qquad +\frac{(-1+i\sqrt{3}) \beta_{3}}{2 \beta_{1} \beta_{4}} Z_{Y,Y}^{X}+3 Z_{Z,1}^{Z}+\frac{6 i \sqrt{3} \beta_{9}}{3 i \beta_{10} \beta_{6}-\sqrt{3} \beta_{10} \beta_{6}} Z_{Z,X}^{Z}+\frac{6 \beta_{9}}{i \beta_{11} \beta_{8}-\sqrt{3} \beta_{11} \beta_{8}} Z_{Z,Y}^{Z}
\Bigr].
\end{split}
\end{equation}

\begin{equation}
\begin{split}
Z_{a,a}&=\frac{1}{12}\Bigl[  Z_{1,1}^{1}- {(-1)}^{1/3} Z_{1,X}^{X}+\frac{1}{2}(-1+i\sqrt{3}) Z_{1,Y}^{Y}+\frac{1}{2} (-1+i\sqrt{3}) Z_{X,1}^{X}\\
&\qquad +\frac{\beta_{2}}{\beta_{1} \beta_{4}} Z_{X,X}^{Y}-\frac{1}{2}(1+ i \sqrt{3}) Z_{X,Y}^{1}- \frac{1}{2} (1+i\sqrt{3}) Z_{Y,1}^{Y}+\frac{1}{2} (-1+i\sqrt{3}) Z_{Y,X}^{1}\\
&\qquad +\frac{\beta_{3}}{\beta_{1} \beta_{4}} Z_{Y,Y}^{X}+\frac{(2 \sqrt{3} \beta_{14}+i (3 i+\sqrt{3}) \beta_{15}) ((3+i \sqrt{3}) \beta_{14}+(3 i+\sqrt{3}) \beta_{15}) \beta_{9}}{2(i+\sqrt{3}) \Omega^{2}} Z_{Z,Z}^{Z;1,1}\\
&\qquad+\frac{3 ((-3 i+\sqrt{3}) \beta_{12}+(3-i \sqrt{3}) \beta_{13}) (2 \sqrt{3} \beta_{14}+i (3 i+\sqrt{3}) \beta_{15}) \beta_{9}}{{(3 i+\sqrt{3})}^{2} \Omega^{2}} Z_{Z,Z}^{Z;1,2}\\
&\qquad+\frac{3 (2 \sqrt{3} \beta_{12}+i (3 i+\sqrt{3}) \beta_{13}) ((-3 i+\sqrt{3}) \beta_{14}+(3-i \sqrt{3}) \beta_{15}) \beta_{9}}{{(3 i+\sqrt{3})}^{2} \Omega^{2}} Z_{Z,Z}^{Z;2,1}\\
&\qquad +\frac{(2 \sqrt{3} \beta_{12}+i (3 i+\sqrt{3}) \beta_{13}) ((3+i \sqrt{3}) \beta_{12}+(3 i+\sqrt{3}) \beta_{13}) \beta_{9}}{2(i+\sqrt{3}) \Omega^{2}} Z_{Z,Z}^{Z;2,2}
\Bigr].
\end{split}
\end{equation}

\begin{equation}
\begin{split}
Z_{a,a^2}&=\frac{1}{12}\Bigl[ Z_{1,1}^{1}-{(-1)}^{1/3} Z_{1,X}^{X}+\frac{1}{2} (-1+i\sqrt{3}) Z_{1,Y}^{Y}-{(-1)}^{1/3} Z_{X,1}^{X}\\
&\qquad +\frac{(-1+i\sqrt{3}) \beta_{2}}{2 \beta_{1} \beta_{4}} Z_{X,X}^{Y}+Z_{X,Y}^{1}+\frac{1}{2} (-1+i\sqrt{3}) Z_{Y,1}^{Y}+Z_{Y,X}^{1}\\
&\qquad -\frac{(\beta_{3}+i \sqrt{3} \beta_{3})}{2 \beta_{1} \beta_{4}} Z_{Y,Y}^{X}+3 Z_{Z,Z}^{1}+\frac{6 \beta_{9}}{i \beta_{10} \beta_{5}+\sqrt{3} \beta_{10} \beta_{5}} Z_{Z,Z}^{X}-\frac{3 (1+i \sqrt{3}) \beta_{9}}{(i+\sqrt{3}) \beta_{11} \beta_{7}} Z_{Z,Z}^{Y}
\Bigr].
\end{split}
\end{equation}

\begin{equation}
\begin{split}
Z_{a^2,e}&=\frac{1}{12}\Bigl[ Z_{1,1}^{1}+\frac{1}{2}(-1+i\sqrt{3}) Z_{1,X}^{X}-\frac{1}{2}(1+i\sqrt{3}) Z_{1,Y}^{Y}+Z_{X,1}^{X}\\
&\qquad +\frac{(-1+i\sqrt{3}) \beta_{2}}{2\beta_{1} \beta_{4}} Z_{X,X}^{Y}-\frac{1}{2}(1+i\sqrt{3}) Z_{X,Y}^{1}+ Z_{Y,1}^{Y}+\frac{1}{2}(-1+i\sqrt{3}) Z_{Y,X}^{1}\\
&\qquad -\frac{(\beta_{3}+i \sqrt{3} \beta_{3})}{2\beta_{1} \beta_{4}} Z_{Y,Y}^{X}+3 Z_{Z,1}^{Z}+\frac{3 i (3 i+\sqrt{3}) \beta_{9}}{(-3 i+\sqrt{3}) \beta_{10} \beta_{6}} Z_{Z,X}^{Z}+\frac{6 i \sqrt{3} \beta_{9}}{(3 i-\sqrt{3})\beta_{11} \beta_{8}} Z_{Z,Y}^{Z}
\Bigr].
\end{split}
\end{equation}

\begin{equation}
\begin{split}
Z_{a^2,a}&=\frac{1}{12}\Bigl[ Z_{1,1}^{1}+\frac{1}{2}(-1+i\sqrt{3}) Z_{1,X}^{X}-\frac{1}{2}(1+i\sqrt{3}) Z_{1,Y}^{Y}+\frac{1}{2}(-1+i\sqrt{3}) Z_{X,1}^{X}\\
&\qquad -\frac{(\beta_{2}+i \sqrt{3} \beta_{2})}{2\beta_{1} \beta_{4}} Z_{X,X}^{Y}+ Z_{X,Y}^{1}-\frac{1}{2}(1+i\sqrt{3}) Z_{Y,1}^{Y}+ Z_{Y,X}^{1}\\
&\qquad +\frac{(-1+i\sqrt{3}) \beta_{3}}{2\beta_{1} \beta_{4}} Z_{Y,Y}^{X}+3 Z_{Z,Z}^{1}+\frac{6 (-i+\sqrt{3}) \beta_{9}}{{(\sqrt{3} i+1)}^{2} \beta_{10} \beta_{5}} Z_{Z,Z}^{X}+\frac{3 (i+\sqrt{3}) \beta_{9}}{(\sqrt{3}i+1) \beta_{11} \beta_{7}} Z_{Z,Z}^{Y}
\Bigr].
\end{split}
\end{equation}

\begin{equation}
\begin{split}
Z_{a^2,a^2}&=\frac{1}{12}\Bigl[  Z_{1,1}^{1}+\frac{1}{2} (-1+i\sqrt{3}) Z_{1,X}^{X}-\frac{1}{2} (1+i\sqrt{3}) Z_{1,Y}^{Y}- {(-1)}^{1/3} Z_{X,1}^{X}\\
&\qquad +\frac{ \beta_{2}}{\beta_{1} \beta_{4}} Z_{X,X}^{Y}+\frac{1}{2} (-1+i\sqrt{3}) Z_{X,Y}^{1}+\frac{1}{2} (-1+i\sqrt{3}) Z_{Y,1}^{Y}-\frac{1}{2}(1+ i \sqrt{3}) Z_{Y,X}^{1}\\
&\qquad +\frac{\beta_{3}}{\beta_{1} \beta_{4}} Z_{Y,Y}^{X}+\frac{(2 \sqrt{3} \beta_{14}+i (3 i+\sqrt{3}) \beta_{15}) ((3+i \sqrt{3}) \beta_{14}+(3 i+\sqrt{3}) \beta_{15}) \beta_{9}}{2(i+\sqrt{3}) \Omega^{2}} Z_{Z,Z}^{Z;1,1}\\
&\qquad+\frac{3 (2 \sqrt{3} \beta_{12}+i (3 i+\sqrt{3}) \beta_{13}) ((-3 i+\sqrt{3}) \beta_{14}+(3-i \sqrt{3}) \beta_{15}) \beta_{9}}{{(3 i+\sqrt{3})}^{2} \Omega^{2}} Z_{Z,Z}^{Z;1,2}\\
&\qquad+\frac{3 ((-3 i+\sqrt{3}) \beta_{12}+(3-i \sqrt{3}) \beta_{13}) (2 \sqrt{3} \beta_{14}+i (3 i+\sqrt{3}) \beta_{15}) \beta_{9}}{{(3 i+\sqrt{3})}^{2} \Omega^{2}} Z_{Z,Z}^{Z;2,1}\\
&\qquad +\frac{(2 \sqrt{3} \beta_{12}+i (3 i+\sqrt{3}) \beta_{13}) ((3+i \sqrt{3}) \beta_{12}+(3 i+\sqrt{3}) \beta_{13}) \beta_{9}}{2(i+\sqrt{3}) \Omega^{2}} Z_{Z,Z}^{Z;2,2}
\Bigr].
\end{split}
\end{equation}

\begin{equation}
\begin{split}
Z_{b,e}&=\frac{1}{12}\Bigl[Z_{1,1}^{1}+Z_{1,X}^{X}+Z_{1,Y}^{Y}-Z_{1,Z}^{Z} +Z_{X,1}^{X}+\frac{\beta_{2}}{\beta_{1} \beta_{4}} Z_{X,X}^{Y}+Z_{X,Y}^{1}-\frac{i \beta_{2}}{\beta_{5} \beta_{8}} Z_{X,Z}^{Z}\\
&\qquad +Z_{Y,1}^{Y}+Z_{Y,X}^{1}+\frac{\beta_{3}}{\beta_{1} \beta_{4}} Z_{Y,Y}^{X}-\frac{i \beta_{3}}{\beta_{6} \beta_{7}} Z_{Y,Z}^{Z} +3 Z_{Z,1}^{Z}+\frac{3 i \beta_{9}}{\beta_{10} \beta_{6}} Z_{Z,X}^{Z}+\frac{3 i \beta_{9}}{\beta_{11} \beta_{8}} Z_{Z,Y}^{Z}-Z_{Z,Z}^{1}\\
&\qquad -\frac{i \beta_{9}}{\beta_{10} \beta_{5}} Z_{Z,Z}^{X}-\frac{i \beta_{9}}{\beta_{11} \beta_{7}} Z_{Z,Z}^{Y}+\frac{({\beta_{14}}^{2}-2 i \beta_{14} \beta_{15}+2 {\beta_{15}}^{2}) \beta_{9}}{\Omega^{2}} Z_{Z,Z}^{Z;1,1}\\
&\qquad +\frac{i (\beta_{13} (\beta_{14}+2 i \beta_{15})+\beta_{12} (i \beta_{14}+\beta_{15})) \beta_{9}}{\Omega^{2}} (Z_{Z,Z}^{Z;1,2}+Z_{Z,Z}^{Z;2,1})\\
&\qquad+\frac{({\beta_{12}}^{2}-2 i \beta_{12} \beta_{13}+2 {\beta_{13}}^{2}) \beta_{9}}{\Omega^{2}} Z_{Z,Z}^{Z;2,2}
\Bigr].
\end{split}
\end{equation}

\begin{equation}
\begin{split}
Z_{b,b}&=\frac{1}{12}\Bigl[ Z_{1,1}^{1}+Z_{1,X}^{X}+Z_{1,Y}^{Y}-Z_{1,Z}^{Z}+Z_{X,1}^{X}+\frac{\beta_{2}}{\beta_{1} \beta_{4}} Z_{X,X}^{Y}+Z_{X,Y}^{1}-\frac{i \beta_{2}}{\beta_{5} \beta_{8}} Z_{X,Z}^{Z}\\
&\qquad +Z_{Y,1}^{Y}+Z_{Y,X}^{1}+\frac{\beta_{3}}{\beta_{1} \beta_{4}} Z_{Y,Y}^{X}-\frac{i \beta_{3}}{\beta_{6} \beta_{7}} Z_{Y,Z}^{Z} -Z_{Z,1}^{Z}-\frac{i \beta_{9}}{\beta_{10} \beta_{6}} Z_{Z,X}^{Z}-\frac{i \beta_{9}}{\beta_{11} \beta_{8}} Z_{Z,Y}^{Z}+3 Z_{Z,Z}^{1}\\
&\qquad +\frac{3 i \beta_{9}}{\beta_{10} \beta_{5}} Z_{Z,Z}^{X}+\frac{3 i \beta_{9}}{\beta_{11} \beta_{7}} Z_{Z,Z}^{Y}+\frac{({\beta_{14}}^{2}-2 i \beta_{14} \beta_{15}+2 {\beta_{15}}^{2}) \beta_{9}}{\Omega^{2}} Z_{Z,Z}^{Z;1,1}\\
&\qquad +\frac{i (\beta_{13} (\beta_{14}+2 i \beta_{15})+\beta_{12} (i \beta_{14}+\beta_{15})) \beta_{9}}{\Omega^{2}} (Z_{Z,Z}^{Z;1,2}+Z_{Z,Z}^{Z;2,1})\\
&\qquad+\frac{({\beta_{12}}^{2}-2 i \beta_{12} \beta_{13}+2 {\beta_{13}}^{2}) \beta_{9}}{\Omega^{2}} Z_{Z,Z}^{Z;2,2}
\Bigr].
\end{split}
\end{equation}

\begin{equation}
\begin{split}
Z_{b,a^2ba}&=\frac{1}{12}\Bigl[Z_{1,1}^{1}+Z_{1,X}^{X}+Z_{1,Y}^{Y}-Z_{1,Z}^{Z}+Z_{X,1}^{X}+\frac{\beta_{2}}{\beta_{1} \beta_{4}} Z_{X,X}^{Y}+Z_{X,Y}^{1}-\frac{i \beta_{2}}{\beta_{5} \beta_{8}} Z_{X,Z}^{Z}\\
&\qquad +Z_{Y,1}^{Y}+Z_{Y,X}^{1}+\frac{\beta_{3}}{\beta_{1} \beta_{4}} Z_{Y,Y}^{X}-\frac{i \beta_{3}}{\beta_{6} \beta_{7}} Z_{Y,Z}^{Z} -Z_{Z,1}^{Z}-\frac{i \beta_{9}}{\beta_{10} \beta_{6}} Z_{Z,X}^{Z}-\frac{i \beta_{9}}{\beta_{11} \beta_{8}} Z_{Z,Y}^{Z}-Z_{Z,Z}^{1}\\
&\qquad -\frac{i \beta_{9}}{\beta_{10} \beta_{5}} Z_{Z,Z}^{X}-\frac{i \beta_{9}}{\beta_{11} \beta_{7}} Z_{Z,Z}^{Y}+\frac{6 ((7+i \sqrt{3}) {\beta_{14}}^{2}+2 (5 i-3 \sqrt{3}) \beta_{14} \beta_{15}+2 (1-i \sqrt{3}) {\beta_{15}}^{2}) \beta_{9}}{{(3 i+\sqrt{3})}^{2} \Omega^{2}} Z_{Z,Z}^{Z;1,1}\\
&\qquad+\frac{((5 i+3 \sqrt{3}) \beta_{12} \beta_{14}+i (7 i+\sqrt{3}) \beta_{13} \beta_{14}+i (7 i+\sqrt{3}) \beta_{12} \beta_{15}+2 (-i+\sqrt{3}) \beta_{13} \beta_{15}) \beta_{9}}{(-i+\sqrt{3}) \Omega^{2}} \\
&\qquad \times (Z_{Z,Z}^{Z;1,2}+Z_{Z,Z}^{Z;2,1}) +\frac{6 ((7+i \sqrt{3}) {\beta_{12}}^{2}+2 (5 i-3 \sqrt{3}) \beta_{12} \beta_{13}+2 (1-i \sqrt{3}) {\beta_{13}}^{2}) \beta_{9}}{{(3 i+\sqrt{3})}^{2} \Omega^{2}} Z_{Z,Z}^{Z;2,2}
\Bigr].
\end{split}
\end{equation}

\begin{equation}
\begin{split}
Z_{b,aba^2}&=\frac{1}{12}\Bigl[Z_{1,1}^{1}+Z_{1,X}^{X}+Z_{1,Y}^{Y}-Z_{1,Z}^{Z} +Z_{X,1}^{X}+\frac{\beta_{2}}{\beta_{1} \beta_{4}} Z_{X,X}^{Y}+Z_{X,Y}^{1}-\frac{i \beta_{2}}{\beta_{5} \beta_{8}} Z_{X,Z}^{Z}\\
&\qquad +Z_{Y,1}^{Y}+Z_{Y,X}^{1}+\frac{\beta_{3}}{\beta_{1} \beta_{4}} Z_{Y,Y}^{X}-\frac{i \beta_{3}}{\beta_{6} \beta_{7}} Z_{Y,Z}^{Z} -Z_{Z,1}^{Z}-\frac{i \beta_{9}}{\beta_{10} \beta_{6}} Z_{Z,X}^{Z}-\frac{i \beta_{9}}{\beta_{11} \beta_{8}} Z_{Z,Y}^{Z}-Z_{Z,Z}^{1}\\
&\qquad -\frac{i \beta_{9}}{\beta_{10} \beta_{5}} Z_{Z,Z}^{X}-\frac{i \beta_{9}}{\beta_{11} \beta_{7}} Z_{Z,Z}^{Y}\\
&\qquad+\frac{6 ((-5-3 i \sqrt{3}) {\beta_{14}}^{2}+2 (-7 i+\sqrt{3}) \beta_{14} \beta_{15}+2 (1-i \sqrt{3}) {\beta_{15}}^{2}) \beta_{9}}{{(3 i+\sqrt{3})}^{2} \Omega^{2}} Z_{Z,Z}^{Z;1,1}\\
&\qquad-\frac{((7 i+\sqrt{3}) \beta_{12} \beta_{14}+(-5+3 i \sqrt{3}) \beta_{13} \beta_{14}+(-5+3 i \sqrt{3}) \beta_{12} \beta_{15}-2 (-i+\sqrt{3}) \beta_{13} \beta_{15}) \beta_{9}}{(-i+\sqrt{3}) \Omega^{2}}\\
&\qquad \times (Z_{Z,Z}^{Z;1,2}+Z_{Z,Z}^{Z;2,1})+\frac{6 ((-5-3 i \sqrt{3}) {\beta_{12}}^{2}+2 (-7 i+\sqrt{3}) \beta_{12} \beta_{13}+2 (1-i \sqrt{3}) {\beta_{13}}^{2}) \beta_{9}}{{(3 i+\sqrt{3})}^{2} \Omega^{2}} Z_{Z,Z}^{Z;2,2}
\Bigr].
\end{split}
\end{equation}

\section{Examples : Infinite $\text{Rep}(G)$ category}

In this section, we will turn to the representation category $\text{Rep}(G)$ where $G$ is a simply connected Lie group. Our main example will be the \text{Rep}(SU(2)), which is an infinite dimensional category whose simple objects (irreducible representations) are labeled by the non-negative half integers $j\in \frac{1}{2}\mathbb{Z}_{\geq 0}$. For $SU(2)$, there is a natural choice of intertwiners and cointertwiners, whose coefficients under the ordered bases are given by the Clebsch-Gordan coefficients and their complex conjugation. The Frobenius algebra is also infinite dimensional, and is understood as the group function on $SU(2)$. We will give the Fourier kernel and inverse Fourier kernel, and comment on gauging the infinite non-invertible symmetry $\text{Rep}(G)$

We emphasize that the construction in this section is a direct analogue of the finite-group gauging construction. In particular, the $G$-gauging should be understood as a \emph{flat} gauging by summing over only flat $G$-connection, which is not the same to the ordinary gauging procedure by summing over all gauge configurations. The continuous symmetry as a symmetric category has been studied recently from both categorical side~\cite{Freed:2009qp,Jia:2025vrj,Stockall:2025ngz} and operator algebraic side~\cite{Jia:2026vcr}, both intended to generalizing the fusion category to incorporate infinitely many simple objects. The Lagrangian algebra for $SU(2)$ has also been studied in~\cite{Jia:2026tsl}. Nevertheless, the treatment of irregular element $g\in G$, whose centralizer $C_G(g)$ is larger than Cartan torus $T$, still remains illusive and requires more analytic input. Accordingly, the distributional manipulations below are understood formally
on the regular stratum and are used only to determine the form of the resulting
integral up to an overall normalization. We do not attempt a complete analytic
treatment of the irregular strata.

\paragraph{Group data}

It is convenient to use the Euler-angle parametrization, where each $g\in SU(2)$ can be written as
    \begin{equation}
        g= g_{\phi} a_{\theta} g_{\psi}\,,
    \end{equation}
with
    \begin{equation}
        g_{\phi} = \left(\begin{array}{cc}
            e^{\frac{i\phi}{2}} & 0 \\
            0 & e^{-\frac{i\phi}{2}} 
        \end{array} \right) \,,\quad a_{\theta}=\left(\begin{array}{cc}
            \cos \frac{\theta}{2} & -\sin \frac{\theta}{2} \\
             \sin \frac{\theta}{2} & \cos \frac{\theta}{2}
        \end{array} \right)\,,
    \end{equation}
and
    \begin{equation}
        0\leq \theta \leq \pi\,,\quad 0 \leq \phi \leq 2\pi\,, \quad -2\pi \leq \psi \leq 2\pi\,.
    \end{equation}
The metric $g_{ij}$ defined by $(4\pi)^{\frac{4}{3}}g_{ij} = -2\textrm{tr} (\partial_i g^{-1} \partial_j g)$ is
    \begin{equation}
        (4\pi)^{\frac{4}{3}}ds^2 = d\theta^2 + \sin^2 \theta d\phi^2 + (d\psi + \cos \theta d \phi)^2\,,
    \end{equation}
so that the Haar measure and group volume are normalized as
    \begin{equation}
        d\mu(\theta,\phi,\psi) = \frac{1}{16\pi^2}\sin(\theta) d\theta d\phi d\psi\,,\quad |SU(2)|\equiv \int_{SU(2)} d\mu(\theta,\phi,\psi)=1\,.
    \end{equation}

The irreducible representations are labeled by the half integer $j=0,\frac{1}{2},1,\frac{3}{2},\cdots$ with dimension $d_j=2j+1$, and the fusion rule is 
    \begin{equation}
        j_1 \otimes j_2= \bigoplus_{\substack{J=|j_1-j_2|\\\text{with step }1}}^{|j_1+j_2|} J\,,
    \end{equation}
where all fusion coefficients are $0$ or $1$. We label the basis for $j$ as
    \begin{equation}
        |jm\rangle\,, \quad m=-j,-j+1,\cdots,j\,,
    \end{equation}
and the tensor product basis as
    \begin{equation}
        |j_1m_1,j_2m_2\rangle \equiv |j_1m_1\rangle \otimes |j_2 m_2\rangle\,.
    \end{equation}

\paragraph{Intertwiners and cointertwiners}

Consider $J$ an irreducible representation decomposed from $j_1\otimes j_2$ with $|j_1-j_2|\leq J \leq |j_1+j_2|$, the Clebsch-Gordan coefficients of $SU(2)$ are defined by
    \begin{equation}
    \begin{split}
        |JM\rangle =& \sum_{m_1=-j_1}^{j_1}\sum_{m_2=-j_2}^{j_2} |j_1m_1,j_2m_2\rangle \langle j_1m_1,j_2m_2|JM\rangle\\
        \equiv& \sum_{m_1=-j_1}^{j_1}\sum_{m_2=-j_2}^{j_2}C_{j_1m_1,j_2m_2}^{JM} |j_1m_1,j_2m_2\rangle \,,
    \end{split}
    \end{equation}
with $C_{j_1m_1,j_2m_2}^{JM}=\langle j_1m_1,j_2m_2|JM\rangle$. Moreover, $C_{j_1m_1,j_2m_2}^{JM}$ can only be non-zero when
    \begin{equation}
        |j_1-j_2|\leq J\leq |j_1+j_2|\,,\quad M=m_1+m_2\,.
    \end{equation}
The inverse is
    \begin{equation}
    \begin{split}
        |j_1m_1,j_2m_2\rangle =& \sum_{J=|j_1-j_2|}^{|j_1+j_2|}\sum_{M=-J}^{J} |JM\rangle \langle JM|j_1m_1,j_2m_2\rangle \\
        \equiv&\sum_{J=|j_1-j_2|}^{|j_1+j_2|}\sum_{M=-J}^{J} \overline{C}_{j_1m_1,j_2m_2}^{JM}|JM\rangle\,,
    \end{split}
    \end{equation}
with $\overline{C}^{JM}_{j_1m_1,j_2m_2}=\langle JM|j_1m_1,j_2m_2\rangle$. They satisfy the orthogonal relations
    \begin{equation}\label{eq:CG_orthogonal_1}
        \sum_{J=|j_1-j_2|}^{|j_1+j_2|}\sum_{M=-J}^J C_{j_1m_1,j_2m_2}^{JM} \overline{C}_{j_1m'_1,j_2m'_2}^{JM} = \delta_{m_1,m'_1} \delta_{m_2,m'_2}\,,
    \end{equation}
and
    \begin{equation}\label{eq:CG_orthogonal_2}
        \sum_{m_1=-j_1}^{j_1} \sum_{m_2=-j_2}^{j_2} \overline{C}_{j_1m_1,j_2m_2}^{JM} C^{J'M'}_{j_1m_1,j_2m_2} = \delta_{J,J'} \delta_{M,M'}\,.
    \end{equation}
They also satisfy the exchange relations
    \begin{equation}\label{eq:CG_exchange_relation}
        C_{j_2m_2,j_1m_1}^{JM}= (-1)^{j_1+j_2-J}C_{j_1m_1,j_2m_2}^{JM}\,.
    \end{equation}

We define the intertwiner $I_{j_1,j_2}^{J}:E^{j_1}\otimes E^{j_2}\rightarrow E^J$ as
    \begin{equation}
        I_{j_1,j_2}^{J} = \sum_{M,m_1,m_2}|JM\rangle\langle JM|j_1m_1,j_2m_2\rangle \langle j_1 m_1,j_2m_2|\,.
    \end{equation}
Acting on the basis $|j_1m_1,j_2m_2\rangle$ gives
    \begin{equation}
        I_{j_1,j_2}^{J}|j_1m_1,j_2m_2\rangle = \sum_M\overline{C}_{j_1m_1,j_2m_2}^{JM} |JM\rangle\,.
    \end{equation}
Similarly, we define the cointertwiner $I_{j_1,j_2}^{\vee J}: E^{J}\rightarrow E^{j_1}\otimes E^{j_2}$ as
    \begin{equation}
        I_{j_1,j_2}^{\vee J} = \sum_{m_1,m_2,M}|j_1m_1,j_2m_2\rangle \langle j_1 m_1,j_2m_2|JM\rangle\langle JM|\,.
    \end{equation}
Acting on the basis $|JM\rangle$ gives
    \begin{equation}
        I^{\vee j_1,j_2}_{J}|JM\rangle = \sum_{m_1,m_2} C_{j_1m_1,j_2m_2}^{JM} |j_1m_1,j_2m_2\rangle\,.
    \end{equation}
Therefore we have
    \begin{equation}
        \mathcal{I}_{(j_1,m_1),(j_2,m_2)}^{(j,m)} = \overline{C}_{j_1m_1,j_2m_2}^{JM}\,,\quad \mathcal{I}^{\vee (j_1,m_1),(j_2,m_2)}_{JM} = C_{(j_1,m_1),(j_2,m_2)}^{JM}\,.
    \end{equation}

\paragraph{Frobenius Algebra}

The Frobenius algebra is considered as the group function 
    \begin{equation}
        f= \int d \mu(g) f(g) v_g\,,
    \end{equation}
where the basis satisfy $v_g(h) = \delta_{g,h}$, where the delta function on $G$ satisfies
    \begin{equation}
        \int d\mu(g) \delta(g,h) f(g) = f(h)\,.
    \end{equation}
The group action is given by the left action
    \begin{equation}
        L_g\circ f = \int d \mu(h) f(h) v_{gh}= \int d\mu(h) f(g^{-1}h) v_h\,,
    \end{equation}
therefore the function $f(g)$ is understood as an infinite dimensional representation, transformed under the regular representation of $G$.

The Peter-Weyl theorem says that the matrix coefficients of irreducible unitary representations of $G$ form an orthonormal basis of $L^2(G)$. For $SU(2)$, denote the matrix coefficient for the $j$-th irreducible representation as $U^j_{\alpha \beta}(g)$, and one has
    \begin{equation}
        U^j_{\alpha \beta}(g^{-1}h) = U^j_{\alpha \gamma} (g^{-1}) U^j_{\gamma \beta}(h)\,,
    \end{equation}
so that for given $\beta=1,\cdots,2j+1$, $U^j_{\gamma \beta}(h)$ transform as a vector under the $j$-th irreducible representation. Thus we can write the Frobenius algebra formally as
    \begin{equation}
        \mathcal{A} = \bigoplus_{j=0,\frac{1}{2},1,\cdots} d_j\, j\,,
    \end{equation}
with $d_j=2j+1$ the dimension of the irreducible representation $j$. There are $d_j$ copy of $j$, labeled by the second index of the representation matrix.

For the chosen spherical basis $|jm\rangle$, the representation matrix of $g(\phi,\theta,\psi)\in SU(2)$ is given by Wigner D-matrix
    \begin{equation}
        D^j_{mm'}(\phi,\theta,\psi)\equiv \langle jm| \hat{U}(\phi,\theta,\psi)|jm'\rangle\,,\quad \hat{U}(\phi,\theta,\psi) = e^{-i \phi J_z} e^{-i \theta J_y}e^{-i\psi J_z}\,,
    \end{equation}
which is unitary. We define the intertwiner $J_{j,r}:E^j \rightarrow \mathcal{A} $ and $J^{\vee}_{j,r}:\mathcal{A} \rightarrow E^j$ by
    \begin{equation}
        J_{j,r}|jm\rangle = \sqrt{d_j} \int d\mu(g) \overline{D}^j_{m r}(g) v_g\,,\quad J^{\vee}_{j,r}\circ v_g= \sqrt{d_j} \sum_{m=-j}^{j} D^j_{mr}(g) |jm\rangle\,,
    \end{equation}
where $r=-j,\cdots,j$ labels the copy of $j$ inside $\mathcal{A}$. The intertwiners defined above satisfy the following nice properties
    \begin{equation}
        J^{\vee}_{j',r'}\circ J_{j,r} = \delta_{j',j} \delta_{r',r} \mathbf{1}_{E^j}\,,\quad \sum_{j=0}^{\infty} \sum_{r=-j}^j J_{j,r} \circ J^{\vee}_{j,r} = \mathbf{1}_{\mathcal{A}}\,.
    \end{equation}

Before moving on, let us proof the two properties. For $J^{\vee}_{j',r'}\circ J_{j,r}$ we have
\begin{equation}
    J^{\vee}_{j',r'}\circ J_{j,r}|jm\rangle = \sqrt{d_j d_{j'}} \int d\mu(g) \sum_{m'=-j'}^{j'} D^{j'}_{m'r'}(g) \overline{D}^j_{mr}(g)|j'm'\rangle\,.
\end{equation}
Using the orthogonal relation
\begin{equation}
    \frac{1}{16\pi^2}\int_0^{2\pi} d\phi \int_0^{\pi} \sin \theta d\theta \int_{0}^{4\pi} d\psi \,\overline{D}_{mr}^j(\phi,\theta,\psi) D_{m'r'}^{j'}(\phi,\theta,\psi) = \frac{1}{d_j} \delta_{j,j'} \delta_{m,m'} \delta_{r,r'}\,,
\end{equation}
one has
    \begin{equation}
        J^{\vee}_{j',r'}\circ J_{j,r}|jm\rangle = \delta_{j,j'} \delta_{r,r'} |jm\rangle\,.
    \end{equation}
On the other hand, for $J_{j,r} \circ J^{\vee}_{j',r'}$ one has
    \begin{equation}
        \sum_{j=0}^{\infty}\sum_{r=-j}^j J_{j,r} \circ J^{\vee}_{j,r}\circ v_g=  \sum_{j=0}^{\infty}\sum_{r=-j}^j\sum_{m=-j}^j \int d\mu(h) d_j D_{mr}^j (g) \overline{D}_{mr}^j(h) v_h\,.
    \end{equation}
To proceed, notice that there exists a general relationship to the spin-weighted spherical harmonics $\,_rY_{j,m}(\theta,\phi)$
    \begin{equation}
        D^j_{mr}(\phi,\theta,\psi)=(-1)^r \sqrt{\frac{4\pi}{d_j}} \,_rY_{j,-m}(\theta,\phi) e^{-\ii r \psi}\,,
    \end{equation}
where the spin-weighted spherical harmonics satisfy
    \begin{equation}
        \sum_{j\geq |r|} \sum_{m=-j}^j\,_r\overline{Y}_{j,m}(\theta,\phi) \,_rY_{j,m}(\theta',\phi') = \delta(\phi-\phi') \delta(\cos\theta - \cos\theta')\,,
    \end{equation}
which implies
    \begin{equation}
    \begin{split}
        &\sum_{j=0}^{\infty} \sum_{r=-j}^j \sum_{m=-j}^j d_j D^j_{mr}(g) \overline{D}^j_{mr}(h)\\ 
        =& \sum_{r=-\infty}^{\infty} \sum_{j\geq |r|}  \sum_{m=-j}^j d_j D^j_{mr}(g) \overline{D}^j_{mr}(h)\\
        =&4\pi \delta(\phi - \phi') \delta(\cos \theta - \cos \theta') \sum_{r=-\infty}^{\infty} e^{-\ii r(\psi-\psi')}\,.
    \end{split}
    \end{equation}
Here the summation of $r$ has step $\frac{1}{2}$, therefore
    \begin{equation}
        \sum_{r=-\infty}^{\infty} e^{-\ii r (\psi - \psi')} = 4\pi \delta(\psi-\psi')\,,
    \end{equation}
and we have
\begin{equation}
    \sum_{j=0}^{\infty} \sum_{r=-j}^j \sum_{m=-j}^j d_j D^j_{mr}(g) \overline{D}^j_{mr}(h) = 16\pi^2 \frac{1}{\sin\theta} \delta(\phi-\phi') \delta(\theta-\theta') \delta(\psi-\psi')\,,
\end{equation}
where we use $\delta(\cos \theta - \cos \theta')=\delta(\theta-\theta')/|\sin \theta|$ and $\sin\theta \geq 0$ for $\theta\in [0,\pi]$. Therefore we have
\begin{equation}
    \sum_{j=0}^{\infty}\sum_{r=-j}^j J_{j,r} \circ J^{\vee}_{j,r}\circ v_g = v_g\,.
\end{equation}

\paragraph{Fourier kernels}

Let us compute the Fourier kernels based on \eqref{eq:Rep(G)_G_kernel}
\begin{equation}
        W_{j_1,j_2}^{J} (g,h) = \textrm{Tr}_{E^{j_1}\otimes E^{j_2}} \left[ P^{J}_{j_1,j_2}\circ \ \hat{U}(h^{-1}) \otimes \hat{U}(g)  \right]\,,
    \end{equation}
where the (signed) projector is
    \begin{equation}
        P^{J}_{j_1,j_2} = S_{j_1,j_2} \circ I^{\vee j_2,j_1}_{J} I^J_{j_1,j_2}\,,
    \end{equation}
and the action on the tensor basis gives
    \begin{equation}
       S_{j_1,j_2}\circ  I^{\vee j_2,j_1}_{J} I^J_{j_1,j_2} |j_1m_1,j_2m_2\rangle= \sum_{M,m'_1,m'_2} \overline{C}^{JM}_{j_1m_1,j_2m_2} C^{JM}_{j_2m_2',j_1m'_1} |j_1m'_1,j_2m'_2\rangle\,,
    \end{equation}
where we use $S_{j_1,j_2}|j_2m'_2,j_1 m'_1\rangle = |j_1m'_1,j_2m'_2\rangle$. In the tensor product basis, one can write the projector as the matrix
    \begin{equation}
        \langle j_1m'_1,j_2m'_2| P^J_{j_1,j_2} |j_1m_1,j_2m_2\rangle = (-1)^{j_1+j_2-J}\sum_{M} \overline{C}^{JM}_{j_1m_1,j_2m_2} C^{JM}_{j_1m'_1,j_2m_2'}\,,
    \end{equation}
and one finds $(P^J_{j_1,j_2})^{\dagger}=P^J_{j_1,j_2}$ is hermitian. Using the orthogonal relations \eqref{eq:CG_orthogonal_1}, one can show that 
    \begin{equation}
        P^J_{j_1,j_2} \times P^K_{j_1,j_2} = (-1)^{j_1+j_2-J}\delta_{J,K}P^K_{j_1,j_2}\,,
    \end{equation}
and the completeness reads
    \begin{equation}
        \sum_J (-1)^{j_1+j_2-J} P^J_{j_1,j_2} = \mathbf{1}_{V_{j_1}\otimes V_{j_2}}\,.
    \end{equation}
Taking the trace gives
    \begin{equation}
        \text{Tr}P^J_{i,j}= (-1)^{j_1+j_2-J}d_J\,.
    \end{equation}

For a compact Lie group, we can always rotate a group element $g$ to lie along $T/W$, where $T$ is a maximal torus and $W$ is the Weyl group. Then the centralizer group $C_G(g)=T$ is the same maximal torus for regular $g$. In the present case, we can choose $T$ to be generated by
    \begin{equation}
        T = \left\{ g_{\psi} | -2\pi \leq \psi < 2\pi\right\}\,,
    \end{equation}
which is diagonal. The Weyl group for $SU(2)$ is $\mathbb{Z}_2$, and we can simply restrict $\psi\in[0,2\pi)$ to parametrize $T/W$.

For simplicity, we will choose a representative of the commutating pair $(g,h)$ such that $g\in T/W$ and $h\in T$, and we write
    \begin{equation}
        g= \left(\begin{array}{cc}
            e^{\frac{i\psi_1}{2}} & 0 \\
            0 & e^{-\frac{i\psi_1}{2}} 
        \end{array} \right)\,,\quad h=\left(\begin{array}{cc}
            e^{\frac{i\psi_2}{2}} & 0 \\
            0 & e^{-\frac{i\psi_2}{2}} 
        \end{array} \right)\,,
    \end{equation}
with $0 \leq \psi_1<2\pi$ and $-2\pi \leq \psi_2 <2\pi$. Then the tensor basis $|j_1m_1,j_2m_2\rangle$ is the eigenvector of $\hat{U}(h^{-1}) \otimes \hat{U}(g)$
    \begin{equation}
        \hat{U}(h^{-1})\otimes \hat{U}(g)|j_1m_1,j_2m_2\rangle =  e^{\ii m_1 \psi_2} e^{-\ii m_2 \psi_1}|j_1m_1,j_2m_2\rangle\,,
    \end{equation}
and one has
    \begin{equation}
    \begin{split}
        &P^{J}_{j_1,j_2}\circ \ \hat{U}(h^{-1}) \otimes \hat{U}(g) |j_1m_1,j_2m_2\rangle\\
        =& \sum_{M,m'_1,m'_2} e^{\ii m_1 \psi_2} e^{-\ii m_2 \psi_1} \overline{C}^{JM}_{j_1m_1,j_2m_2} C^{JM}_{j_2m_2',j_1m'_1} |j_1m'_1,j_2m'_2\rangle\,.
    \end{split}
    \end{equation}
Therefore the kernel reads
    \begin{equation}
        W^J_{j_1,j_2}(g,h) = \sum_{M,m_1,m_2} e^{\ii m_1 \psi_2} e^{-\ii m_2 \psi_1} \overline{C}^{JM}_{j_1m_1,j_2m_2} C^{JM}_{j_2m_2,j_1m_1}\,.
    \end{equation}

We can write the defect partition function $Z_{j_1,j_2}^J$ as
    \begin{equation}
        Z_{j_1,j_2}^J = \int_{\mathcal{M}} d\mu(g,h) \, W_{j_1,j_2}^J(g,h) Z_{g,h}\,,
    \end{equation}
where $\mathcal{M}$ is the moduli space of commuting pair $(g,h)$
\begin{equation}
    \mathcal{M} = \{(g,h) | gh=hg \}\,,
\end{equation}
and $d\mu(g,h)$ is the measure induced from $G\times G$ onto $\mathcal{M}_{g,h}$, the submanifold of the commuting pair. However, $\mathcal{M}$ is not smooth. When $g$ is regular, then $h\in C_G(g) = T$, where $T$ is the Cartan torus. However, when $g$ approach the irregular elements of $G$, $C_G(g)$ will change discontinuously. For example, $C_G(e) = G$ is the whole group $G$. As mentioned at the beginning of this section, we will not try to give an explicit expression of $d\mu(g,h)$ all over $\mathcal{M}$, instead, we will consider the Haar measure $d\mu(g)d\mu(h)$ on $G\times G$ and regularize $\mathcal{M}$ using a smooth signed density function over $G\times G$ in the following. We will come to that later.

\paragraph{Inverse Fourier kernels}
    
Lastly, we discuss the inverse Fourier kernels related to gauging $\text{Rep}(SU(2))$. In particular, we need the expansion of multiplication and comultiplication with dressing
\begin{equation}
\begin{split}
    m_{(j_1,r_1),(j_2,r_2)}^{(J,R)} (y) =&  J^{\vee}_{J,R} \circ m\circ (R_{y}\otimes \text{id}_{\mathcal{A}})\circ   (J_{j_1,r_1} \otimes J_{j_2,r_2})\\
    =&M_{(j_1,r_1),(j_2,r_2)}^{(J,R)} (y) I_{j_1,j_2}^{J}\,,
\end{split}
\end{equation}
and
\begin{equation}
\begin{split}
m^{\vee (j_1,r_1),(j_2,r_2)}_{(J,R)}(x) =& (J^{\vee}_{j_1,r_1}\otimes J^{\vee}_{j_2,r_2})\circ \left( R_{x^{-1}}\otimes \text{id}_{\mathcal{A}} \right) \circ m^{\vee} \circ J_{J,R}\\
=&M^{\vee (j_1,r_1),(j_2,r_2)}_{(J,R)}(x) I^{\vee j_1,j_2}_{J}\,,
\end{split}
\end{equation}
and we shall evaluate $M$ and $M^{\vee}$ directly in the following.

For the multiplication, one has
    \begin{equation}
    \begin{split}
        &m^{(j_1,r_1),(j_2,r_2)}_{(J,R)}(y) |j_1m_1,j_2m_2\rangle\\
        =&J^{\vee}_{J,R} \circ m\circ (R_{y}\otimes \text{id}_{\mathcal{A}})\circ   (J_{j_1,r_1} \otimes J_{j_2,r_2})|j_1m_1,j_2m_2\rangle\\
        =&\sqrt{d_{j_1}d_{j_2}} \int d\mu(g)d\mu(h) \overline{D}^{j_1}_{m_1r_1}(g) \overline{D}^{j_2}_{m_2r_2}(h) J^{\vee}_{J,R} \circ m\circ (R_{y}\otimes \text{id}_{\mathcal{A}}) \circ (v_g \otimes v_h)\\
        =&\sqrt{d_{j_1}d_{j_2}} \int d\mu(g)d\mu(h) \overline{D}^{j_1}_{m_1r_1}(g) \overline{D}^{j_2}_{m_2r_2}(h) J^{\vee}_{J,R} \circ m \circ (v_{gy} \otimes v_h)\\
        =&\sqrt{d_{j_1}d_{j_2}} \int d\mu(g)d\mu(h) \overline{D}^{j_1}_{m_1r_1}(g) \overline{D}^{j_2}_{m_2r_2}(h) J^{\vee}_{J,R} \circ v_h \delta_{gy,h}\\
        =&\sqrt{d_{j_1}d_{j_2}} \int d\mu(g) \overline{D}^{j_1}_{m_1r_1}(g) \overline{D}^{j_2}_{m_2r_2}(gy) J^{\vee}_{J,R} \circ v_{gy}\\
        =&\sqrt{d_{j_1}d_{j_2}d_{J}}\int d\mu(g) \sum_{M=-J}^J \overline{D}^{j_1}_{m_1r_1}(g) \overline{D}^{j_2}_{m_2r_2}(gy) D_{MR}^{J}(gy) |JM\rangle\,.
    \end{split}
    \end{equation}
For simplicity, we will set $y=g_{\psi_y}$ lying along the Cartan torus, and one has
    \begin{equation}
        D_{m_2r_2}^{j_2}(gy) = D_{m_2r_2}^{j_2}(g) e^{-\ii r_2 \psi_y}\,, \quad D_{MR}^J(gy) = D_{MR}^J(g) e^{-\ii R\psi_y}\,.
    \end{equation}
The Kronecker product of Wigner D-matrices satisfy
    \begin{equation}
        D^{j_1}_{m_1r_1}(\phi,\theta,\psi) D^{j_2}_{m_2r_2}(\phi,\theta,\psi)=\sum_{J=|j_1-j_2|}^{|j_1+j_2|} C_{j_1m_1,j_2m_2}^{J(m_1+m_2)} C_{j_1r_1,j_2r_2}^{J(r_1+r_2)} D_{(m_1+m_2)(r_1+r_2)}^{J}(\phi,\theta,\psi)\,,
    \end{equation}
and we proceed with
    \begin{equation}
    \begin{split}
        &m^{(j_1,r_1),(j_2,r_2)}_{(J,R)}(y) |j_1m_1,j_2m_2\rangle\\
        =&\sqrt{d_{j_1}d_{j_2}d_{J}}\int d\mu(g) \sum_{M=-J}^J e^{-\ii (R-r_2)\psi_y}\overline{D}^{j_1}_{m_1r_1}(g) \overline{D}^{j_2}_{m_2r_2}(g) D_{MR}^{J}(g) |JM\rangle\\
        =&\sqrt{d_{j_1}d_{j_2}d_{J}}\sum_{J'=|j_1-j_2|}^{|j_1+j_2|} \sum_{M=-J}^{J} e^{-\ii (R-r_2)\psi_y}\overline{C}_{j_1m_1,j_2m_2}^{J'(m_1+m_2)} \overline{C}_{j_1r_1,j_2r_2}^{J'(r_1+r_2)} 
        \\ &\times \int d\mu(g) \overline{D}_{(m_1+m_2)(r_1+r_2)}^{J'}(g)  D_{MR}^{J}(g) |JM\rangle\\
        =&\sqrt{d_{j_1}d_{j_2}d_{J}}\sum_{J'=|j_1-j_2|}^{|j_1+j_2|} \sum_{M=-J}^J e^{-\ii (R-r_2)\psi_y}\overline{C}_{j_1m_1,j_2m_2}^{J'(m_1+m_2)} \overline{C}_{j_1r_1,j_2r_2}^{J'(r_1+r_2)} \frac{1}{d_J} \delta_{J,J'} \delta_{m_1+m_2,M} \delta_{r_1+r_2,R}|JM\rangle\\
        =&\sqrt{\frac{d_{j_1}d_{j_2}}{d_J}} e^{-\ii r_1 \psi_y}\overline{C}_{j_1m_1,j_2m_2}^{J(m_1+m_2)} \overline{C}_{j_1r_1,j_2r_2}^{J(r_1+r_2)} \delta_{r_1+r_2,R} |J(m_1+m_2)\rangle
    \end{split}
    \end{equation}
Since the intertwiner is
    \begin{equation}
        I_{j_1,j_2}^{J}|j_1m_1,j_2m_2\rangle = \sum_M\overline{C}_{j_1m_1,j_2m_2}^{JM} |JM\rangle = \overline{C}_{j_1m_1,j_2m_2}^{J(m_1+m_2)} |J(m_1+m_2)\rangle\,,
    \end{equation}
then one reads
    \begin{equation}
        m^{(j_1,r_1),(j_2,r_2)}_{(J,r_1+r_2)} = \sqrt{\frac{d_{j_1}d_{j_2}}{d_J}}e^{-\ii r_1 \psi_y}\overline{C}_{j_1r_1,j_2r_2}^{J(r_1+r_2)} I_{j_1,j_2}^{J}\,,
    \end{equation}
and the coefficient is
    \begin{equation}
        M^{(j_1,r_1),(j_2,r_2)}_{(J,r_1+r_2)}=\sqrt{\frac{d_{j_1}d_{j_2}}{d_J}}e^{-\ii r_1 \psi_y}\overline{C}_{j_1r_1,j_2r_2}^{J(r_1+r_2)}\,.
    \end{equation}

Similarly, for comultiplication one has
\begin{equation}
\begin{split}
&m^{\vee (j_1,r_1),(j_2,r_2)}_{(J,R)}(x)\circ |JM\rangle \\
=& (J^{\vee}_{j_1,r_1}\otimes J^{\vee}_{j_2,r_2})\circ \left( R_{x^{-1}}\otimes \text{id}_{\mathcal{A}} \right) \circ m^{\vee} \circ J_{J,R}|JM\rangle\\
=&\sqrt{d_J} \int d\mu(g) \overline{D}_{MR}^J(g)(J^{\vee}_{j_1,r_1}\otimes J^{\vee}_{j_2,r_2})\circ \left( R_{x^{-1}}\otimes \text{id}_{\mathcal{A}} \right) \circ m^{\vee}\circ v_g\\
=&\sqrt{d_J} \int d\mu(g) \overline{D}_{MR}^J(g)(J^{\vee}_{j_1,r_1}\otimes J^{\vee}_{j_2,r_2})\circ \left( R_{x^{-1}}\otimes \text{id}_{\mathcal{A}} \right) \circ v_g\otimes  v_g\\
=&\sqrt{d_J} \int d\mu(g) \overline{D}_{MR}^J(g)(J^{\vee}_{j_1,r_1}\otimes J^{\vee}_{j_2,r_2}) \circ v_{gx^{-1}}\otimes  v_g\\
=&\sqrt{d_J d_{j_1}d_{j_2}}\int d\mu(g) \sum_{m_1=-j_1}^{j_1} \sum_{m_2=-j_2}^{j_2}\overline{D}_{MR}^J(g) D_{m_1r_1}^{j_1}(gx^{-1}) D_{m_2r_2}^{j_2}(g) |j_1m_1,j_2m_2\rangle\,.
\end{split}
\end{equation}
Setting $x=g_{\psi_x}$ lying along the Cartan torus, and use the Kronecker product formula gives
\begin{equation}
\begin{split}
&m^{\vee (j_1,r_1),(j_2,r_2)}_{(J,R)}(x)\circ |JM\rangle \\
=&\sqrt{d_J d_{j_1}d_{j_2}}\int d\mu(g) \sum_{m_1=-j_1}^{j_1} \sum_{m_2=-j_2}^{j_2} \overline{D}_{MR}^J(g) D_{m_1r_1}^{j_1}(gx^{-1}) D_{m_2r_2}^{j_2}(g) |j_1m_1,j_2m_2\rangle\\
=&\sqrt{d_J d_{j_1}d_{j_2}} \sum_{m_1=-j_1}^{j_1} \sum_{m_2=-j_2}^{j_2} \sum_{J'=|j_1-j_2|}^{|j_1+j_2|} e^{\ii r_1 \psi_x}C^{J'(m_1+m_2)}_{j_1m_1,j_2m_2} C^{J'(r_1+r_2)}_{j_1r_1,j_2r_2}\\
&\times \int d\mu(g)\overline{D}_{MR}^J(g) D_{(m_1+m_2)(r_1+r_2)}^{J'}(g)  |j_1m_1,j_2m_2\rangle\\
=&\sqrt{\frac{d_{j_1}d_{j_2}}{d_J}} \sum_{m_1=-j_1}^{j_1} \sum_{m_2=-j_2}^{j_2} \sum_{J'=|j_1-j_2|}^{|j_1+j_2|}e^{\ii r_1 \psi_x} C^{J'(m_1+m_2)}_{j_1m_1,j_2m_2} C^{J'(r_1+r_2)}_{j_1r_1,j_2r_2}\delta_{J,J'} \delta_{M,m_1+m_2}\delta_{R,r_1+r_2}|j_1m_1,j_2m_2\rangle\\
=&\sqrt{\frac{d_{j_1}d_{j_2}}{d_J}} \sum_{m_1=-j_1}^{j_1} \sum_{m_2=-j_2}^{j_2} e^{\ii r_1 \psi_x} C^{J(m_1+m_2)}_{j_1m_1,j_2m_2} C^{J(r_1+r_2)}_{j_1r_1,j_2r_2}\delta_{M,m_1+m_2}\delta_{R,r_1+r_2}|j_1m_1,j_2m_2\rangle
\end{split}
\end{equation}
Compared to the action of cointertwiner
    \begin{equation}
        I^{\vee j_1,j_2}_{J}|JM\rangle = \sum_{m_1,m_2} C_{j_1m_1,j_2m_2}^{JM} |j_1m_1,j_2m_2\rangle=\sum_{m_1,m_2} C_{j_1m_1,j_2m_2}^{J(m_1+m_2)} |j_1m_1,j_2m_2\rangle\,,
    \end{equation}
we have
    \begin{equation}
        m^{\vee (j_1,r_1),(j_2,r_2)}_{(J,r_1+r_2)} = \sqrt{\frac{d_{j_1}d_{j_2}}{d_J}} e^{\ii r_1 \psi_x}C^{J(r_1+r_2)}_{j_1r_1,j_2r_2} I^{\vee j_1,j_2}_J\,,
    \end{equation}
and the coefficient is 
    \begin{equation}
        M^{\vee (j_1,r_1),(j_2,r_2)}_{(J,r_1+r_2)} = \sqrt{\frac{d_{j_1}d_{j_2}}{d_J}} e^{\ii r_1 \psi_x}C^{J(r_1+r_2)}_{j_1r_1,j_2r_2}\,.
    \end{equation}

Combined $M$ and $M^{\vee}$, the inverse kernel for $Z_{e,e}$ in \eqref{eq:gauge_Rep_tivial_coefficient} is given by\footnote{Compared to \eqref{eq:gauge_Rep_tivial_coefficient}, we do not include the $|\mathcal{A}|=\dim \mathcal{A}$, which is divergent for Lie groups.}
\begin{equation}
        C_{j_1,j_2}^{J} = \sum_{r_1,r_2,R} M^{\vee (j_2,r_2),(j_1,r_1)}_{(J,R)} M_{(j_1,r_1),(j_2,r_2)}^{(J,R)} = \frac{d_{j_1}d_{j_2}}{d_J} \sum_{r_1,r_2,R} e^{\ii (r_2\psi_x -r_1\psi_y)} C^{JR}_{j_2r_2,j_1r_1}\overline{C}_{j_1r_1,j_2r_2}^{JR} \,.
    \end{equation}
Using the relation
    \begin{equation}
        C_{j_2m_2,j_1m_1}^{JM} = (-1)^{j_1+j_2-J} C_{j_1m_1,j_2m_2}^{JM}\,,
    \end{equation}
one has
    \begin{equation}
    \begin{split}
        C_{j_1,j_2}^{J}(g,h) =& \frac{d_{j_1}d_{j_2}}{d_J} (-1)^{j_1+j_2-J}\sum_{R,r_1,r_2} e^{\ii (r_2\psi_x -r_1\psi_y)} C^{JR}_{j_1r_1,j_2r_2}\overline{C}_{j_1r_1,j_2r_2}^{JR}\,.
    \end{split}
    \end{equation}
Compared to the Fourier kernel, one finds
    \begin{equation}\label{eq:SU(2)_Inverse_Kernel}
        C_{j_1,j_2}^J(g,h) = \frac{d_{j_1}d_{j_2}}{d_J} \overline{W}_{j_1j_2}^J(g,h)\,.
    \end{equation}
Although we assume $g,h$ lying along the Cartan torus, the relation actually holds for arbitrary commuting $g,h$. 
Recall that
\begin{equation}
        W_{j_1,j_2}^{J} (g,h) = \textrm{Tr}_{E^{j_1}\otimes E^{j_2}} \left[ P^{J}_{j_1,j_2}\circ \ \hat{U}(h^{-1}) \otimes \hat{U}(g)  \right]\,,
    \end{equation}
we then have
    \begin{equation}
        C^J_{j_1,j_2}(g,h) = \frac{d_{j_1} d_{j_2}}{d_J} \textrm{Tr}_{E^{j_1}\otimes E^{j_2}} \left[ P^{J}_{j_1,j_2}\circ \ \hat{U}(h) \otimes \hat{U}(g^{-1})  \right]\,,
    \end{equation}
where we use the fact that $P^J$ is hermitian.

As a final remark, in the present case we have $P^J_{j_1,j_2} = S^{-1}_{j_1,j_2} P^J_{j_2,j_1} S_{j_1,j_2}$ due to the exchange relation of the CG-coefficient, therefore we also have
    \begin{equation}
        C^J_{j_1,j_2}(g,h) = \frac{d_{j_1} d_{j_2}}{d_J} \textrm{Tr}_{E^{j_2}\otimes E^{j_1}} \left[ P^{J}_{j_2,j_1}\circ \ \hat{U}(g^{-1}) \otimes \hat{U}(h)  \right] = \frac{d_{j_1} d_{j_2}}{d_J}W^J_{j_2,j_1} (h,g)\,,
    \end{equation}
which agrees with \eqref{eq:gauge-fixed-inverse-kernel} in the finite group case.

We will show that gauging the $\text{Rep}(SU(2))$ symmetry indeed recovers the original partition function $Z_{g,h}$. To compute that, we will follow the same method in \ref{sec:nice_gauge} and first prove the following identity. For a given pair of spin $(j_1,j_2)$, consider the representation matrix in the tensor product basis $R(g) \equiv U^{j_1}(g)\otimes U^{j_2}(g)$. Then for any endomorphism $B$ of $V_{j_1}\otimes V_{j_2}$, one has
    \begin{equation}\label{eq:The_Identity}
        \Lambda(B)\equiv \int d\mu(k) R(k) B R(k^{-1}) = \sum_J \frac{\text{Tr}(P^J B)}{d_J} P^J\,,
    \end{equation}
where $P^J=P^J_{j_1,j_2}$ is the projection matrix, where we omit the $(j_1,j_2)$ label. Notice that the LHS commute with all $R(g)$ since the Haar measure is left invariant. In the coupled basis, each $V_J \in V_{j_1}\otimes V_{j_2}$ appears only once, we can write the $\Lambda(B)$ matrix in terms of the block form $\Lambda(B)_{IJ}$ with $|j_1-j_2| \leq I,J\leq |j_1+j_2|$. Each block $\Lambda(B)_{IJ} : V_J\rightarrow V_I$ is a intertwiner between two irreducible representations. By Schur lemma, we have
    \begin{equation}
        \Lambda(B)_{IJ} = \left\{\begin{array}{l}
        0\,,\quad I\neq J\,,\\
        \lambda_J(B) \mathbf{1}_{V_J}\,,\quad I=J\,,
        \end{array} \right.
    \end{equation}
for some independent scalar $\lambda_J(B)$. As a result, we can expand such $\Lambda(B)$ using the projection matrix as
    \begin{equation}
        \Lambda(B) = \sum_{K} \alpha_K(B) (-1)^{j_1+j_2-K}P^K\,,
    \end{equation}
where $P^J$ commutes with $R(g)$ by definition. Multiply $P^J$ and take a trace on both side, one has
    \begin{equation}
        \text{Tr}(P^J B)= (-1)^{j_1+j_2-J}\alpha_J d_J \quad \rightarrow \quad \alpha_J = (-1)^{j_1+j_2-J} \frac{\text{Tr}(P^J B)}{d_J}\,,
    \end{equation}
and we have proved the identity. Multiply $A$ on both side of \eqref{eq:The_Identity} and take the trace, we have
    \begin{equation}\label{eq:The_Identity_2}
    \int d\mu(k)  \text{Tr} (R(k)BR(k^{-1})A) = \sum_J\frac{\text{Tr}(P^J B) \text{Tr}(P^JA)}{d_J}\,.
    \end{equation}

Let us consider re-gauging the $\text{Rep}(SU(2))$
    \begin{equation}
    \begin{split}
        \widetilde{Z}/\text{Rep}(SU(2))(g',h') =& \frac{1}{\dim \mathcal{A}}\sum_{j_1,j_2,J} C_{j_1,j_2}^J(g',h') \widetilde{Z}_{j_1,j_2}^J\\
        =& \frac{1}{\dim \mathcal{A}}\sum_{j_1,j_2,J} \int_{\mathcal{M}} d\mu(g,h) C_{j_1,j_2}^J(g',h') W^J_{j_1,j_2}(g,h) Z_{g,h}\,,
    \end{split}
    \end{equation}
where $\dim \mathcal{A} = \sum_{j} d_j^2$ is divergent and need regularization. We have
    \begin{equation}
    \begin{split}
        &\sum_{j_1,j_2,J} C_{j_1,j_2}^J(g',h') W^J_{j_1,j_2}(g,h)\\
        =& \sum_{j_1,j_2,J}\frac{d_{j_1} d_{j_2}}{d_J} \textrm{Tr}_{E^{j_1}\otimes E^{j_2}} \left[ P^{J}_{j_1,j_2}\circ \ \hat{U}(h') \otimes \hat{U}(g'{}^{-1})  \right] \textrm{Tr}_{E^{j_1}\otimes E^{j_2}} \left[ P^{J}_{j_1,j_2}\circ \ \hat{U}(h^{-1}) \otimes \hat{U}(g)  \right]\,.
    \end{split}
    \end{equation}
Setting
    \begin{equation}
        B=\hat{U}(h') \otimes \hat{U}(g'{}^{-1})\,,\quad A=\hat{U}(h^{-1}) \otimes \hat{U}(g)\,,
    \end{equation}
in \eqref{eq:The_Identity_2}, we have
    \begin{equation}
    \begin{split}
        &\sum_{j_1,j_2,J} C_{j_1,j_2}^J(g',h') W^J_{j_1,j_2}(g,h)\\
        =& \sum_{j_1,j_2} d_{j_1}d_{j_2} \int d\mu(k) \text{Tr}\left( R(k) \circ (\hat{U}(h') \otimes \hat{U}(g'{}^{-1})) \circ R(k^{-1}) \circ (\hat{U}(h^{-1}) \otimes \hat{U}(g))\right)\\
        =&\sum_{j_1,j_2} d_{j_1}d_{j_2} \int d\mu(k) \text{Tr} \left( \hat{U}(k h' k^{-1} h^{-1}) \otimes \hat{U}(k {g'}^{-1} k^{-1}g)\right)\\
        =& \sum_{j_1,j_2} d_{j_1}d_{j_2} \int d\mu(k) \chi_{j_1}(k h' k^{-1} h^{-1}) \chi_{j_2}(k {g'}^{-1} k^{-1}g)\,.
    \end{split}
    \end{equation}
Since $d_j=\chi_j(e)$,  using the orthogonal identity for characters (see, for example, the paragraph above (4.15) in \cite{Witten:1992xu}),
    \begin{equation}
        \sum_{\rho} \chi_{\rho}(g) \chi_{\rho}(h^{-1}) = \int dk \delta_{gkh^{-1}k^{-1},e}\,,
    \end{equation}
we have separately
    \begin{equation}
        \sum_{j_1}d_{j_1} \chi_{j_1}(k h' k^{-1} h^{-1})  = \delta_{h,kh'k^{-1}}\,,\quad \sum_{j_2}d_{j_2}\chi_{j_2}(k {g'}^{-1} k^{-1}g) = \delta_{g,kg'k^{-1}}\,.
    \end{equation}
Similar to the finite group, the Dirac $\delta$-function on $SU(2)$ will restrict $(g,h)\in \text{Conj}(g',h')$, and we have 
    \begin{equation}
        \widetilde{Z}/\text{Rep}(SU(2))(g',h') = \frac{1}{\mathcal{N}}\int_{\mathcal{M}} d\mu(g,h) \int d\mu(k) \delta_{h,kh'k^{-1}} \delta_{g,kg'k^{-1}} Z_{g,h}\,,
    \end{equation}
where $\mathcal{N}$ is a normalization factor which will be determined in the following.

\paragraph{Measure on the commuting pair moduli space $\mathcal{M}$}

To proceed, we need to define the measure $d\mu(g,h)$ on the submanifold $\mathcal{M}$. One straightforward way is to define it as
    \begin{equation}
        \int_{\mathcal{M}}d\mu(g,h) = \Lambda \int_{G\times G} d\mu(g) d\mu(h) \delta_{ghg^{-1}h^{-1},e}\,,
    \end{equation}
where we relax the integral from $\mathcal{M}$ to $G\times G$, and insert $\delta_{ghg^{-1}h^{-1},e}$, $\Lambda$ is a normalization constant. Therefore $d\mu(g,h)$ is the Haar measure on $G\times G$ induced onto $\mathcal{M}$ with the proper Jacobian. Moreover, we will replace $\delta_{ghg^{-1}h^{-1},e}$ using the character property as
    \begin{equation}
        \delta_{ghg^{-1}h^{-1},e} = \sum_{j} d_j\chi_j(ghg^{-1}h^{-1})\,.
    \end{equation}
Introduce the integer cutoff of spin $N\gg 1$ and regularize
    \begin{equation}
        \Delta_N(g,h) = \sum_{j=0,\frac{1}{2},\cdots,N} d_j \chi_j(ghg^{-1}h^{-1})\, 
    \end{equation}
and rewrite the regularized measure as
    \begin{equation}
        \frac{\Lambda_N}{\mathcal{N}_N}\int_{G\times G} d\mu(g) d\mu(h) \Delta_N(g,h) \,,
    \end{equation}
where we will also regularize the two normalization constant as follows. 

First of all, for any character $\chi_j(ghg^{-1}h^{-1})$, we have
\begin{equation}
\begin{split}
    &\int_{G\times G} d\mu(g) d\mu(h) \chi_j(ghg^{-1}h^{-1})\\
    =& \int_G d\mu(h)  \text{Tr}_{j}\left( \int_G  d\mu(g) U(g) U(h) U(g^{-1})\right) U(h^{-1})\\
    =&\int_G d\mu(h) \text{Tr}_j \left(\frac{\chi_j(h)}{d_j} U(h^{-1}) \right)\\
    =&\int_G d\mu(h) \frac{\chi_j(h) \chi_j(h^{-1})}{d_j}=\frac{1}{d_j}\,,
\end{split}
\end{equation}
where we use the identity
\begin{equation}
    \int d \mu(h) \chi_{\rho}(h) \chi_{\nu}(h^{-1}) = \delta_{\rho,\nu}\,.
\end{equation}
and therefore
    \begin{equation}
        \int_{G\times G} d\mu(g) d\mu(h) \Delta_N(g,h) = 2N+1\,.
    \end{equation}
In order to have a measure which gives finite integral in the large $N$ limit, we will choose
    \begin{equation}
        \Lambda_N = \frac{1}{2N+1}\,,
    \end{equation}
so that
    \begin{equation}
        \lim_{N\rightarrow \infty} \Lambda_N\int_{G\times G} d\mu(g) d\mu(h)\Delta_N(g,h)=1\,.
    \end{equation}
    
Actually, the reason that $\lim_{N\rightarrow \infty}\int_{G\times G} d\mu(g) d\mu(h) \Delta_N(g,h) \rightharpoonup \infty$ is simple. For $gh=hg$ with $g,h\in SU(2)$, if we fix $g$ to be regular, then $h\in C_{SU(2)}(g) = U(1)$, and therefore the regular part of the manifold $\mathcal{M}$ is a four-dimensional subspace in $SU(2)\times SU(2)$. On the other hand, $\delta_{ghg^{-1}h^{-1},e}$ is a three-dimensional Dirac $\delta$-function over $SU(2)$. For any $h_0$ satisfying $gh_0g^{-1}h_0^{-1}=e$, we can parametrize $h=h_0e^{i\psi H} e^{i\theta_+J_++i\theta_-J_-}$ near $h_0$ with the local coordinates $\psi,\theta_{\pm}\ll 1$. Here $H$ is chosen as the Cartan generator of the centralizer $U(1)$, and $J_{\pm}$ are the two root vectors. Then we can decompose locally
    \begin{equation}
        \delta_{ghg^{-1}h^{-1},e}= \delta^{H}(0) \delta^{J_+}(\theta_+)\delta^{J_-}(\theta_-)\,.
    \end{equation}
The $\delta^{J_+}(\theta_+)\delta^{J_-}(\theta_-)$ will restrict $SU(2)\times SU(2)$ to $\mathcal{M}$, but the first Dirac $\delta$-function is divergent since $\psi$ is a flat coordinate. Therefore the divergence of the integral comes from the singular Dirac $\delta$-function from the $U(1)$ part $\delta^{U(1)}_{e,e}$. On the other hand, notice that we can write $1/\Lambda_N$ as
    \begin{equation}
        \frac{1}{\Lambda_N}=2N+1 = \lim_{\psi\rightarrow 0} \sum_{n=-N}^N e^{in \psi}\,,
    \end{equation}
where $e^{in \psi}$ is the character of the $U(1)$ and we have $\sum_{n=-\infty}^{+\infty} e^{in\psi} \sim \delta^{U(1)}_{e^{i\psi},e}$. Therefore $1/\Lambda_N$ is the cutoff scaling assigned to the flat-direction divergence.

Then we can integrate $d\mu(g)$ and $d\mu(h)$ first and get
    \begin{equation}
        \widetilde{Z}/\text{Rep}(SU(2))(g',h') = \frac{1}{\mathcal{N_N}}\frac{\Delta_N(g',h')}{2N+1} Z_{g',h'}\,,
    \end{equation}
where we used $Z_{kg'k^{-1},kh'k^{-1}}=Z_{g',h'}$ and $\int d\mu(k)=1$. Notice that $\Delta_N(g,h) = \sum_{j=0}^{N} d_j^2$ if $gh=hg$, in order to get a finite limit, we will further choose
    \begin{equation}
        \mathcal{N}_N = \frac{\sum_{j=0}^{N} d_j^2}{2N+1}\,,
    \end{equation}
to make $\frac{1}{\mathcal{N_N}}\frac{\Delta_N(g,h)}{2N+1}=1$ for $gh=hg$, and decay rapidly as $(g,h)$ move away from $\mathcal{M}$. Sending the regulator $N\rightarrow \infty$, we then have
    \begin{equation}
        \widetilde{Z}/\text{Rep}(SU(2))(g',h') = \left\{ \begin{array}{l}
            Z_{g',h'}\,,\quad \text{if}\quad g'h'=h'g'\,,\\
            0\,,\quad \text{others}\,,
        \end{array} \right.
    \end{equation}
as expected.

Finally, let us evaluate the $SU(2)$ gauging
    \begin{equation}
        Z/SU(2) = \int_{\mathcal{M}} d\mu(g,h) Z_{g,h}\,,
    \end{equation}
with $\mu(g,h)$ regularized as above. We have
    \begin{equation}
        Z/SU(2) = \lim_{N\rightarrow \infty} \frac{1}{2N+1}\int_{G\times G} d\mu(g) d\mu(h)\Delta_N(g,h) Z_{g,h}\,.
    \end{equation}
Use the Weyl integration formula,
    \begin{equation}
        \int_G d\mu(h) f(h) = \frac{1}{|W|} \int_T |\Delta(t)|^2 dt \int_{G/T} d(hT) f(hth^{-1})\,, 
    \end{equation}
where 
    \begin{equation}
        |\Delta(t)|^2 = \det \left(\text{Ad}_{t}-I \right)|_{\mathfrak{g}/\mathfrak{t}} = \prod_{\alpha>0} (1-e^{i\alpha(t)})(1-e^{-i\alpha(t)})\,,
    \end{equation}
and $G/T$ is the coset of $T$ in $G$. We can write the integral as
    \begin{equation}
        \begin{split}
            &\lim_{N\rightarrow \infty} \frac{1}{|W|(2N+1)}\int_{G\times G} d\mu(g) d\mu(h)\Delta_N(g,h) Z_{g,h}\\
            =&\lim_{N\rightarrow \infty} \frac{1}{|W|(2N+1)}\int_{G} d\mu(g) \int_{T}|\Delta(t)|^2 dt \int_{G/T} d(hT)\Delta_N(g,hth^{-1}) Z_{g,hth^{-1}}\,.
        \end{split}
    \end{equation}
Shifting $g\rightarrow hgh^{-1}$ and assuming the Haar measure $d\mu(g)$ is invariant under both left and right action, which is true for all semisimple Lie groups. Moreover, one has
    \begin{equation}
        \Delta_N(hgh^{-1},hth^{-1})=\Delta_N(g,t)\,,\quad Z_{hgh^{-1},hth^{-1}}=Z_{g,t}\,.
    \end{equation}
Normalize $\int_{G/T} d(hT)=1$, the integral is then
    \begin{equation}
        Z/SU(2) = \lim_{N\rightarrow \infty} \frac{1}{|W|(2N+1)}\int_{G} d\mu(g) \int_{T}|\Delta(t)|^2 dt \Delta_N(g,t) Z_{g,t}\,.
    \end{equation}

Now we take the large $N$ limit with $\lim_{N\rightarrow \infty} \Delta_N(g,t) = \delta_{gtg^{-1}t^{-1},e}$. Since $t\in T$ and we assume $t$ is regular, one should also have $g\in T$ for $\delta_{gtg^{-1}t^{-1},e}$ to be non-vanishing. Let us expand
    \begin{equation}
        g=s e^{X }\,, \quad s\in T\,,\quad X\in \mathfrak{g}/\mathfrak{t}\,,
    \end{equation}
where $X$ is an infinitesimal Lie algebra and $\mathfrak{g}/\mathfrak{t}$ is the transverse space $\mathfrak{t}^{\bot}$ identified by a standard choice of Killing metric. Then we have
    \begin{equation}
        gtg^{-1}t^{-1} = s e^{X}te^{-X}t^{-1}s^{-1} \sim  \exp \left(\text{Ad}_s \circ (I-\text{Ad}_{t})\circ X +\mathcal{O}(X^2)\right)\,,
    \end{equation}
and we can factorize the Dirac $\delta$-function locally as
    \begin{equation}
        \delta_{gtg^{-1}t^{-1},e} \sim \frac{1}{|\det (I-\text{Ad}_t)|_{\mathfrak{g}/\mathfrak{t}}|} \delta^{\mathfrak{t}}(0) \delta^{\mathfrak{g}/\mathfrak{t}}(X)\,,
    \end{equation}
where $\delta^{\mathfrak{g}/\mathfrak{t}}(\text{Ad}_s\circ X)=\delta^{\mathfrak{g}/\mathfrak{t}}(X)$ since the positive and negative roots cancel with each other. Notice that the Jacobian $\det (I-\text{Ad}_t)|_{\mathfrak{g}/\mathfrak{t}}$ exactly cancel the determinant $|\Delta(t)|^2$ in the Weyl integration formula, and the singular Dirac $\delta$-function $\delta^{\mathfrak{t}}(0)$ also regularize to $2N+1$ and compensate the $\frac{1}{2N+1}$ factor as mentioned above. On the other hand, the Haar measure $ds^2 \sim \text{Tr}(gdg^{-1} gdg^{-1})$ is expanded near the locus as
    \begin{equation}
        ds^2 \sim \text{Tr} \left[( sds^{-1} -\text{Ad}_s\circ dX)( sds^{-1} -\text{Ad}_s\circ dX)\right] + \mathcal{O}(X)
    \end{equation}
where we use $gdg^{-1} \sim sds^{-1} -\text{Ad}_s\circ dX+\mathcal{O}(X)$. Therefore for the standard choice of Killing metric, the Haar measure is locally $d\mu(g)=J(s,X) ds dX$ with $J(s,0)=1$.

Gather everything together and use $\int dX \delta^{\mathfrak{g}/\mathfrak{t}}(X) f(X)=f(0)$, we then have
    \begin{equation}
        Z/SU(2) \sim \frac{1}{|W|} \int_{T\times T} ds dt Z_{s,t}\,,
    \end{equation}
which reduces to a uniform integral along two Cartan tori $T\times T$, and is agree with the gauging in \cite{Gaberdiel:2011aa} up to some finite normalization constant\footnote{In \cite{Gaberdiel:2011aa}, the authors consider flat gauging of $SO(3)$ instead of $SU(2)$. Nevertheless, we can extend $SO(3)$ to the double cover $SU(2)$ providing the $\mathbb{Z}_2$ center acts trivially. }.

Lastly, we would like to comment that when we apply the Weyl integration formula to $h$ and expand $g$ along the commuting pair locus, we have assumed both $h$ and $g$ are regular, and omit the irregular region which has measure zero. The calculation determines only the regular-stratum contribution and does not control possible regulator-dependent contributions from singular strata.

\section{Conclusion}

In this work we have constructed a direct Fourier transform between partition
functions of a two-dimensional theory $\mathfrak{T}_G$ in flat $G$
backgrounds and those of its orbifold
$\widetilde{\mathfrak{T}}_{\text{Rep}(G)}=\mathfrak{T}_G/G$ in resolved
$\text{Rep}(G)$ defect backgrounds.  For the fixed-cut torus networks
considered here, the forward kernel is obtained by transporting Wilson lines
through a commuting pair of holonomies and contracting the resulting
representation matrices with the intertwiners at the two junctions.  The
fusion channel and all junction multiplicities therefore remain explicit, and
the coefficients are computed locally from the representation data of $G$
without using modular covariance as an input.  We derived the inverse
transform from the regular Frobenius algebra
$\mathcal{A}=\text{Fun}(G)$ by pairing its irreducible embeddings with dual
projections and dressing its multiplication and co-multiplication by commuting right-regular actions.  These dressings resolve the simultaneous-conjugacy orbits of flat holonomies and realize the regauging relation sector by sector. We worked out both transforms for the multiplicity-free examples $S_3$, $D_4$, and $Q_8$, and for $A_4$, where the two-dimensional junction space in $Z\otimes Z\to Z$ is retained throughout the calculation.  In every finite example, composing the inverse with the forward transform gives the identity on the physical space of $G$-backgrounds, while the opposite composition is a projector when the resolved defect partition functions form a redundant set. We further developed the compact-group counterpart for $\text{Rep}(SU(2))$, using Clebsch-Gordan coefficients for the junction data, Peter-Weyl matrix elements for the regular-algebra decomposition, and Haar integrals over commuting holonomies to obtain the forward and inverse kernels. On the regular stratum, this
construction reproduces the expected Cartan torus form of flat gauging at the
level of partition functions, up to an overall normalization.

There are several natural directions to extend this construction.
One is to incorporate discrete torsion and other twisted gaugings.  Stacking
with a $G$-SPT phase changes the topological weights assigned to flat
backgrounds, and it would be useful to determine how the resulting junction
phases deform both Fourier kernels and the compatible Frobenius-algebra data
\cite{Dijkgraaf:1989pz,Chen:2012ctz}.
A second direction is to go beyond gauging the full group $G$ and the regular
algebra of $\text{Rep}(G)$.  Gauging a subgroup $H\subset G$, or condensing a
different separable Frobenius algebra object in $\text{Rep}(G)$, should lead
to transforms between a larger family of orbifold phases and make their
relations under sequential gauging explicit
\cite{Bhardwaj:2017xup,Gaiotto:2020iye,Diatlyk:2023fwf,Chen:2024ulc}.  More
generally, one can ask for partition functions of generalized orbifolds of
theories with an arbitrary non-invertible symmetry category, further dressed
by additional topological defect lines and their junctions.  The resulting
symmetry-resolved sectors and the action of the remaining defect lines are
naturally expected to be organized by an appropriate generalized tube
algebra, suggesting a categorical extension of the present Fourier transform
to symmetry-resolved partition functions
\cite{Choi:2024tri}.

\section*{Acknowledgments}
J.~Chen is supported by the Fujian Provincial Natural Science Foundation of China (No.2025J01004), and the National Natural Science Foundation of China (Grants No.12247103). QJ is supported by National Research Foundation of Korea (NRF) Grant No. RS-2024-00405629 and Jang Young-Sil Fellow Program at the Korea Advanced Institute of Science and Technology.

\bibliographystyle{JHEP}
\bibliography{biblio.bib}

\end{document}